\documentclass[%
 reprint,
superscriptaddress,
nofootinbib,
 amsmath,amssymb,
 aps,
]{revtex4-1}

\usepackage{graphicx}% Include figure files
\usepackage{dcolumn}% Align table columns on decimal point
\usepackage{bm}% bold math
\usepackage{lipsum}
\usepackage{amsmath,amssymb}
\usepackage{caption}
\usepackage{subcaption}
\usepackage{comment}
\usepackage{verbatim}
\usepackage[left=2cm, right=2cm, top=2cm]{geometry}

\usepackage[utf8]{inputenc}
\usepackage{epigraph}
\usepackage{siunitx}
\usepackage[english]{babel}
\usepackage{listings}
\usepackage{graphicx}
\usepackage[utf8]{inputenc}
\usepackage{listings}
\usepackage{matlab-prettifier}
\usepackage{xcolor}
\usepackage{amsfonts}
\usepackage{amsmath}
\usepackage{amssymb}
\usepackage{amsthm}
\usepackage{array}
\newcolumntype{P}[1]{>{\centering\arraybackslash}p{#1}}
\newcolumntype{M}[1]{>{\centering\arraybackslash}m{#1}}
\usepackage{caption}
\usepackage{subcaption}
\usepackage{epigraph}
\usepackage{csquotes}
\usepackage{braket}
\usepackage{xcolor}
\usepackage[colorlinks = true,
            linkcolor = blue,
            urlcolor  = blue,
            citecolor = blue,
            anchorcolor = blue]{hyperref}
\usepackage{hyperref}
\usepackage{color}

\newcommand{\eps}{\epsilon}

\newcommand{\dd}{\text{d}}

\newcommand{\tr}{\tilde{r}}
\newcommand{\ttau}{\tilde{\tau}}
\newcommand{\tE}{\tilde{E}}
\newcommand{\tL}{\tilde{L}}
\newcommand\isc{{\rm isco}}

\begin{document}

\preprint{APS/123-QED}

\title{Transition from Inspiral to Plunge:\\ A Complete Near-Extremal Trajectory and Associated Waveform}

\author{Ollie Burke}
\email[]{ollie.burke@aei.mpg.de}
\affiliation{%
Max Planck Institute for Gravitational Physics (Albert Einstein Institute), Am M\"{u}hlenberg 1, Potsdam-Golm 14476, Germany}
\affiliation{%
  School of Mathematics, University of Edinburgh, James Clerk Maxwell Building, Peter Guthrie Tait Road, Edinburgh EH9 3FD, UK}
\author{Jonathan Gair}
%\email[]{jonathan.gair@aei.mpg.de}
\affiliation{%
Max Planck Institute for Gravitational Physics (Albert Einstein Institute), Am M\"{u}hlenberg 1, Potsdam-Golm 14476, Germany}
\affiliation{%
  School of Mathematics, University of Edinburgh, James Clerk Maxwell Building, Peter Guthrie Tait Road, Edinburgh EH9 3FD, UK}
\author{Joan Sim\'on}
%\email[]{j.simon@ed.ac.uk}
\affiliation{%
  School of Mathematics, University of Edinburgh, James Clerk Maxwell Building, Peter Guthrie Tait Road, Edinburgh EH9 3FD, UK}

\begin{abstract}
We extend the Ori and Thorne (OT) procedure to compute the transition from an adiabatic inspiral into a geodesic plunge \emph{for any} spin, with emphasis on near-extremal ones. Our analysis revisits the validity of the approximations made in OT. In particular, we discuss possible effects coming from eccentricity and non-geodesic past-history of the orbital evolution. We find three different scaling regimes according to whether the mass ratio is much smaller, of the same order or much larger than the near extremal parameter describing how fast the primary black hole rotates. Eccentricity and non-geodesic past-history corrections are always sub-leading, indicating that the quasi-circular approximation applies throughout the transition regime. However, we show that the OT assumption that the energy and angular momentum evolve linearly with proper time must be modified in the near-extremal regime. Using our transition equations, we describe an algorithm to compute the full worldline in proper time for an extreme mass ratio inspiral (EMRI) and the resultant gravitational waveform in the high spin limit.

%\begin{description}
%\end{description}
\end{abstract}

\pacs{Valid PACS appear here}% PACS, the Physics and Astronomy
\maketitle

%\tableofcontents
\section{Introduction}
The LIGO observation of the transient gravitational wave (GW) signal from the collision of two stellar mass black holes \cite{PhysRevLett.116.061102} in September 2015 spectacularly opened the new field of gravitational wave astronomy. By the end of the O2 observing run in August 2017, the LIGO/Virgo detectors had observed ten binary black hole mergers and a single binary neutron star inspiral~\cite{scientific1811gwtc}. This handful of observations has already had a profound impact on our understanding of the astrophysics of compact objects and ruled out a number of modified theories of gravity~\cite{baker2017strong,langlois2018scalar,ezquiaga2017dark,sakstein2017implications,boran2018gw170817}. During the ongoing O3 observing run new events are being reported at the rate of one per week, so these constraints are rapidly improving. However, the masses of the objects being observed are all in the range of $1$--$100M_\odot$, which is determined by the frequency sensitivity of the instruments~\cite{sathyaprakash2009physics}. Black holes with much higher masses are expected to exist in the centres of most galaxies~\cite{croton2006many} and will be even stronger sources of GWs, but these waves will be at millihertz frequencies which are inaccessible to ground-based detectors due to the seismic noise background.

The launch of the Laser Interferometer Space Antennae (LISA)~\cite{2017arXiv170200786A}, scheduled for 2034, will open the millihertz band from $10^{-4}$--$10^{-1}$Hz for the first time. Expected sources in this frequency band include massive black hole binaries, cosmic strings and extreme mass ratio inspirals (EMRIs). Detection of these sources, and estimation of their parameters, will rely on the comparison of accurate theoretical models of the expected gravitational waveforms to the observed data. Building these models for LISA is extremely challenging, in particular for EMRIs, which are expected to have a very rich structure and to be observed for hundreds of thousands of waveform cycles prior to merger with the central object~\cite{amaro2007intermediate}. In this paper we focus on modelling of a particular class of EMRIs, in which the central black hole has very large angular momentum (spin). All of the LIGO observations to date are consistent with zero or small spin~\cite{scientific1811gwtc}, but the massive black holes that will be probed by LISA are a different population. These black holes are observed in high accretion states as quasars, and accretion tends to spin the black holes up. Semi-analytic models predict that the typical spins of these objects are $a \gtrsim 0.95$~\cite{2014ApJ...794..104S}. 

The maximum spin of massive black holes is a quantity of fundamental interest for understanding the origin of black holes in the Universe. It was shown by Thorne~\cite{1974ApJ...191..507T} that the angular momentum of black holes being spun up through thin disc accretion saturates at a limit of $a=0.998$ where an equilibrium is reached between spin up by accreted material and spin down by captured retrograde photons. Black holes with higher spin could in principle be formed directly in the early Universe and for sufficiently high mass these black holes can retain spins above the Thorne limit for a Hubble time~\cite{arbey2019any}. A direct observation of a system with spin above the Thorne limit would thus have profound implications for our understanding of the origin and growth of black holes. It is therefore important to understand how well observations of EMRIs can constrain the spin of near-extremal black holes and to determine this we first need to build accurate representations of the gravitational waves emitted by such systems.

The near extremal limit is also relevant for more theoretical considerations. Indeed, as the primary rotates faster, its Hawking's temperature decreases because the distance between the inner $(\tr_-)$ and outer $(\tr_+)$ horizons in Boyer-Lindquist (BL) coordinates reduces according to
\begin{equation}
  \tilde{r}_{\pm} = 1 + \sqrt{1 - a^2} = 1 + \epsilon\,,
\end{equation}
where $\tilde{r} = r/M$ and $a$ the dimensionless Kerr spin parameter. The existence of a double pole in the function determining the black hole horizons in this limit is responsible for an enhancement of symmetry in the near horizon geometry of the Kerr black hole \cite{Bardeen:1999px}, a feature that remains true for \emph{any} extremal black hole \cite{Kunduri:2007vf}. This enhancement of symmetry from time translations to the conformal group has allowed several groups to analytically solve the master Teukolsky equation in the presence of the in spiraling probe particle leading to an analytic expression for the energy fluxes carried by the gravitational waves generated by this source 
\cite{Porfyriadis:2014fja,Hadar:2014dpa,Hadar:2015xpa,2015PhRvD..92f4029G,2015PhRvD..92f4029G,van2015near,Hadar:2016vmk,2016CQGra..33o5002G,2016CQGra..33o5002G,2018arXiv180403704C,compere2018_NHEK}. This provides a very exciting opportunity where analytic tools developed in the high energy theoretical physics community can provide accurate predictions to generate gravitational waveform templates. Future observations using such templates will be directly testing these theoretical predictions.

It has already been shown that gravitational waveforms emitted by these sources contain unique qualitative features that provide a smoking gun for the existence of near-extremal systems \cite{2016CQGra..33o5002G}. The amplitude of an EMRI waveform (averaged over a suitable amount of orbits) typically increases linearly in time for moderate spin $ a \approx 0.9$. It was shown in~\cite{2016CQGra..33o5002G} that the amplitudes of these signals \emph{dampen} in the high spin limit due to behaviour of the flux close to the horizon. There has been progress in modelling the inspiral from radial infinity to the innermost stable circular orbit (ISCO) \cite{2016CQGra..33o5002G},
by integrating the geodesic equations in the near horizon geometry of the Kerr black hole \cite{kapec2019particle} and exploiting the enhanced set of symmetries to compute the energy fluxes for more source trajectories~\cite{compere2018_NHEK}. However, no one has focused on providing a model which encapsulates the inspiral \emph{and} plunge in the limit of high spins. This is precisely what this paper seeks to do.

In this work, we build such a model for an EMRI comprised of a small compact object of mass $\mu$ gravitationally bound to a supermassive Kerr black hole of mass $M$ and study the transition from an adiabatic inspiral into a geodesic plunge \emph{for any} spin of the primary black hole. This transition to plunge was originally discussed by Ori and Thorne (OT)~\cite{2000PhRvD..62l4022O} for moderate values of the spin in the limit of small mass ratios. A similar but independent analysis conducted by Buonanno and Damour in~\cite{buonanno2000transition} solved the problem for Schwarzschild black holes with arbitrary (reduced) mass ratio. The technical reason why high spins require a separate discussion is because of the existence of a second independent small parameter competing with the mass ratio $\eta=\mu/M\ll 1$. This new parameter is the near-extremal parameter $\epsilon = \sqrt{1-a^{2}}$ encoding the distance of the spin parameter $a$ from its upper/lower bound, since Kerr black holes have spin parameters $a \in [-1,+1]$. Since the dynamical equations describing the transition depend on the spin, the near extremal limit, i.e. $\epsilon\to 0$, modifies the original scaling discussed by OT. The transition to plunge for near-extremal EMRIs was previously considered in~\cite{kesden2011transition} and our work clarifies and extends those results in a number of ways. We point out the physical interpretation of the mathematical procedure used in that paper, identify a missing term in the near-extremal regime and incorporate recent analytic results for the near-extremal energy flux for the first time. 

In this paper, we will first review the treatment of the transition regime given by OT in \cite{2000PhRvD..62l4022O}. We analyse their methodology and approximations and carefully estimate the scaling of terms that are being omitted. In each of \cite{2000PhRvD..62l4022O,sundararajan2008transition,Transition_Inspiral_Scott_Hughes} the notion of eccentricities and non-circular motion was ignored. We discuss the potential growth of eccentricities before and during the transition regime and find that corrections to our equations due to eccentric motion are sub-leading \emph{for any} spin. We identify three separate transition regimes, each with a slightly different equation of motion: $\eta \ll \epsilon$, $\epsilon \sim \eta$ and $\epsilon \ll \eta$.  %In the case of $\epsilon \sim \eta$, we highlight that there are extra terms in the transition equation of motion which have to be taken into account and are not present in $OT$. 
We then discuss a numerical algorithm to generate full inspiral trajectories in Boyer-Lindquist coordinates, alongside the corresponding evolution of the integrals of motion $E(\tau)$ and $L(\tau)$. Finally, we extend the waveform from the inspiral only results of \cite{2016CQGra..33o5002G} to include the plunge in the regime $\epsilon\sim\eta$.

This paper is organised as follows. In section~\ref{CH:8 Sec: Preliminaries}, we review the properties of equatorial and circular orbits in the Kerr black hole and, in section ~\ref{ssec:GW:Flux}, we review and compare the results describing gravitational fluxes emitted by circular EMRIs as a function of the spin.  In section~\ref{sec:master} we set-up the master transition equation of motion in general and, in subsection~\ref{sec:ecc}, we estimate corrections due to eccentricity and non-geodesic past-history of the orbital evolution. The transition equations of motion in the three different sxcaling regimes are described in subsections~\ref{ssec:OT},~\ref{ssec:near-extremal} and~\ref{ssec:very-near-extremal} respectively. The numerical scheme to integrate our transition equations of motion for the $\epsilon \sim \eta$ regime is presented in section~\ref{ssec:numerical_integration}. We describe how to generate a near-extremal EMRI gravitational waveform encapsulating inspiral and plunge in subsections~\ref{ssec:BL_coords} and~\ref{subsec:waveform}. We finish with a summary of our main results in section~\ref{sec:conc}. 

\textbf{Notation:} Any quantity carrying a tilde refers to a dimensionless quantity in units of the primary mass M, i.e.  $\tilde{r} = r/M$, $\tilde{\tau} = \tau/M$, $\tilde{t} = t/M$, $\tilde{E} = E/\mu$ and $\tilde{L} = L/M\mu$, but we keep $a$ as the dimensionless Kerr spin parameter with \emph{no} tilde.  Dotted quantities (eg $\dot{\tilde{E}}$) denote coordinate time derivatives of that quantity. Finally, expressions $A \sim \mathcal{O}(B)$ or, for brevity, $A \sim B$ stress that both $A$ and $B$ scale in the same way with the small parameters under consideration. We impose geometrized units by setting the constants $G = c = 1$. 

\vspace{0.5cm}
\noindent \emph{Note added.} In the final stages of this work, we became aware of overlapping results that were independently obtained in \cite{Geoffrey_Transition}.

\section{Preliminaries}
\label{CH:8 Sec: Preliminaries}

In Boyer-Lindquist (BL) coordinates $(\tr,\phi,\theta,\tilde{t})$, the motion of a point particle with mass $\mu$ in a Kerr black hole on the equatorial plane  ($\theta = \pi/2$) is given by \cite{Schmidt:2002qk}
\begin{align}
\left(\frac{d\tilde{r}}{d\tilde{\tau}}\right)^{2} &= \frac{[\tilde{E}(\tilde{r}^2 + a^2) - a\tilde{L}]^{2} - \Delta [(\tilde{L} - a\tilde{E})^2 + \tilde{r}^2]}{\tilde{r}^4} \nonumber \\ &= \tilde{E}^{2} - V_{\text{eff}}(\tilde{r},\tilde{E},\tilde{L},a) =  G(\tilde{r},\tilde{E},\tilde{L},a) \label{CH:8 Radial Geodesic Equation}\\
\frac{d\phi}{d\tilde{\tau}} &= \frac{-\left(a\tilde{E} - \tilde{L}\right) + a(\tilde{E}[\tilde{r}^2 + a^2] - a\tilde{L})/\tilde{r}}{\tilde{r}^2} \nonumber \\ &= \Phi(\tilde{r},\tilde{E},\tilde{L},a) \label{CH:8 Phi Geodesic Equation}\\
\frac{d\tilde{t}}{d\tilde{\tau}} & = \frac{-\Delta a(a\tilde{E} -\tilde{L}) + (\tilde{r}^2 + a^2)(\tilde{E}[\tilde{r}^2 + a^2] - a\tilde{L})}{\Delta\tilde{r}^2} \nonumber\\ & = T(\tilde{r},\tilde{E},\tilde{L},a), \label{CH:8 Time Geodesic Equation}
\end{align}
where the largest root of $\Delta = \tilde{r}^{2} - 2\tilde{r} + a^{2}$ corresponds to the outer horizon $\tr_+$
\begin{equation*}
\tilde{r}_{+} = 1 + \sqrt{1-a^2}\,,
\end{equation*}
$\ttau=\tau/M$ denotes proper time in units of the Kerr black hole mass $M$ and $a$ is the \emph{dimensionless} spin parameter $a \in [-1,1]$. 

The particle is on a prograde (retrograde) orbit if it follows the same (opposite) direction as the rotation of the primary hole. Prograde (retrograde) orbits correspond to $a > 0 \ (a < 0)$ while keeping the azimuthal component of the angular momentum $\tL>0$. Since retrograde orbits do not reach the near horizon geometry of the primary black hole in the near-extremal limit (see appendix \ref{App:Retrograde_Orbits}), while \emph{prograde} orbits do, we only consider the latter from here on.

For an equatorial orbit to be circular,  the BL radial coordinate $\tilde{r}$ must be constant and to be stable, the latter must be at a minimum of the potential $V_{\text{eff}}$ in \eqref{CH:8 Radial Geodesic Equation} so that 
\begin{equation*}
G = \frac{\partial{G}}{\partial \tilde{r}}= 0, \ \text{and} \ \frac{\partial^{2}G}{\partial \tilde{r}^{2}} \geq 0\,.
\end{equation*}
These conditions determine the energy $\tilde{E}$ and angular momentum $\tilde{L}$ of these orbits to be \cite{Chandrasekhar:579245} 
\begin{align}
\tilde{E} &= \frac{1 - 2/\tilde{r} + a/\tilde{r}^{3/2}}{\sqrt{1 - 3/\tilde{r} + 2a/\tilde{r}^{3/2}}}\,, \label{CH:8 Circular Energy} \\
\tilde{L} &= \tilde{r}^{1/2}\frac{1 - 2a/\tilde{r}^{3/2} + a^{2}/\tilde{r}^{2}}{\sqrt{1 - 3/\tilde{r} + 2a/r^{3/2}}}\,.
\label{CH:8 Circular Ang Mom}
\end{align}

Substituting \eqref{CH:8 Circular Energy}-\eqref{CH:8 Circular Ang Mom} into \eqref{CH:8 Phi Geodesic Equation}-\eqref{CH:8 Time Geodesic Equation} gives rise to
\begin{eqnarray}
  \frac{d\phi}{d\tilde{\tau}} &=& \frac{1}{\tilde{r}^{3/2}\sqrt{1 - 3/\tilde{r} + 2a/\tilde{r}^{3/2}}}\,, \\
  \frac{d\tilde{t}}{d\tilde{\tau}} &=& \frac{1 + a/\tilde{r}^{3/2}}{\sqrt{1 - 3/\tilde{r} + 2a/\tilde{r}^{3/2}}}\,,
 \label{CH:8 Circular Geodesic Equation Time}
\end{eqnarray}
whose ratio defines the angular velocity $\tilde{\Omega}$ of the particle
\begin{equation}\label{CH:8 Orbital Velocity}
\frac{d\phi}{d\tilde{t}} =\tilde{\Omega} = (\tilde{r}^{3/2} + a)^{-1}\,.
\end{equation}
  
Equatorial circular orbits are also known to satisfy the identity~\cite{1998PhRvD..58f4012K}
\begin{equation}
  \frac{\partial G}{\partial\tE}(\tr)\,\tilde{\Omega}(\tr) + \frac{\partial G}{\partial \tL}(\tr) = 0\,,
\label{eq:circular-1}
\end{equation} 
where we stress the equality holds for any circular orbit labelled by $(\tr, \tE, \tL)$. Differentiating \eqref{eq:circular-1} with respect to $\tr$, we can derive further equalities satisfied for any such orbits. The ones below
\begin{eqnarray}
  \frac{\partial^2G}{\partial\tr\partial\tE}\,\tilde{\Omega} + \frac{\partial^2G}{\partial\tr\partial\tL} &=& -\frac{\partial\tilde{\Omega}}{\partial\tr} \frac{\partial G}{\partial \tE}\,, 
\label{eq:circular-2}\\
  -\frac{1}{2}\left(\frac{\partial^3G}{\partial\tr^2\partial\tE}\,\tilde{\Omega} + \frac{\partial^3G}{\partial\tr^2\partial\tL}\right) &=& 
\frac{\partial\tilde{\Omega}}{\partial\tr} \frac{\partial^2 G}{\partial \tr\partial \tE} \nonumber \\
& & + \frac{1}{2} \frac{\partial^2\tilde{\Omega}}{\partial\tr^2} \frac{\partial G}{\partial \tE}
\label{eq:circular-3}
\end{eqnarray}
will play a role in our analysis later on.

The \emph{innermost stable circular orbit} (ISCO) is the marginal circular stable orbit satisfying
\begin{equation}\label{CH:8 Circular_Motion_Constraints}
G\rvert_{\isc} = \frac{\partial{G}}{\partial \tilde{r}}\bigg\rvert_{\isc}= \frac{\partial^{2}G}{\partial \tilde{r}^{2}}\bigg\rvert_{\isc} = 0\,.
\end{equation}
The last equality, describing marginality, allows to solve for its radius as a function of the spin \cite{bardeen1972rotating}
\begin{align}\label{CH:8 ISCO location}
\tilde{r}_{\isc} &= 3 + Z_{2} - [(3-Z_{1})(3+Z_{1}+2Z_{2})]^{1/2} \\
Z_{1} &= 1 + (1-a^2)^{1/3}[(1+a)^{1/3} + (1-a)^{1/3}]\nonumber \\
Z_{2} &= (3a^{2} + Z_{1}^{2})^{1/2}\,.\nonumber
\end{align}
This is the last radii before plunging into the horizon occurs. In appendix \ref{app:A:ISCO}, we derive general formulas for \eqref{CH:8 Circular_Motion_Constraints} and higher order derivatives of $G(\tr,\tE,\tL)$, which are valid for any spin $a>0$, when evaluated at ISCO that will be relevant in the rest of this work. 

For \emph{near-extremal} Kerr black holes, it is natural to introduce the near extremal parameter
\begin{equation}\label{CH:8_Extremality_Parameter}
\epsilon = \sqrt{1-a^2} \quad \text{for} \ \epsilon \ll 1\,,
\end{equation}
to mathematically capture the large spin limit $a\to 1$. For these black holes, the ISCO location can be expanded in $\epsilon$
\begin{equation}\label{CH:8 ISCO Expansion}
\tilde{r}_{\isc} = 1 + 2^{1/3}\epsilon^{2/3} + \mathcal{O}(\epsilon^{4/3})
\end{equation}
and the physical parameters of this marginal orbit reduce to
\begin{align}
\tilde{E}_{\isc} &\to \frac{1}{\sqrt{3}}\left(1 + 2^{1/3}\epsilon^{2/3}\right)  + \mathcal{O}(\epsilon^{4/3})\,, \label{CH:8 Eng Expansion}\\
\tilde{L}_{\isc} &\to \frac{2}{\sqrt{3}}\left(1 + 2^{1/3}\epsilon^{2/3}\right)  + \mathcal{O}(\epsilon^{4/3})\,, \label{CH:8 Ang Expansion}\\
\tilde{\Omega}_{\isc} &\to \frac{1}{2}\left(1 -\frac{3}{ 2^{5/3}}\epsilon ^{2/3}\right) + \mathcal{O}(\epsilon^{4/3}) \,. \label{CH:8 Omega Expansion}
\end{align}
Notice $|\tr_{\isc} - \tilde{r}_{+}|\sim\mathcal{O}(\epsilon^{2/3})$ for prograde orbits, whereas it is $\mathcal{O}(1)$ for retrograde ones, as mentioned below \eqref{eq:isco-retro}. This further justifies our interest in prograde orbits in the near extremal limit.

\subsection{Gravitational Wave Flux}\label{ssec:GW:Flux}

For equatorial orbits, the motion of a (point) particle on the Kerr spacetime background generates gravitational waves carrying energy and angular momentum, either escaping towards infinity or being absorbed by the horizon of the primary hole. 

Due to energy and angular momentum conservation, the orbit averaged rates of change $\langle\dot{\tE}\rangle$ and $\langle\dot{\tL}\rangle$ satisfy $\langle\dot{\tE}\rangle = -\langle\dot{\tE}_{\text{GW}}\rangle$ and $\langle\dot{\tL}\rangle = - \langle\dot{\tL}_{\text{GW}}\rangle$, where the averaged gravitational wave dissipative fluxes are 
\begin{equation}
\label{Energy_Flux}
\begin{aligned}
   \langle\dot{\tilde{E}}_{\text{GW}}\rangle &= \langle \dot{\tilde{E}}_{\text{GW},H}\rangle + \langle\dot{\tilde{E}}_{\text{GW},\infty}\rangle = \langle f^{\text{diss}}_{t} / u^{t}\rangle\,, \\
  \langle\dot{\tilde{L}}_{\text{GW}}\rangle &= \langle\dot{\tilde{L}}_{\text{GW},H}\rangle +\langle\dot{\tilde{L}}_{\text{GW},\infty}\rangle = -\langle f^{\text{diss}}_{\phi} / u^{t}\rangle\,,
\end{aligned}
\end{equation}
determined in terms of the time-averaged $t$ and $\phi$ components of the dissipative self-force $\tilde{f}^{\text{diss}}$ normalised by the $t-$component of the four velocity $u^{t}$. These quantities are discussed in more depth in the next subsection \ref{sec:self-force} (see appendix~\ref{App:D_Osculating_Elements_Equations}, in particular the discussion around \eqref{eq:dEdtau}, for a derivation of relations such as \eqref{Energy_Flux}). Notice we split these fluxes into their horizon and asymptotic infinity contributions in the right hand side. We stress here this flux balance law only holds for \emph{adiabatically} evolving binaries forcing the small mass ratio limit $\eta \to 0$. The (orbit-averaged and dissipative) gravitational wave fluxes are determined by solving the Teukolsky equation in the presence of the point particle source~\cite{teuk73,sasaki82,hu1,2002PhRvD..66d4002G,dras}.  From hereon, to avoid cumbersome notation, we shall drop the angled-brackets to denote time-averaging and simply write, for example, $\langle\dot{\tilde{E}}\rangle = \dot{\tilde{E}}$.

In \cite{2000PhRvD..62l4021F}, Finn and Thorne (F\&$\text{T}$) parameterise the energy flux as the (Peters and Mathews~\cite{peters1963gravitational}) leading order post newtonian correction with an extra general relativistic correction $\dot{\mathcal{E}}$ factor
\begin{equation}\label{CH:8 Flux_Thorne}
\frac{d\tilde{E}_{GW}}{d\tilde{t}}= \frac{32}{5}\eta\,\tilde{\Omega}^{10/3}\dot{\mathcal{E}}(\tilde{r})\,.
\end{equation}
These fluxes are spin dependent and are typically computed through numerical means. See the tables in \cite{2000PhRvD..62l4021F} for some of the values of these relativistic corrections.

As we increase the spin of the black hole, the two roots $\tr_{\pm} = 1 \pm \epsilon$ of the function $\Delta$ determining the outer and inner horizons of the rotating black hole coincide in the extremal limit $\epsilon=0$. In this limit, the geometry close to the horizon of the black hole, which can be isolated using the change of coordinates
\begin{equation}
\tr - \tr_+ = \lambda\,\rho\,, \quad \quad \tilde{t} = \frac{T}{\lambda}\,, \quad \quad \tilde{\phi} = \phi + \frac{\tilde{t}}{2\lambda}
\label{eq:nh}
\end{equation}
with $\lambda\to 0$, has an enhancement of symmetry from $\mathbb{R}\times \mathbb{U}(1)$, i.e. time translations and rotational symmetry, to $\mathbb{SL}(2,\mathbb{R}) \times \mathbb{U}(1)$. The resulting near horizon geometry is warped AdS$_2$ over a 2-sphere. The enhanced $\mathbb{SL}(2,\mathbb{R})$, the isometry group of AdS$_2$, includes the scaling symmetry $\rho\to c\rho$ and $T\to T/c$. This was already observed in the original work \cite{Bardeen:1999px} and it is true for any extremal black hole \cite{Kunduri:2007vf}.

Larger symmetry in physics implies larger kinematic constraints which can provide further analytic control over the given problem, in this case, the calculation of the gravitational wave fluxes \eqref{Energy_Flux}. It is precisely the emergence of this conformal group $(\mathbb{SL}(2,\mathbb{R}))$ and the use of asymptotic expansion matching methods that allowed to find analytic expressions for these energy and angular momentum fluxes for equatorial circular orbits close to the horizon \cite{Porfyriadis:2014fja,Hadar:2014dpa,Hadar:2015xpa,2015PhRvD..92f4029G,Hadar:2016vmk,2016CQGra..33o5002G,compere2018_NHEK}. This body of work led to the simple relationship for the flux given in \cite{2016CQGra..33o5002G}
\begin{equation}\label{CH:8 Flux_Near_Extremal}
\frac{d\tilde{E}_{GW}}{d\tilde{t}} \approx \eta(\tilde{C}_{H} + \tilde{C}_{\infty})\frac{\tilde{r} - \tilde{r}_{+}}{\tilde{r}_{+}}\,.
\end{equation}
The quantities $\tilde{C}_{H}$ and $\tilde{C}_{\infty}$ are constants representing how much wave emission goes towards the horizon and infinity respectively. These constants are given analytically in equations (76) and (77) of \cite{2015PhRvD..92f4029G}. Numerically evaluating them and summing the contribution of the first $|m| \leq l = 30$ modes gives $\tilde{C}_{H} = 0.987$ and $\tilde{C}_{\infty} = -0.133$.

\begin{figure*}
\centering
\includegraphics[height = 6cm, width = 8cm]{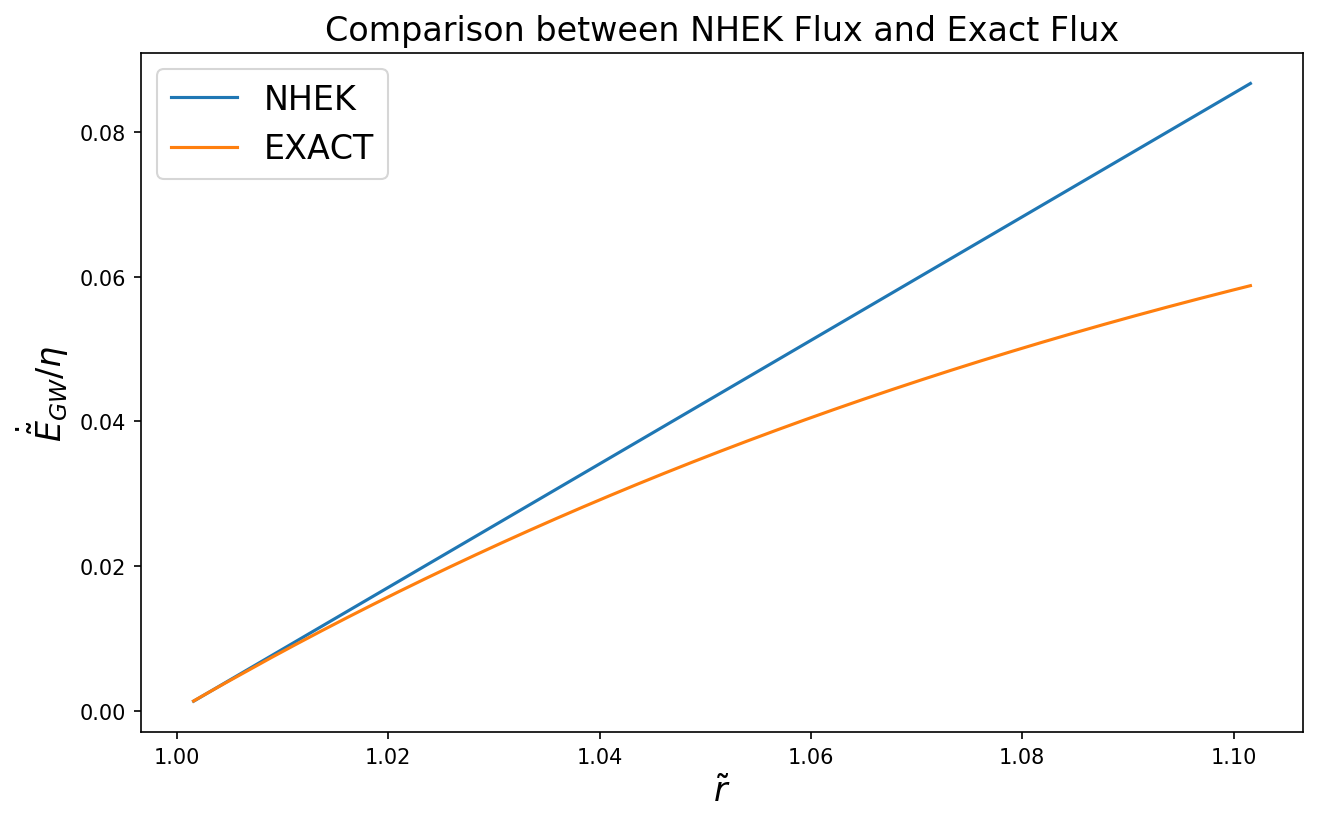}
\includegraphics[height = 6cm, width = 8cm]{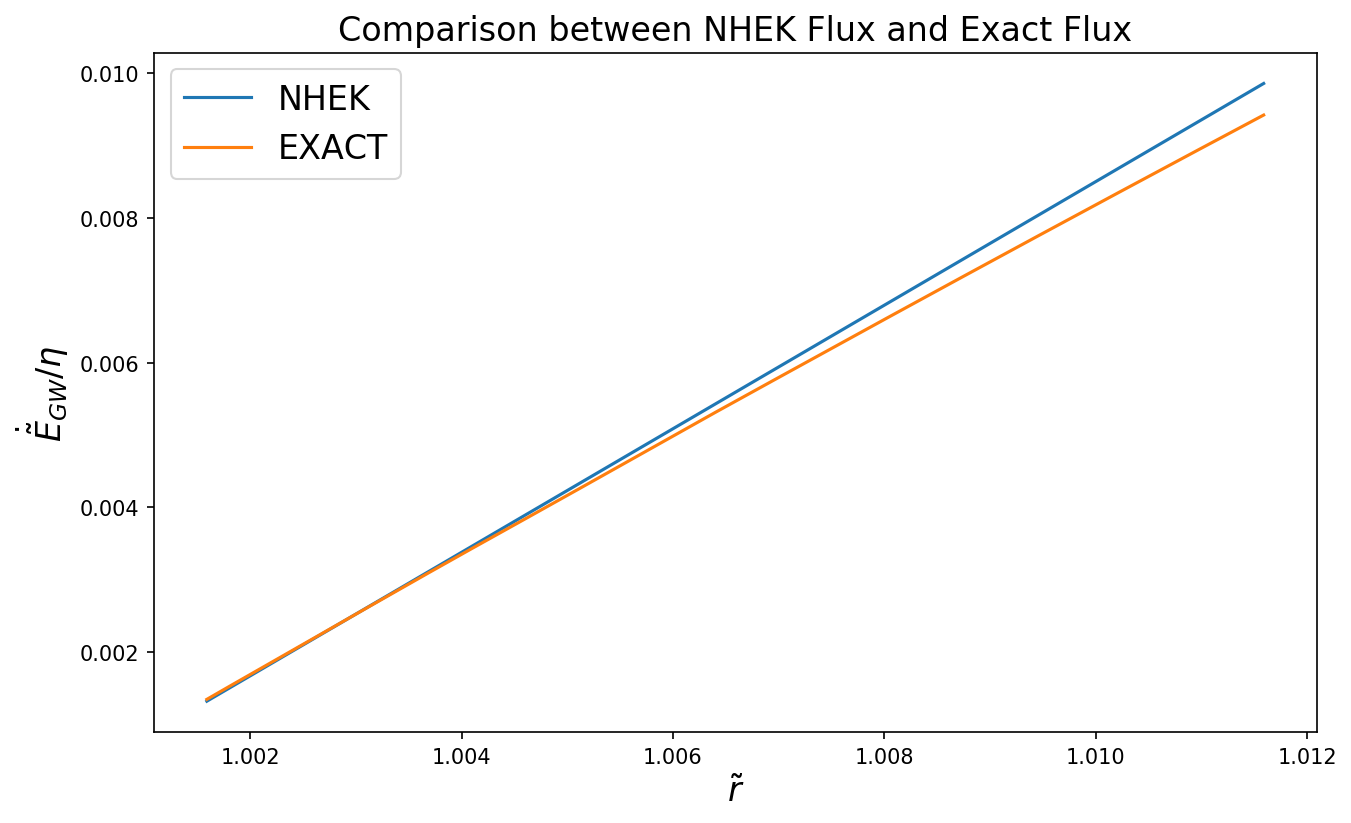}
\includegraphics[height = 6cm, width = 8cm]{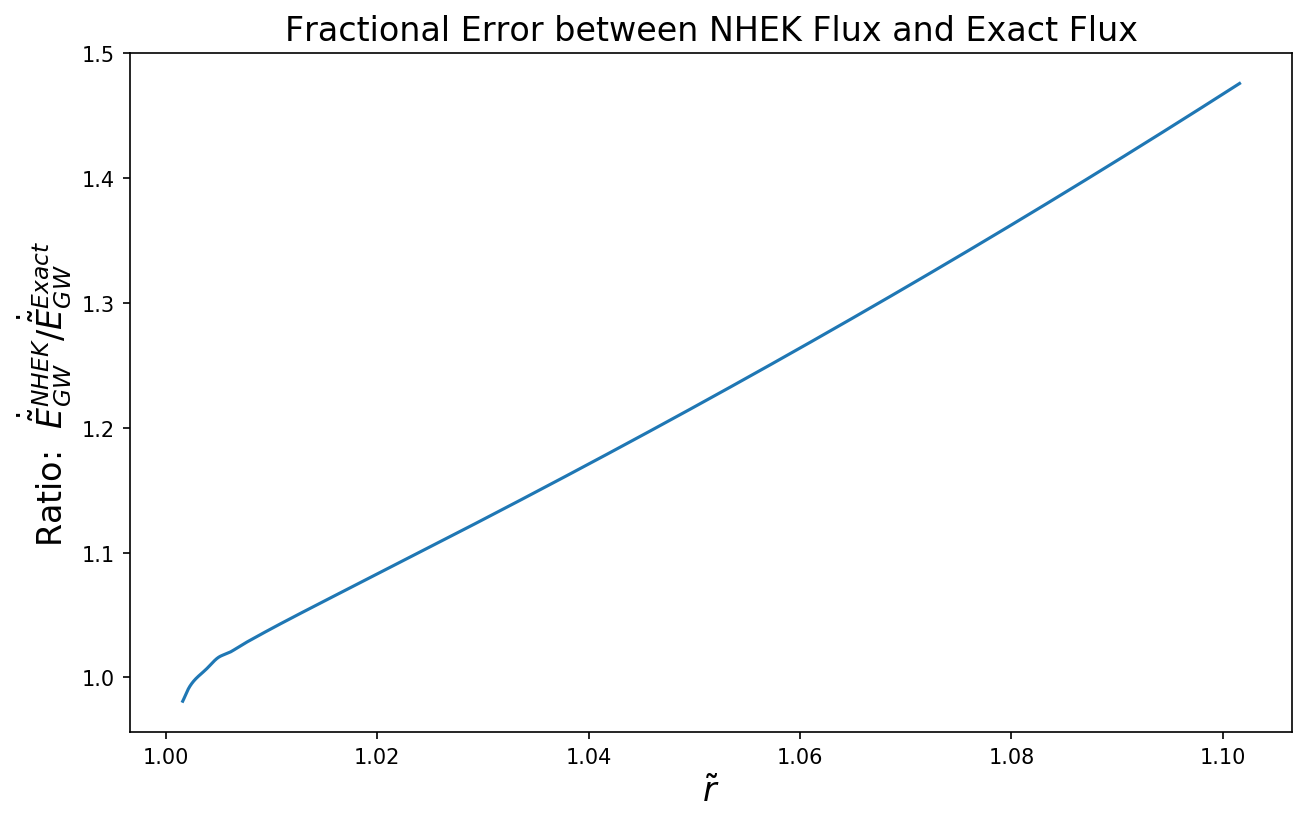}
\includegraphics[height = 6cm, width = 8cm]{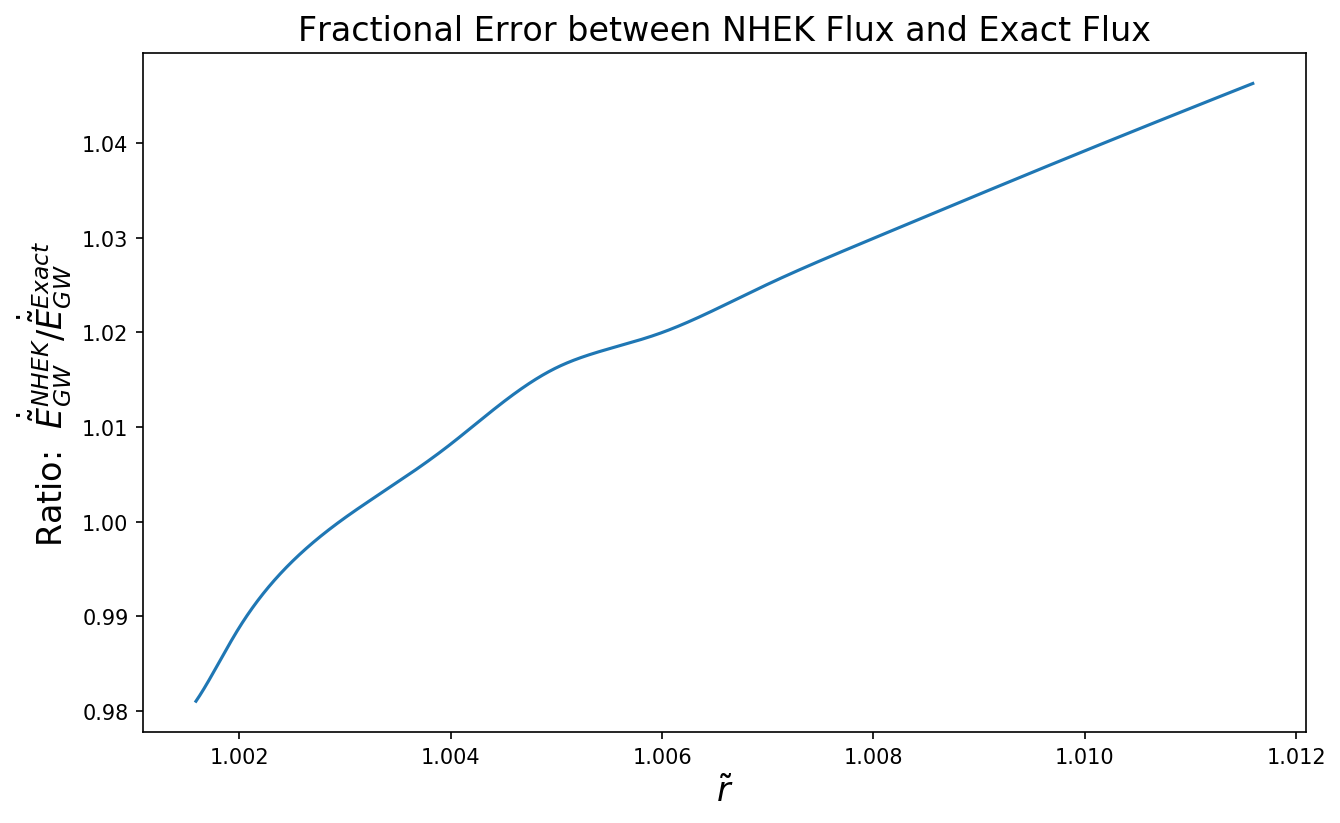}
\caption{These plots show the deviation between using the exact results for the flux \eqref{CH:8 Flux_Thorne} and the near extremal approximation given in \eqref{CH:8 Flux_Near_Extremal}. Notice that, to keep the error $< 5\%$, we require $\tr \lesssim 1.01$. For each of these plots, we used a spin parameter $a = 1-10^{-9}$. Similar plots can be found in \cite{2015PhRvD..92f4029G}.}
\label{CH:8 fig:Compare_Flux_Formulas}
\end{figure*}

The flux \eqref{CH:8 Flux_Near_Extremal} is only reliable when working with near extremal black holes and interested in near horizon physics.  This fact can be checked by comparing the exact fluxes \eqref{CH:8 Flux_Thorne}, using exact results found in the black hole perturbation toolkit (\href{http://bhptoolkit.org}{BHPT}), with the near extremal approximation \eqref{CH:8 Flux_Near_Extremal}. This comparison is shown in figure \ref{CH:8 fig:Compare_Flux_Formulas}. Fixing the radial coordinate to
$\tilde{r}=\tilde{r}_{\isc}$ and varying the spin parameter $a$, we observe in Table (\ref{CH:8 Table 2:ComparisonNHEKFormula}) that as $a \to 1$, the NHEK flux \eqref{CH:8 Flux_Near_Extremal} converges towards the exact value computed using the \href{http://bhptoolkit.org}{BHPT}. Furthermore, fixing the spin parameter to $a = 1-10^{-9}$, as in figure (\ref{CH:8 fig:Compare_Flux_Formulas}), the NHEK flux \eqref{CH:8 Flux_Near_Extremal} provides a nearly-perfect agreement up to a coordinate radii $\tilde{r} \approx 1.012$. The reason for the (extremely small) discrepancy at the ISCO is because Eq.(\ref{CH:8 Flux_Near_Extremal}) is only valid for $\epsilon \to 0$ and we consider $\epsilon \approx 10^{-5}$.  
\begin{table}
\centering
\begin{tabular}{ccccc}
   &  &  \\ \hline
 $a$ & $\dot{\tilde{E}}_{\text{Exact}}/\eta$ & $\dot{\tilde{E}}_{\text{NHEK}}/\eta$ &$|\dot{E}_{\text{NHEK}}-\dot{E}_{\text{Exact}}|/\eta$  \\ \hline
 $1-10^{-5}$ & 0.0264197 & 0.0261523  & 0.0002674 \\
 $1-10^{-6}$ & 0.0129344 & 0.0125200 & 0.0004143 \\
 $1-10^{-7}$ & 0.0061516 & 0.0059484 & 0.0002031 \\
 $1-10^{-8}$ & 0.0028875 & 0.0028082 & 0.0000793 \\
 $1-10^{-9}$ & 0.0013472 & 0.0013193 & 0.0000280\\
 $1-10^{-10}$ &0.0006273 & 0.0006176  & 0.0000097\\
 $1-10^{-11}$ &0.0002915 & 0.0002883 & 0.0000031\\
 $1-10^{-12}$& 0.0001354 & 0.0001344 & 0.0000009 \\ \hline
\end{tabular}
\caption{Comparing the NHEK flux \eqref{CH:8 Flux_Near_Extremal} with exact flux data found in the \href{http://bhptoolkit.org}{BHPT}. We fix the radial coordinate at $\tr = \tilde{r}_{\isc}$ and change the spin parameter $a$. This data can also be found in \cite{2015PhRvD..92f4029G}.}
\label{CH:8 Table 2:ComparisonNHEKFormula}
\end{table}
Thus we can use \eqref{CH:8 Flux_Thorne} to build a trajectory throughout the adiabatic inspiral regime. Then, as we near the ISCO, we can use the powerful analytic result given by Eq.\eqref{CH:8 Flux_Near_Extremal}. Using Eq.\eqref{CH:8 Flux_Near_Extremal} allows for a more analytic treatment of the analysis of the transition regime.

\section{The Transition Equation of Motion}
\label{sec:master}

In this section we revisit the earlier work by OT~\cite{2000PhRvD..62l4022O} describing how a small body following an initial equatorial circular orbit around the large black hole inspirals and eventually transitions into a plunging trajectory falling into the black hole. 

Our discussion is organised as follows. First, we analyse in section~\ref{sec:self-force} the effects arising from the radial self-force in the vicinity of the ISCO on the dynamics of this small body, justifying the starting point in OT. Second, assuming the dissipative fluxes of energy and angular momentum for quasi-circular and equatorial orbits are still related as in circular orbits \cite{apostolatos1993gravitational,1998PhRvD..58f4012K} 
\begin{equation}
\label{eq:circular-flux}
\dot{\tilde{E}} = \tilde{\Omega}(\tilde{r})\dot{\tilde{L}}\,,
\end{equation}
we derive in section~\ref{sec:general} the transition equation for arbitrary black hole spins \emph{without} the OT assumption that both energy $\tilde{E}$ and angular momenta $\tilde{L}$ evolve \emph{linearly} in proper time $\tilde{\tau}$. Third, given the quasi-circular nature of our assumed orbits, we argue in section~\ref{sec:ecc} there can be corrections to \eqref{eq:circular-flux} of the form
\begin{equation}
  \dot{\tilde{E}} - \tilde{\Omega}(\tilde{r})\dot{\tilde{L}} \sim \eta \dot{\tilde{r}} + \eta e^{2}\,,
\label{eq:correction-flux}
\end{equation}
whose scaling behaviour on the trajectory of the small body is determined. Finally, in sections~\ref{ssec:OT}-\ref{ssec:very-near-extremal}, we discuss in great detail the existence of three different scaling regimes in our transition equation, depending on the black hole spin, paying special attention to the near-extremal ones which contain new physics. We show the corrections due to \eqref{eq:correction-flux} are subleading in all the regimes.

\subsection{The self-force}
\label{sec:self-force}

 This subsection shows both that quasi-circular and equatorial orbits have vanishing dissipative effects and the conservative piece of the radial self-force can be neglected close to the ISCO. 

Consider the radial geodesic equation, Eq.~(\ref{CH:8 Radial Geodesic Equation}). Differentiating it with respect to proper time, one obtains
\begin{equation}\label{eq:acceleration}
\frac{d^{2}\tilde{r}}{d\tilde{\tau}^{2}} - \frac{1}{2}\frac{\partial G}{\partial \tilde{r}} = \frac{1}{2}\left(\frac{d\tE}{d\tilde{\tau}}\frac{\partial G}{\partial \tE} + \frac{d\tL}{d\tilde{\tau}}\frac{\partial G}{\partial \tilde{L}}\right)\left(\frac{d\tilde{r}}{d\tilde{\tau}}\right)^{-1}\,.
\end{equation}
It is shown in Appendix~\ref{App:D_Osculating_Elements_Equations} this is equivalent to
\begin{equation}
\label{eq:acceleration_comparison}
\frac{d^{2}\tr}{d\tilde{\tau}^{2}} + \Gamma^{\tr}_{\rho\sigma}\frac{d\tilde{x}^{\rho}}{d\tilde{\tau}}\frac{d\tilde{x}^{\sigma}}{d\tilde{\tau}} = \tilde{f}^{\tr}_{\text{diss}}\,.
\end{equation}
Hence, the terms on the left hand side of eq.~\eqref{eq:acceleration} correspond to the usual ones for geodesic motion, whereas the one on the right hand side can be understood as the perturbing force $\tilde{f}^{\tilde{r}}_{\text{diss}}$ exerted on the particle driving energy $\left(d\tilde{E}/d\tilde{\tau}\right)$ and angular momentum $\left(d\tilde{L}/d\tilde{\tau}\right)$ loss due to gravitational wave emission. 

%arising due to the evolution of the orbit under radiation reaction, 
%. \ans{The fluxes $\dot{\tE}_{\text{GW}} = -(u^{\tilde{t}})^{-1} F_{\tilde{t}}$ and $\dot{\tL}_{\text{GW}} = %(\tilde{u}^{\tilde{t}})^{-1}F_{\phi}$ for $F_{\tilde{t}}$ and $F_{\phi}$ the $\tilde{t}$ and $\phi$ components of the %gravitational self-force and $u^{\tilde{t}}$ the $\tilde{t}$ component of the four velocity.  Due to conservation of the %velocity norm, $\tilde{u}_{\tilde{\alpha}} \tilde{u}^{\tilde{\alpha}}=-1$, we find a relationship between %$\tilde{u}_{\tilde{\alpha}}$ and acceleration $\tilde{f}^{\tilde{\alpha}}$ given by $\tilde{u}_{\tilde{\alpha}} %\tilde{f}^{\tilde{\alpha}} = 0$}. It is shown in Appendix~\ref{App:D_Osculating_Elements_Equations} that
%\begin{equation}
%\begin{aligned}\label{eq:acceleration_comparison}
%\frac{d^{2}\tr}{d\tilde{\tau}^{2}} + %\Gamma^{\tr}_{\rho\sigma}\frac{d\tilde{x}^{\rho}}{d\tilde{\tau}}\frac{d\tilde{x}^{\sigma}}{d\tilde{\tau}} &= %\tilde{f}^{\tr}\,, \\
%\Longleftrightarrow \ \frac{d^{2}\tilde{r}}{d\tilde{\tau}^{2}} - \frac{1}{2}\frac{\partial G}{\partial \tilde{r}} &= %\frac{1}{2}\left(\frac{d\tE}{d\tilde{\tau}}\frac{\partial G}{\partial \tE} + \frac{d\tL}{d\tilde{\tau}}\frac{\partial %G}{\partial \tilde{L}}\right)\left(\frac{d\tilde{r}}{d\tilde{\tau}}\right)^{-1}.
%\end{aligned}
%\end{equation}
%\ans{which demonstrates that the quantity $\tilde{f}^{\tilde{r}}$ can be understood as the force exerted on the particle, %due to energy loss coming from gravitational wave emission, encoded through $\dot{\tilde{E}}$ and $\dot{\tilde{L}}.$}

In~\cite{Mino:2003yg}, Mino recognised that the forcing term for general geodesic motion can perturbatively be split into a radiative reactive dissipative and a conservative piece at first order in the mass ratio $\eta$
\begin{equation}\label{splitting_self_force}
    \tilde{f}^{\tilde{r}} = \eta\left(\tilde{f}^{\tilde{r}}_{(1)\text{diss}} + \tilde{f}^{\tr}_{(1)\text{cons}}\right) + \mathcal{O}(\eta^{2})\,.
\end{equation}
More details on this splitting can be found in \cite{barack2018self}.

For circular orbits, as considered by OT, the dissipative fluxes of energy and angular momentum are related by
\cite{apostolatos1993gravitational,1998PhRvD..58f4012K} 
\begin{equation}\label{app:OT_Flux_Balance_Law}
\dot{\tilde{E}} = \tilde{\Omega}(\tilde{r})\dot{\tilde{L}}\,.
\end{equation}
Hence, the dissipative part of the self-force $\tilde{f}^{\tilde{r}}_{(1)\text{diss}}$ vanishes, leading to
\begin{equation}
    \frac{d^{2}\tilde{r}}{d\tilde{\tau}^{2}} - \frac{1}{2}\frac{\partial G}{\partial \tilde{r}} = \eta\tilde{f}^{\tilde{r}}_{(1)\text{cons}} + \mathcal{O}(\eta^{2})
\end{equation}
which is precisely Eq.(3.10) of OT in~\cite{2000PhRvD..62l4022O}. A gauge invariant way to quantify the leading order in $\eta$ effect on the trajectory due to $\tilde{f}^{\tilde{r}}_{(1)\text{cons}}$ is to study the orbital velocity $\tilde{\Omega}$ (see \cite{barack2018self} for a review). This generates a shifted orbital velocity $\tilde{\Omega}^{\text{shifted}}_{\isc}$ with respect to the Kerr orbital velocity $\tilde{\Omega}_{\isc}$ at ISCO given by \cite{barack2009gravitational, van2017self}
\begin{equation}\label{orb_vel_shift}
    (1 + \eta)\tilde{\Omega}^{\text{shifted}}_{\isc} = \tilde{\Omega}_{\isc}(1 + \eta C_{\tilde{\Omega}}(a)) + \mathcal{O}(\eta^{2})
\end{equation}
with the quantity $C_{\tilde{\Omega}}(a)$ discussed in depth and independently (numerically) calculated in both \cite{isoyama2014gravitational, van2017self}. According to \cite{van2017self}, $C_{\tilde{\Omega}}(a)\in(1.24,1.39)$. Hence, $C_{\tilde{\Omega}}(a)\sim \mathcal{O}(1)$ for all spins, and since $\tilde{\Omega}_{\isc}\sim\mathcal{O}(1)$, it follows that for $\eta \ll 1$
\begin{equation}
|\tilde{\Omega}^{\text{shifted}}_{\isc} - \tilde{\Omega}_{\isc}| \approx \eta (C_{\tilde{\Omega}}(a) - 1)\tilde{\Omega}_{\isc} \sim \eta.
\end{equation}
It is further shown in \cite{van2017self} that  $C_{\tilde{\Omega}}(a) \to 1 + 1/2\sqrt{3}$ as $a \to 1$ in an averaged sense\footnote{$C_{\tilde{\Omega}}(a)$ is shown to actually oscillate around this limiting value as $a\to 1$. This phenomenon is non-trivial and still not well understood today. See \cite{van2017self} for more details.}. Using Eq.\eqref{CH:8 Omega Expansion} in the high spin limit and $\eta \ll 1$, equation \eqref{orb_vel_shift} becomes
\begin{equation}\label{eq:shifted_orb_vel}
    |\tilde{\Omega}^{\text{shifted}}_{\isc} - \tilde{\Omega}_{\isc}| = \frac{\eta}{4\sqrt{3}} + \mathcal{O}(\eta^{2}) + \mathcal{O}(\eta\epsilon^{2/3}).
\end{equation}
This implies that the change in the orbital velocity at the ISCO due to conservative self-force effects is an $\mathcal{O}(\eta)$ quantity. 

Since $\tilde{\Omega}^{\text{shifted}}_{\isc}$ is related to the shifted Boyer-Lindquist radial coordinate at the ISCO by
\begin{equation}
\tilde{r}^{\text{shifted}}_{\isc} = \left(\frac{1}{\tilde{\Omega}^{\text{shifted}}_{\isc}} - a\right)^{2/3}\,,   
\end{equation}
it follows, using eq.\eqref{eq:shifted_orb_vel}, that the ``radial thickness"
\begin{equation}
    \tilde{r} - \tilde{r}^{\text{shifted}}_{\isc} \sim \tilde{r} - \tilde{r}_{\isc} + \mathcal{O}(\eta)\,,
\label{eq:sf-radial}
\end{equation}
differs by an $\mathcal{O}(\eta)$ quantity when including the conservative self-force effects.

It will be shown in this paper that there are three different transition regimes depending on the ratio of $\epsilon$ and $\eta$. The ``radial thickness" of the transition in each regime scales according to
\begin{itemize}
    \item For $\eta \ll \epsilon \Rightarrow \tilde{r} - \tilde{r}_{\isc} \sim\eta^{2/5}$,
    \item For $\eta \sim \epsilon \Rightarrow \tilde{r} - \tilde{r}_{\isc} \sim (\eta/\epsilon)^{2/5}\epsilon^{2/3} \sim \eta^{2/3}$,
    \item For $\eta \gg \epsilon \Rightarrow \tilde{r} - \tilde{r}_{\isc} \sim \eta^{2/3}$.
\end{itemize}
Thus, the effect of the conservative piece of the self-force is subleading in all regimes. For this reason, like in the original OT analysis, we shall ignore these effects.

\subsection{Transition Equation - Generalities} 
\label{sec:general}

To discuss the evolution of the orbit, we pursue the following strategy : we assume the corrections in \eqref{eq:correction-flux} are subleading, and once the scaling behaviour of the different dynamical regimes is identified, we double check the consistency of our original assumption.

To evolve the orbit, OT used the circular flux relationship \eqref{app:OT_Flux_Balance_Law} and additionally assumed that the energy $\tilde{E}$ and angular momentum $\tilde{L}$ evolve \emph{linearly} in proper time $\tilde{\tau}$ throughout the transition regime
\begin{equation}
\begin{aligned}\label{Linear_Time_Ang_Momenta}
\tilde{E} - \tE_{\isc} &=  \tilde{\Omega}_{\isc}\frac{d\tilde{L}_{\isc}}{d\tilde{\tau}}(\tilde{\tau} - \tilde{\tau}_{\isc})\,, \\
\tilde{L} - \tL_{\isc} &=  \frac{d\tilde{L}_{\isc}}{d\tilde{\tau}}(\tilde{\tau} - \tilde{\tau}_{\isc}). 
\end{aligned}
\end{equation}
\newline
\\
In our analysis of the transition, we will {\it not} assume a strict equality in Eq.\eqref{Linear_Time_Ang_Momenta}. Instead, we will keep track of the evolution of $\tilde{\Omega}_\isc ^{-1}(\tilde{E} - \tE_{\isc}) - (\tilde{L} - \tL_{\isc})$, as also considered in \cite{kesden2011transition}. 

OT proposed to analyse the transition to the plunging geodesic by expanding \eqref{eq:acceleration} around the 
ISCO trajectory $(\tr_\isc,\tE_\isc,\tL_\isc)$, since the latter provides the natural starting point for the plunging trajectory for equatorial and circular orbits. It is physically natural to introduce the new variables $\delta E$, $\delta L$ and $R$
\begin{equation}
\begin{aligned}
 \delta E &= \tilde{\Omega}_{\isc}^{-1}(\tilde{E} - \tE_{\isc})   \\
\delta L &= \tilde{L} - \tL_{\isc}  \label{eq:per-variables}\\
R &= \tilde{r} - \tr_{\isc}
\end{aligned}    
\end{equation}

to study the inspiral evolution of the small body perturbatively around the primary. The presence of $\tilde{\Omega}_{\isc}$ is for technical convenience.

Instead of expanding \eqref{eq:acceleration}, we find it more convenient to expand \eqref{CH:8 Radial Geodesic Equation}. Our conclusions do not depend on this choice. The latter is given by
\begin{widetext}
\begin{multline}
\left(\frac{d\tr}{d\tilde{\tau}}\right)^{2} = G(\tilde{r}_\isc,\tE_\isc,\tL_\isc)  + \sum_{i=1}^{\infty}\frac{1}{i!}\frac{\partial^{i} G}{\partial \tilde{r}^{i}}\bigg\rvert_{\isc}(\tilde{r} - \tr_{\isc})^{i}  \\
+ \sum_{i=0}^{\infty}\frac{1}{i!} \left(\frac{\partial^{i+1} G}{\partial \tilde{r}^{i} \partial \tE}\bigg\rvert_{\isc} (\tE - \tE_{\isc}) + \frac{\partial^{i+1} G}{\partial \tilde{r}^{i} \partial \tilde{L}}\bigg\rvert_{\isc} (\tilde{L} - \tL_{\isc})\right)  (\tilde{r} - \tr_{\isc})^{i} \\
\frac{1}{2}\sum_{i=0}^{\infty}\frac{1}{i!} \left(\frac{\partial^{i+2} G}{\partial \tilde{r}^{i} \partial \tE^{2}}\bigg\rvert_{\isc} (\tE - \tE_{\isc})^{2} + 2\frac{\partial^{i+2} G}{\partial \tilde{r}^{i} \partial \tilde{L}\partial \tilde{E}}\bigg\rvert_{\isc} (\tilde{E} - \tE_{\isc})(\tilde{L} - \tL_{\isc}) + \frac{\partial^{i+2} G}{\partial \tilde{r}^{i} \partial \tL^{2}}\bigg\rvert_{\isc} (\tilde{L} - \tL_{\isc})^{2}\right)  (\tilde{r} - \tr_{\isc})^{i}\,.
\label{eq:gexpansion}
\end{multline}
Since $G(\tr,\tE,\tL)$ is quadratic in $\tE$ and $\tL$, we have ignored the derivatives
\begin{equation}\label{app:energy_angularmomentum_vanish}
\frac{\partial^{n}G}{\partial \tilde{E}^{n}} = \frac{\partial^{n}G}{\partial \tilde{L}^{n-k}\partial \tilde{E}^{k}} = \frac{\partial^{n}G}{\partial\tilde{E}^{n-k}\partial\tilde{L}^{k}} = \frac{\partial^{n}G}{\partial{\tL}^{n}} = 0 \quad \text{for} \ n\geq 3 \ \text{and} \ k<n.  
\end{equation}
Plugging the perturbative variables \eqref{eq:per-variables}, using the definition of the coefficients \eqref{eq:Taylor-coe} and the results in \eqref{App:General_Radial_Derivative}-\eqref{App:General_Mixed_Radial_First_E2_EL_LL},
one can rewrite the general transition equation as
\begin{equation}\label{app:B:General_Master_Equation}
\left(\frac{dR}{d\tilde{\tau}}\right)^{2} =  \sum_{n=3}^{\infty}\frac{1}{n!} A_n\,R^{n} + \delta L 
\sum_{n=1}^{\infty}\frac{1}{n!} B_n\,R^{n} + \frac{\delta L^{2}}{2}\sum_{n=0}^{\infty}\frac{1}{n!} C_n R^{n}  
+ \Gamma_{\odot}\,,
\end{equation}
where $\Gamma_{\odot}$ is defined by
\begin{equation}
\label{app:B_Gamma_odot_definition}
\begin{aligned}
\Gamma_{\odot} &= \frac{1}{2}\sum_{n=0}^{\infty}\frac{1}{n!}\tilde{\Omega}_{\isc}(\delta E - \delta L)\left(2\frac{\partial^{n+1}G}{\partial \tilde{r}^{n} \partial \tE}\bigg\rvert_{\isc} + 2\left(\frac{\partial^{n+2}G}{\partial \tilde{r}^{n}\partial \tilde{E} \partial \tilde{L}}\bigg\rvert_{\isc}+ \tilde{\Omega}_{\isc}\frac{\partial^{n+2}G}{\partial \tilde{r}^{n}\partial \tilde{E}^{2}}\bigg\rvert_{\isc} \right) \delta L \right. \\
& \left.+  \tilde{\Omega}_{\isc}\frac{\partial^{n+2}G}{\partial \tilde{r}^{n}\partial \tilde{E}^{2}}\bigg\rvert_{\isc}( \delta E - \delta L) \right)R^{n}\,.
\end{aligned}
\end{equation}
\end{widetext}
Notice that $\Gamma_{\odot}\propto \delta E - \delta L$ at leading order in $R$. Hence, it encodes the deviations from the OT approximation \eqref{Linear_Time_Ang_Momenta}.

The time evolution of $\delta E - \delta L$ near $\tr_{\isc}$ is controlled by the fluxes and the angular velocity. Throughout a quasi-circular inspiral far from ISCO, the compact object inspirals on a sequence of circular geodesics defined by the constants of motion $\tilde{E}(\tilde{r}_{circ}) = \tilde{E}_{circ}$ and $\tilde{L}(\tilde{r}_{circ}) = \tilde{L}_{circ}$, as given in Eq (\ref{CH:8 Circular Energy}) and Eq.(\ref{CH:8 Circular Ang Mom}) respectively. The evolution of the constants of motion is linked through Eq.~\eqref{app:OT_Flux_Balance_Law} above, which simply states that circular geodesics evolve into circular geodesics. It can be shown that solutions to the Teukolsky equation for circular orbits obey this condition~\cite{CircularOrbitsOri,1998PhRvD..58f4012K}. For circular evolutions we therefore see that
\begin{equation}
\label{app:A_Proper_Time_Evolution_dE}
\begin{aligned}
\frac{d}{d\ttau}(\delta E - \delta L) &= \tilde{\Omega}_{\isc}^{-1}\frac{d\tilde{E}}{d\tilde{\tau}} - \frac{d\tilde{L}}{d\tilde{\tau}}  \\
&= (\tilde{\Omega}_{\isc}^{-1}\tilde{\Omega}(\tilde{r}) - 1)\frac{d\tilde{L}}{d\tilde{\tau}}  \\
&\approx -\frac{\partial \log\tilde{\Omega}}{\partial \tilde{r}}\bigg\rvert_{\isc}\eta\kappa R \\
\Longrightarrow (\delta E - \delta L) & \sim \eta R\ttau \,, 
\end{aligned}
\end{equation}
where we expanded $\tilde{\Omega}(r)$ to first order in $R$ and approximated $d\tilde{L}/d\tilde{\tau} \approx (d\tilde{L}/d\tau)_{\isc} = -\eta\kappa$ for $\kappa$ constant defined by
\begin{equation}
\kappa = \left(\tilde{\Omega}^{-1}\frac{d\tilde{t}}{d\ttau}\frac{d\tE_{GW}}{d\tilde{t}}\right)_{\isc} \sim \mathcal{O}(1)\, \quad  \text{for} \ a\in[0,1].  
\end{equation}
Thus we deduce that $\delta E - \delta L \sim \eta R\tilde{\tau}$ for circular inspirals close to $\tr_\isc$. We shall see that these corrections are indeed subleading in the regime considered by OT~\cite{2000PhRvD..62l4022O}. However, they will not be negligible for near-extremal black holes.

\subsection{Corrections arising from deviations from adiabatic nearly-circular inspiral}
\label{sec:ecc}

Given our assumption that the orbit is nearly circular when it reaches the transition regime, one expects corrections to the relation \eqref{app:OT_Flux_Balance_Law} between the fluxes of energy and angular momentum satisfied for an exactly equatorial circular adiabatic inspiral. We discuss below two possible physical effects giving rise to such corrections : \emph{eccentricity} and the \emph{non-geodesic past-history of the orbital evolution}. These will give rise to the corrections \eqref{eq:correction-flux}.

\emph{Eccentricity} can lead to corrections to the transition equation which we will discuss further below, but eccentricity corrections to the fluxes tend to be suppressed during the transition regime. This is because the transition, for an arbitrary eccentric inspiral, corresponds to the orbit passing over the maximum of the effective potential given by Eq.\eqref{CH:8 Radial Geodesic Equation}. 
%\jg{NOTE: This was correct because we were talking about eccentric inspirals here [which is now clarified] . Later we say it is even less important for near-circular inspirals because in that case it is a point of inflection.} 
The radial velocity throughout the transition regime is therefore always small, while the angular velocity remains $\mathcal{O}(1)$. Hence the orbit looks very much like a circular orbit, even if it is technically eccentric or even plunging. For nearly-circular transitions, the orbit is passing over a point of inflection of the effective potential and corrections to this approximately-circular assumption are even smaller. 

Corrections from \emph{non-geodesic past-history} enter because the self-force acting on the small object at a particular time is generated by the intersection of the particle world line with gravitational perturbations generated by the orbital motion in the immediate past~\cite{LRP}. The self-force acting on the orbit when it is at a particular radius will therefore have corrections that depend on how far, in radius, the orbit has moved over the relevant past-history. The latter is determined by the dominant, azimuthal, timescale, and is an $\mathcal{O}(1)$ quantity, when expressed in coordinate time~\footnote{If we are more conservative, we could assume that the timescale for radial oscillations is the appropriate averaging timescale. This is not $\mathcal{O}(1)$, but $\mathcal{O}(T)$, the scaling of the time coordinate in the transition zone. While this condition is more restrictive we will see below that even this condition does not change the conclusion that past-history corrections can be ignored in the transition zone.}. The orbital radius therefore changes by an amount of $\mathcal{O}(\dot{\tilde{r}})$ over the relevant past-history. This is the scaling of the fractional change in the fluxes, and since $\dot{\tilde{E}} \sim \mathcal{O}(\eta \epsilon^{\frac{2}{3}})$, the non-geodesic past-history corrections to the coordinate-time fluxes thus scale like $\eta \epsilon^{2/3} \dot{\tilde{r}}$. In the regime $\eta \ll  \dot{\tilde{r}}$, considered by OT, and discussed in Section~\ref{ssec:OT}, $\epsilon$ can be considered $\mathcal{O}(1)$ and so the scaling of this correction is $\eta \dot{\tilde{r}}$. This is the first type of correction in Eq.\eqref{eq:correction-flux}. In the adiabatic inspiral phase, these corrections are $\mathcal{O}(\eta^2)$ and form part of the second-order component of the self-force. However, in the transition phase these corrections can be larger.

We have argued above that eccentricity corrections to the fluxes should be suppressed in the transition regime. We now make this more concrete. Eccentricity corrections to the fluxes enter as fractional corrections of $\mathcal{O}(e^2)$, since corrections to the orbit at linear order in eccentricity are oscillatory and average to zero over a complete orbit~\cite{1998PhRvD..58f4012K}. The corrections to the coordinate time fluxes thus scale like $\eta \epsilon^{\frac{2}{3}} e^2$ (which is $\eta e^2$ in the OT regime discussed in Section~\ref{ssec:OT}). This is the second type of correction in Eq.\eqref{eq:correction-flux}. If these corrections are to be small relative to the non-geodesic past-history corrections, we need $e^2 < \dot{\tilde{r}}$. In the transition zone we will see that the proper time scales like $R^{-1/2}$, where $R = \tr - \tr_{\isc}$ is the distance from the ISCO, regardless of the spin of the primary. For non near-extremal black holes, i.e., those with $\eta \ll \epsilon$, proper time and coordinate time scale in the same way and the scaling of $\dot{\tilde{r}}$ is therefore the same as that of $R^{3/2}$. The constraint we obtain on eccentricity is therefore $e < R^{3/4}$. However, there is also a geometric constraint, which is that the variation in the orbital radius due to eccentricity should be small compared to the variation due to radiation reaction through the transition zone. The latter is the scaling of $R$, while the former is a quantity of $\mathcal{O}(e)$, so we deduce an additional constraint $e < R < R^{3/4}$, the latter inequality following from the fact that $R$ is a small quantity throughout the transition. We deduce that the geometrical constraint is stronger than the flux-correction constraint in the regime $\eta \ll \epsilon$. In the near-extremal regime, $\eta \lesssim \epsilon$, $d\tilde{t}/d\tilde{\tau} \sim \epsilon^{-2/3}$ and so the constraint on the eccentricity changes to $e < \epsilon^{1/3}R^{3/4}$ if these corrections are to be subleading. This is then more stringent than the geometric constraint. However, in this regime we will see below that eccentricity cannot grow until deep inside the transition zone, so even the more stringent constraint is easily satisfied.
%In the transition zone we will see that time scales like $R^{-1/2}$, where $R = \tr - \tr_{\isc}$ is the distance from the ISCO as in \eqref{eq:per-variables}. The scaling of $\dot{\tilde{r}}$ is therefore the same as that of $R^{3/2}$, and so the constraint we obtain on eccentricity is $e < R^{3/4}$. However, there is also a geometric constraint, which is that the variation in the orbital radius due to eccentricity should be small compared to the variation due to radiation reaction through the transition zone. The latter is the scaling of $R$, while the former is a quantity of $\mathcal{O}(e)$, so we deduce an additional constraint $e < R < R^{3/4}$, the latter inequality following from the fact that $R$ is a small quantity throughout the transition. We deduce that the geometrical constraint is always stronger than the flux-correction constraint, and so it is this that we must verify in deriving the transition equations of motion.

Eccentricity during the transition can arise either from the presence of residual eccentricity prior to the start of the transition zone, or due to the excitation of eccentricity during the transition. The latter manifests itself as additional terms in the transition equation, the existence of which we will check for carefully in our analysis. To understand the former, we need to analyse the growth of eccentricity during the adiabatic inspiral. We will assume that at the beginning of the inspiral the orbit is nearly circular. It was shown in~\cite{1998PhRvD..58f4012K} that, for small eccentricity, the evolution of eccentricity under radiation reaction takes the form $\dot{e} = f(\tilde{r}_0) e$, where $\tilde{r}_0$ is the mean orbital radius and $e$ is an eccentricity defined such that the orbital apoapsis is at $\tilde{r}=\tilde{r}_0(1+e)$. For large $\tilde{r}_0$, $f(\tilde{r}_0) < 0$ and so the eccentricity decreases. In this regime any small eccentricity that is excited by small perturbations arising due to inspiral evolution or other effects is damped away and does not grow. However, for all spins $a < 1$, as the innermost stable circular orbit (or separatrix) is approached the sign of $f(\tilde{r}_0)$ changes and is greater than zero in the vicinity of the ISCO. This means that orbits near to the separatrix are unstable to eccentricity growth. We would therefore expect any eccentricity that is excited to begin to grow.

Denoting $\tilde{v}^2=1/\tilde{r}_0$, Kennefick~\cite{1998PhRvD..58f4012K} showed that the evolution of the orbital parameters, for small eccentricity, was governed by equations of the form
\begin{align}
\frac{\dot{\tilde{r}}_0}{\tilde{r}_0} &=-\frac{2(1-3\tilde{v}^2+2a\tilde{v}^3)^{3/2}}{\tilde{v}^2(1-6\tilde{v}^2+8a\tilde{v}^3-3a^2\tilde{v}^4)}\dot{\tE}_0\\
\frac{\dot{e}}{e} &= \frac{1}{e^2} \left(\dot{\tE}_0 - \Omega(\tilde{v}) \dot{\tL}_0\right) - j(\tilde{v}) \left[\Gamma - h(\tilde{v}) \dot{\tE}_0)\right] \label{kenneccev} \\
\mbox{where } &\nonumber \\
j(\tilde{v}) &= \frac{(1+a\tilde{v}^3)(1-2\tilde{v}^2+a^2 \tilde{v}^4) (1-3\tilde{v}^2+2a\tilde{v}^3)^{1/2}}{\tilde{v}^2 (1-6\tilde{v}^2+8a\tilde{v}^3-3a^2 \tilde{v}^4)}\nonumber \\
h(\tilde{v}) &= \frac{{\cal H}(\tilde{v})(1+a\tilde{v}^3)^{-1}(1-2\tilde{v}^2+a^2\tilde{v}^4)^{-2}}{2(1-6\tilde{v}^2+8a\tilde{v}^3-3a^2\tilde{v}^4)}\nonumber \\
{\cal H}(\tilde{v})&=1-12\tilde{v}^2+66\tilde{v}^4-108\tilde{v}^6 + a\tilde{v}^3+8a^2\tilde{v}^4\nonumber \\&\hspace{0.3cm}-72a\tilde{v}^5-20a^2\tilde{v}^6+204a\tilde{v}^7+38a^3\tilde{v}^7-42a^2\tilde{v}^8\nonumber \\
&\hspace{0.3cm}-9a^4\tilde{v}^8-144a^3\tilde{v}^9+116a^4\tilde{v}^{10}-27a^5\tilde{v}^{11} \nonumber.
\end{align}
Both $\Gamma$ and $\dot{\tE}_0$ are components of the self-force, which can be evaluated by solving the Teukolsky equation. The quantity $\Gamma$ is in fact a linear combination of quantities that are time derivatives and so the above equation takes the same form for any choice of time coordinate with respect to which to evaluate the fluxes. Kennefick's analysis used coordinate time and so we make the same choice in the following discussion. An explicit expression for $\Gamma$ is given in \cite{1998PhRvD..58f4012K} and the quantity $\dot{\tilde{E}}_{0}$ is the energy flux given in \eqref{Energy_Flux}. Numerical calculations show that these are finite quantities of $\mathcal{O}(1)$ throughout parameter space. The first term in the eccentricity evolution equation vanishes for evolution driven by gravitational radiation reaction, while the quantity $h(\tilde{v})$ is singular at the ISCO. Therefore, close to ISCO the eccentricity evolution takes the form
\begin{align}
\frac{\dot{e}}{e} &\approx j(\tilde{v}) h(\tilde{v}) \dot{\tE}_0 \nonumber \\
\Rightarrow \quad r_0 \frac{{\rm d}\ln e}{{\rm d} r_0} &\approx - \frac{\tilde{v}^2(1-6\tilde{v}^2+8a\tilde{v}^3-3a^2\tilde{v}^4) j(\tilde{v}) h(\tilde{v})}{2 (1-3\tilde{v}^2+2a\tilde{v}^3)^{3/2}}.
\end{align}
Notice that the expression is entirely geodesic and independent of the energy flux $\dot{\tE}_{0}$. For non-extremal spin, both $j(\tilde{v})$ and $h(\tilde{v})$ have simple poles at $\tilde{r}=\tilde{r}_\isc$ and there is a simple zero in the term $(1-6\tilde{v}^2+8a\tilde{v}^3-3a^2\tilde{v}^4)$ in the numerator. Therefore as ISCO is approached the eccentricity evolves as
\begin{equation}
\frac{{\rm d}\ln e}{{\rm d}R} \approx - \frac{k(a)}{R} \quad \Rightarrow \quad e = e_0 \left( \frac{R_0}{R}\right)^{k(a)} \label{eq:expecgrth}
\end{equation}
with  $R = \tilde{r}-\tilde{r}_\isc$ as before, and $e_0$ denotes the eccentricity when $R=R_0$ and $\tr_{0} \gg \tr_{\isc}$. The exponent $k(a)$ is given by
\begin{align}
    k(a) &= {\cal H}(\tilde{v}_\isc)/{\cal D}(\tilde{v}_\isc) \nonumber\\
    \mbox{where } {\cal D}(\tilde{v}) &= 2 \tilde{v}^2 (1 -2\tilde{v}^2 + a \tilde{v}^4) \nonumber \\
    & \hspace{0.3cm} \times (12 \tilde{v} - 24 a \tilde{v}^2 + 12 a^2 \tilde{v}^3) \nonumber \\
    & \hspace{0.3cm} \times(1-3\tilde{v}^2+2a\tilde{v}^3) 
\end{align}
and $\tilde{v}^2_\isc=1/\tr_\isc $. We find that $k(a) = 1/4$ for all $a < 1$. The behaviour for near-extremal black holes is slightly different, which we will discuss further below.

For extremal black holes the various factors in the expression for ${\rm d}\ln e/{\rm d}\tilde{r}_0$ have repeated roots at the ISCO. To understand the behaviour for near-extremal black holes we therefore need to do an expansion in both $R$ and $\epsilon$. This takes the form
\begin{equation}
\frac{{\rm d}\ln e}{{\rm d}R} = \frac{a_0 \epsilon^4 + a_1  \epsilon^4 R+ \sum_{i=2}^5 a_i \epsilon^{\frac{2(6-i)}{3}} R^i + a_6 R^6 + \cdots}{ \sum_{i=1}^6 b_i \epsilon^{\frac{2(7-i)}{3}} R^i + b_7 R^7 + \cdots}
\end{equation}
The terms omitted from both the numerator and denominator above are $\mathcal{O}(1)$ in $\epsilon$. The ratio $a_0/b_1 = -1/4$, agreeing with the result for $k(a)$ found above. However, for $\epsilon \ll R$, the behaviour is not dominated by this term, but by the terms from $a_6$ in the numerator and from $b_7$ in the denominator. The leading order behaviour in this regime is therefore 
%\begin{equation}
%\frac{{\rm d}\ln e}{{\rm d}\tilde{r}_0} = \frac{a_6}{b_7} \frac{1}{R}.
%\end{equation}
\begin{equation}
\frac{{\rm d}\ln e}{{\rm d}R} = \frac{a_6}{b_7} \frac{1}{R}.
\end{equation}
This is also exponential, but we find the ratio $a_6/b_7 = 3/2$, i.e., it is greater than zero and therefore the eccentricity \emph{decreases} exponentially until we reach the regime $R \sim \epsilon$. This is the statement that the critical curve, where the sign of the eccentricity evolution changes, is in the near-horizon region, which is consistent with results in~\cite{2002PhRvD..66d4002G}. We conclude that for near-extremal black holes, eccentricity can only grow once the inspiraling object is already very close to the ISCO, which is typically already inside the transition zone.

To complete this discussion we need to determine the scaling of the initial eccentricity $e_0$. If the orbit is truly circular then the eccentricity remains zero, so there must be some mechanism to excite an initial eccentricity which can then grow. Eccentricity can be excited by other physical processes, such as the presence of perturbing material, e.g., dust, or gravitational interactions with third bodies. Those processes are important, but in the pure-vacuum case eccentricity could still in principle be excited by the evolution under radiation reaction. We argued earlier that corrections to the fluxes far from the horizon scale like $\eta \dot{\tilde{r}}$ which is $\eta^2$ during the adiabatic inspiral. These corrections mean that the first term in Eq.~(\ref{kenneccev}) is no longer exactly zero. Setting that term to $\eta^2$ we find an evolution equation of the form ${\rm d}e^2/{\rm d}\tilde{t} \sim \eta^2$. After a few orbits the eccentricity is then $\mathcal{O}(\eta)$. This eccentricity induced by second order corrections to the evolution is damped by the process described above, until we reach the critical curve where it grows, eventually exponentially near the ISCO. This suggests appropriate initial conditions are $e_0 \sim \eta$ and $R_0 \sim \mathcal{O}(1)$.\footnote{A natural continuation of this argument would be to say that the second-order self-force induced corrections continue to drive eccentricity growth, over the whole of the inspiral, lasting a coordinate time $\sim \eta^{-1}$, leading to a final eccentricity of $\mathcal{O}(\eta^{1/2})$, which can be larger than the eccentricity grown through the mechanism discussed here. However, this assumes that the eccentricity grows coherently and monotonically. In practice, once the eccentricity is $\mathcal{O}(\eta)$, the radial motion due to eccentricity becomes larger than the amount the radius evolves over the relevant past-history that determines the self-force and so the argument that the latter is the dominant contribution to corrections no longer applies. Knowledge of the second-order self-force would be required to fully explore the further evolution of the eccentricity and this is not currently available. However, we expect that the growth of initial eccentricities of $\cal{O}(\eta)$ through the instability mechanism will be the dominant contributor to the residual eccentricity in the transition zone.} We note that this mechanism could also excite eccentricity during the transition zone itself, but this would be of order $e^2 \sim \eta \dot{\tilde{r}} \epsilon^{\frac{2}{3}}$ and hence no larger than the non-geodesic past-history corrections described above. If eccentricity grew coherently throughout the transition zone, the eccentricity induced by this process would be no larger than $e^2 \sim \eta \dot{\tilde{r}} \epsilon^{\frac{2}{3}} T$, where $T$ is the coordinate time elapsed through the transition zone, which is typically smaller than the eccentricity grown during adiabatic inspiral prior to the start of the transition zone.

To summarise, we expect corrections to the evolution equations that arise from higher-order terms in the flux to scale like $\eta \epsilon^{\frac{2}{3}} \dot{\tr}$ (which is $\eta \dot{\tr}$ in the OT regime discussed in Section~\ref{ssec:OT}), and we expect residual eccentricity in the transition zone to be no more than $e \sim \eta R^{-k(a)}$. In the non-near extreme case, these eccentricity corrections will be important when $e > R$, which implies $R < \eta^{1/(1+k(a))}$. In the near-extreme case, the corrections only become important when $R\sim \epsilon$, so we simply need to check that this is well inside the transition zone. In the analysis that follows we will evaluate the scaling of these terms and show that they are sub-dominant for inspirals into near-extremal black holes.

\subsection{Ori and Thorne regime}
\label{ssec:OT}

Consider non-extremal black holes, i.e. rotating black holes where the extremality parameter $\epsilon$ is not close to zero so that $\eta\ll \epsilon$. In this regime of spins and according to the discussion below \eqref{App:General_Radial_Extremal}-\eqref{app:ISCO_EE_Expansion}, 
 all the coefficients controlling the general transition equation \eqref{app:B:General_Master_Equation} and \eqref{app:B_Gamma_odot_definition} are $\mathcal{O}(1)$. This is the regime originally discussed in \cite{2000PhRvD..62l4022O}. 

Omitting coefficients of order one, the dominant contributions to the transition equation are
\begin{equation}
\begin{aligned}
  \left(\frac{dR}{d\tilde{\tau}}\right)^{2} &\sim R^3 + R\,\delta L + \Gamma_\odot \\
  \Gamma_\odot &\sim \delta E- \delta L\,,
\end{aligned}
\end{equation}
where we also omitted any further terms from \eqref{app:B:General_Master_Equation} and \eqref{app:B_Gamma_odot_definition} since they are subleading. Looking for a scaling solution $R\sim \eta^p$ and $\ttau\sim \eta^q$, it follows, using equation \eqref{Linear_Time_Ang_Momenta} that $\delta L\sim \eta^{1+q}$. Requiring all dominant terms to have the same scaling fixes $p=2/5$ and $q=-1/5$, so that
\begin{equation}
R = \eta^{2/5}\mathcal{R}\,, \quad
\tau = \eta^{-1/5}\mathcal{T} \,, \quad
\delta L =  \eta^{4/5}\delta \mathcal{L}\,.
\label{app:small_eta_variables_OT}
\end{equation}

Notice the overall scaling of the transition equation is $(d\tilde{r}/d\tilde{\tau})^2\sim \eta^{6/5}$. The remaining question is whether the dominant term in $\Gamma_\odot \sim  \delta E- \delta L$ is subleading or not. 
From \eqref{app:A_Proper_Time_Evolution_dE}, it follows $\delta E - \delta L \sim \eta^{6/5}$ in this regime, suggesting the change of variables
\begin{equation}
  \Gamma_\odot = \eta^{6/5}\mathcal{Y}\,.
\end{equation}
This allows to write the schematic transition equation as
\begin{equation}
  \left(\frac{d\mathcal{R}}{d\mathcal{T}}\right)^{2} \sim \mathcal{R}^3 + \mathcal{R}\delta\mathcal{L} + \mathcal{Y}\,.
\end{equation}
Terms in Eqs.~\eqref{app:B:General_Master_Equation} and 
\eqref{app:B_Gamma_odot_definition} that have been dropped can be seen to scale like the above terms multiplied by additional powers of $\mathcal{R}$ or $\delta\mathcal{L}$. Since both $\mathcal{R}$ and $\delta\mathcal{L}$ are small quantities in the transition zone, these terms are sub-leading and we can ignore them.

%\js{perhaps we can comment on the size of the corrections dropped in \eqref{app:B:General_Master_Equation} and  \eqref{app:B_Gamma_odot_definition} to tie with the discussion below}

The above scaling analysis proves the dominant terms in \eqref{app:B:General_Master_Equation} in the regime $\eta\ll \epsilon$ are captured by%\ob{You have a $\gamma\delta L R^{2}$ term here which we show through the above analysis is subleading. I'm going to remove it for now but if you have an argument for keeping it here then let me know.}
\begin{comment}
\begin{equation}
\label{eq:regime-1}
\left(\frac{dR}{d\tilde{\tau}}\right)^{2} \simeq -\frac{2}{3}\alpha R^{3} + 2\beta\,\delta L\,R + \gamma\,\delta L\,R^{2} + \Gamma_{\odot} + \dots
\end{equation}
\end{comment}
\begin{equation}
\label{eq:regime-1}
\left(\frac{dR}{d\tilde{\tau}}\right)^{2} \simeq -\frac{2}{3}\alpha R^{3} + 2\beta\,\delta L\,R + \Gamma_{\odot} + \dots
\end{equation}
where we neglected all subleading corrections, kept the same original notation as in OT \cite{2000PhRvD..62l4022O} for the coefficients 
\begin{comment}
\begin{align}
\alpha &= -\frac{1}{4}\frac{\partial^{3}G}{\partial \tilde{r}^{3}}\bigg\rvert_{\isc} \label{eq:alpha} \\
\beta &= \frac{1}{2}\left(\frac{\partial^{2}G}{\partial\tilde{r}\partial{\tilde{E}}}\tilde{\Omega} + \frac{\partial^{2}G}{\partial\tilde{r}\partial{\tilde{L}}}\right)_{\isc} \label{eq:beta} \\
\gamma &= \frac{1}{2}\left(\frac{\partial^{3}G}{\partial\tilde{r}^{2}\partial{\tilde{E}}}\tilde{\Omega} + \frac{\partial^{3}G}{\partial\tilde{r}^{2}\partial{\tilde{L}}} \right)_{\isc}\,,\label{eq:gamma}
\end{align}
\end{comment}
\begin{align}
\alpha &= -\frac{1}{4}\frac{\partial^{3}G}{\partial \tilde{r}^{3}}\bigg\rvert_{\isc} \label{eq:alpha} \\
\beta &= \frac{1}{2}\left(\frac{\partial^{2}G}{\partial\tilde{r}\partial{\tilde{E}}}\tilde{\Omega} + \frac{\partial^{2}G}{\partial\tilde{r}\partial{\tilde{L}}}\right)_{\isc} \label{eq:beta} 
\end{align}
and the dominant contribution to \eqref{app:B_Gamma_odot_definition} reduces to
\begin{equation}\label{odot-1}
\Gamma_{\odot} \simeq \tilde{\Omega}_{\isc}(\delta E - \delta L) \frac{\partial G}{\partial \tilde{E}}\bigg\rvert_{\isc} + \dots
\end{equation}

Keeping all coefficients of order one, the natural scaled variables to introduce are
\begin{equation}
\begin{aligned}\label{scalings_OT}
R &= \eta^{2/5}\alpha^{-3/5}(\beta\kappa)^{2/5}X \\
\tilde{\tau} - \tilde{\tau}_{\isc} &= \eta^{-1/5}(\alpha\beta\kappa)^{-1/5}T \\
\delta E - \delta L &= \eta^{6/5} Y  \\
\delta L &= -\eta^{4/5}(\alpha\beta)^{-1/5} \kappa^{4/5} T  
\end{aligned}
\end{equation}
where
\begin{equation}
\label{eq:kappa}
\kappa = \left(\tilde{\Omega}^{-1}\frac{d\tilde{t}}{d\ttau}\frac{d\tilde{E}_{\text{GW}}}{d\tilde{t}}\right)_{\isc}\,.
\end{equation}
Plugging this into \eqref{eq:regime-1}, one obtains
\begin{equation*}
\left(\frac{dX}{dT}\right)^{2} = -\frac{2}{3}X^{3} - 2XT + C_{0}\left(\tilde{\Omega}\frac{\partial G}{\partial E}\right)_{\isc}Y + \mathcal{O}(\eta^{2/5})
\end{equation*}
where we defined $C_{0} = \alpha^{4/5}(\kappa\beta)^{-6/5}$. From now on, we ignore the subleading corrections.

The analogue of the acceleration equation \eqref{eq:acceleration} reduces to 
\begin{equation}
\label{Eq:Almost_There_OT_Eqn}
\begin{aligned}
\frac{d^{2}X}{dT^{2}} &= -X^{2} - T \\
& - \frac{1}{2(dX/dT)}\left(2X - C_0\left[\tilde{\Omega}\frac{\partial G}{\partial \tilde{E}}\right]_{\isc} \frac{dY}{dT}\right)
\end{aligned}
\end{equation}
This depends on the time evolution of the circularity deviation parameter $Y$, whose dominant contribution is derived in \eqref{app:A_Proper_Time_Evolution_dE}. Inserting the re-scaled variables \eqref{scalings_OT} into Eq.\eqref{app:A_Proper_Time_Evolution_dE}
\begin{equation}
\frac{dY}{dT} = -\frac{\partial \log\tilde{\Omega}}{\partial \tilde{r}}\bigg\rvert_{\isc}(\beta C_0)^{-1}X\,,
\label{Eq:OT_eqn_Y}
\end{equation}
leads to a transition equation 
\begin{equation*}
\frac{d^{2}X}{dT^{2}} = -X^{2} - T - \frac{1}{2(dX/dT)}\left(2X + \beta^{-1}\left[\frac{\partial \tilde{\Omega}}{\partial \tilde{r}}\frac{\partial G}{\partial \tilde{E}}\right]_{\isc}X\right)\,.
\end{equation*}
Evaluating \eqref{eq:circular-2} at ISCO, we find the term in square brackets equals
\begin{equation}\label{Low_Spin_Derivative_Identity}
 \left(\frac{\partial \tilde{\Omega}}{\partial \tilde{r}}\frac{\partial G}{\partial \tilde{E}}\right)_{\isc} = -2\beta
\end{equation}
and so the last term vanishes. This was inevitable, since this term is precisely the term that arises from the dissipative part of $f^{\tilde{r}}$ from Eq.\eqref{splitting_self_force}, as identified earlier. The leading order evolution of $Y$ is driven by maintaining the circularity of the orbit and so with this condition we expect the radial self-force corrections to be subleading.
%This is not surprising since imposing circularity through the flux balance law \eqref{Eq:Circ_Go_Circ} cancels the radial self-force term on the right hand side of \eqref{eq:acceleration} using the identity \eqref{eq:circular-1}. 

The resulting transition equation of motion in the regime of low spins $\eta \ll \epsilon$ is
\begin{equation}\label{Eq:OT_Eqn}
\frac{\dd ^{2}X}{\dd T^{2}} = -X^{2} - T 
\end{equation}
and $Y$ is evolved through the ODE \eqref{Eq:OT_eqn_Y}. We note that the transition equation does not depend on $Y$ in this regime. Corrections to this equation arising from evolution of $Y$ enter at an order $\eta^{2/5}$ higher than leading and so are subdominant. As discussed earlier the evolution of $Y$ is related to deviations from the linear-in-proper-time evolution of energy and angular momentum and so the fact that these corrections do not enter the transition equation for $\eta \ll \epsilon$ demonstrate that the linear evolution assumed by OT is appropriate in this regime.

Let us check the self-consistency of the transition equation \eqref{Eq:OT_Eqn} by verifying that all neglected corrections to it are indeed smaller when evaluated on the scaling regime \eqref{scalings_OT}. First, as discussed in section~\ref{sec:self-force}, the corrections to the orbit due to the conservative piece in the self-force are order $\mathcal{O}(\eta)$, see \eqref{eq:sf-radial}. This is indeed smaller than the "radial thickness" $R\sim \eta^{2/5}$ in \eqref{scalings_OT}. Second, corrections due to $\eta \dot{\tr}$, appearing in \eqref{eq:correction-flux}, are $\mathcal{O}(\eta^{8/5})$. Hence, these corrections are $\mathcal{O}(\eta^{3/5})$ smaller than the dominant $\delta L$ and $\delta E$ scaling in \eqref{scalings_OT}~\footnote{Using the more conservative assumption that the averaging timescale is determined by the period of radial oscillations, which scales with $T\sim \eta^{-\frac{1}{5}}$, the corrections are still suppressed by a factor of $\eta^{\frac{2}{5}}$. Third, 
corrections to $d\Gamma_{\odot}/d\tilde{\tau}$ arising from non-geodesic past history corrections to the fluxes scale like $\eta^{8/5}$ and those arising from additional terms in the expansion of the azimuthal frequency as a function of radius scale as $\eta^{9/5}$, which are both subdominant to the leading $\eta^{7/5}$ scaling, albeit only by a factor of $\eta^{1/5}$.}

Finally, corrections arising from eccentricity are subleading provided $e < \tr-\tr_\isc$, as discussed in Section~\ref{sec:ecc}. In the non-extremal case we therefore need $e < \eta^{\frac{2}{5}}$, due to \eqref{eq:expecgrth}. This yields the constraint
\begin{equation*}
\eta^{1-2k/5} < \eta^{2/5} \quad \Rightarrow \quad 3 - 2 k > 0 \quad \Rightarrow \quad k < \frac{3}{2}.
\end{equation*}
We saw previously that $k=1/4$ for all spins $a<1$, which satisfies this bound. We deduce that eccentricity corrections are subdominant in the non-near-extremal regime.

\subsection{General Transition Equation of Motion - Near-Extremal}\label{ssec:near-extremal}

Let us consider rapidly rotating black holes with spin parametrized by $a = \sqrt{1-\epsilon^{2}}$ for $\epsilon \ll 1$, as in \eqref{CH:8_Extremality_Parameter}. The discussion below equations \eqref{App:General_Radial_Extremal}-\eqref{app:ISCO_EE_Expansion} allows to identify the a priori dominant contributions to the transition equation 
\eqref{app:B:General_Master_Equation} as 
\begin{equation}
\begin{aligned}\label{Master_Transition_Equation_Motion}
  \left(\frac{dR}{d\tilde{\tau}}\right)^{2} &\sim R^3 + R\,\delta L\,\epsilon^{2/3} + R^{2}\delta L + \delta L^2\,\epsilon^{4/3} + \Gamma_\odot \\
  \Gamma_\odot &\sim (\delta E- \delta L) \left(\epsilon^{2/3} + R + \delta L\,\epsilon^{2/3}\right)\,.
\end{aligned}
\end{equation}

Since the functional dependence of the above equation does not depend on $\eta$, we learn the $\eta$ scaling should be the same as before if we keep the $R^3$ and $R\,\delta L$ terms. Hence, we are left to determine any possible $\epsilon$ scaling. Proceeding as before, we look for scalings of the form $R \sim \eta^{2/5}\epsilon^p$ and $\ttau \sim \eta^{-1/5}\epsilon^q$. We learn from equation \eqref{Linear_Time_Ang_Momenta} that $\delta L \sim \eta^{4/5}\epsilon^{q}$. Requiring these dominant terms to scale in the same way determines $p=4/15$ and $q=-2/15$, so that
\begin{equation}
\begin{aligned}
  R &= \eta^{2/5}\epsilon^{4/15}\mathcal{R}\,, \quad
\tilde{\tau} = \eta^{-1/5}\epsilon^{-2/15}\mathcal{T}\,, \\
\delta L &=  \eta^{4/5}\epsilon^{-2/15}\delta \mathcal{L}\,.
\label{eq:extremal-scaling}
\end{aligned}
\end{equation}
Hence, if $\eta\sim \epsilon$, the term $R^2\delta L$ scales like the velocity squared $(d\tr/d\tilde{\tau})^2 \sim \eta^{6/5}\epsilon^{4/5} \sim \epsilon^2$ and must be kept in the transition equation, whereas the term $\delta L^2 \epsilon^{4/3}$ is $\mathcal{O}(\epsilon^{2/3})$ smaller and, consequently, subdominant.

The only remaining question is whether $\Gamma_\odot$ is relevant in this regime or not. Using \eqref{app:A_Proper_Time_Evolution_dE} and the scalings \eqref{eq:extremal-scaling}, we infer $(\delta E - \delta L)\sim\eta^{6/5}\epsilon^{2/15}$.  Since in the regime $\eta\sim\epsilon$, $R\sim \delta L \sim \epsilon^{2/3}$ we conclude
$\Gamma_{\odot}\sim\eta^{6/5}\epsilon^{4/5}\sim (d\tr/d\tilde{\tau})^{2}$ and must be kept in the transition equation. Introducing the finite variable $\mathcal{Y}$
\begin{equation}
\Gamma_\odot = \eta^{6/5}\epsilon^{4/5}\mathcal{Y}\,,
\label{eq:odot-near}
\end{equation}
the general transition equation in the $\eta\sim\epsilon$ regime reduces to
\begin{equation}
  \left(\frac{d\mathcal{R}}{d\mathcal{T}}\right)^{2} \sim \mathcal{R}^3 + \mathcal{R}\delta\mathcal{L} + \mathcal{R}^{2}\delta\mathcal{L} + \mathcal{Y}\,.
\end{equation}
 Notice the radial velocity throughout the transition regime scales like $d\tr/d\tilde{\tau} \sim \eta^{3/5}\epsilon^{2/5} \sim \eta$ in the regime $\epsilon \sim \eta$. This is as in the adiabatic regime, but smaller than in the OT regime where $d\tr/d\tilde{\tau}\sim\eta^{3/5}$.

As a self-consistency check, we can write the radial geodesic equation using the change of variables \eqref{eq:extremal-scaling} and \eqref{eq:odot-near}
\begin{multline}\label{app:BTransition_Equation_OT}
\left(\frac{d\mathcal{R}}{d\mathcal{T}}\right)^{2} \sim \sum_{i=3}^{\infty}\eta^{2(i-3)/5}\epsilon^{4(i-3)/15}\mathcal{R}^{i} + \delta \mathcal{L}\mathcal{R} + \\ \sum_{m=2}^{\infty}\left(\frac{\eta}{\epsilon}\right)^{2(m-1)/5}\epsilon^{2(m-2)/3}\mathcal{R}^{m}\delta\mathcal{L} + \eta^{2/5}\epsilon^{4/15}\delta\mathcal{L}^{2}  \\ + \sum_{n=1}^{\infty}\left(\frac{\eta}{\epsilon}\right)^{2(n+1)/5}\epsilon^{2(5n - 1)/15}\delta \mathcal{L}^{2}\mathcal{R}^{n} + \mathcal{Y}.
\end{multline}
It is apparent that the dominant terms are the $i=3$ and $m=2$ terms, all others being subleading. 

The above scaling analysis proves the dominant terms in \eqref{app:B:General_Master_Equation} in the regime $\eta\sim\epsilon$ are captured by
\begin{equation}
\label{eq:regime-2}
\left(\frac{dR}{d\tilde{\tau}}\right)^{2} \simeq -\frac{2}{3}\alpha R^{3} + 2\beta\,\delta L\,R + \gamma\, \delta L\,R^{2} + \Gamma_{\odot} + \ldots
\end{equation}
where $\alpha$ and $\beta$ are defined as in \eqref{eq:alpha}-\eqref{eq:beta} and $\gamma = B_{2}$ in \eqref{app:B:General_Master_Equation}. As shown in appendix \ref{app:A:ISCO}, they are approximated by 
\begin{equation}
\alpha \to 1\,,\quad
\beta \to 2^{-2/3}\sqrt{3}\epsilon^{2/3}\equiv \hat{\beta}\,\epsilon^{2/3}\,, \quad
\gamma \to \sqrt{3}\,.
\label{near_extremal_beta}
\end{equation}
Furthermore, the dominant contributions to $\Gamma_{\odot}$ are 
\begin{equation}
\label{odot-2}
  \Gamma_{\odot} = \tilde{\Omega}_{\isc}(\delta E - \delta L) \left(\frac{\partial G}{\partial \tilde{E}}\bigg\rvert_{\isc} + \frac{\partial^{2} G}{\partial\tilde{r}\partial \tilde{E}}\bigg\rvert_{\isc} R + \ldots \right)\,.
\end{equation}

Keeping all coefficients of order one, the natural scaled variables to introduce are
\begin{equation}
\begin{aligned}\label{near-extremal-scalings}
R &= \eta^{2/5}\epsilon^{4/15}\alpha^{-3/5}(\hat{\beta}\kappa)^{2/5}X\,, \\
\tilde{\tau} - \tilde{\tau}_{\isc} &= \eta^{-1/5}\epsilon^{-2/15}(\alpha\hat{\beta}\kappa)^{-1/5}T \,, \\
\delta E - \delta L &= \eta^{6/5}\epsilon^{2/15} Y \\
\delta L &= -\eta^{4/5}\epsilon^{-2/15}(\alpha\hat{\beta})^{-1/5} \kappa^{4/5} T\,.
\end{aligned}
\end{equation}
Since $\eta \sim \epsilon$, it follows $R\sim \epsilon^{2/3}$. Hence, the near ISCO expansion corresponds to the near horizon geometry of the primary black hole since, in Boyer-Lindquist coordinates, $|\tr_{\isc} - \tilde{r}_{+}| \sim \epsilon^{2/3}$. As a result, we will be able to use the (leading order and analytic) expression for the energy flux due to gravitational radiation in \eqref{CH:8 Flux_Near_Extremal}. This allows to compute $\kappa$ in \eqref{eq:kappa} in this regime as 
\begin{align}
\kappa &= \left(\tilde{\Omega}^{-1}\frac{d\tilde{t}}{d\tilde{\tau}}\frac{d\tilde{E}}{d\tilde{t}}\right)_{\isc} \rightarrow \frac{8}{\sqrt{3}}(\tilde{C}_{H} + \tilde{C}_{\infty}).
\end{align}
Notice $\kappa \sim \mathcal{O}(1)$ since $\tilde{C}_{H} + \tilde{C}_{\infty}\sim \mathcal{O}(1)$. 

Ignoring subleading terms, the general transition equation \eqref{eq:regime-2} reduces to
\begin{equation}\label{Square_Vel_X_NHEK}
\left(\frac{dX}{dT}\right)^{2} = -\frac{2}{3}X^{3} - 2XT -(\eta/\epsilon)^{2/5}C_{1}TX^{2} + \tilde{\Gamma}_{\odot} 
\end{equation}
with
\begin{align}
C_{1} &= \gamma (\alpha\hat{\beta}\kappa)^{-3/5}\kappa \label{C1} \\
\tilde{\Gamma}_{\odot} &= \epsilon^{-4/5}\eta^{-6/5}\alpha^{4/5}(\hat{\beta}\kappa)^{-6/5}\Gamma_{\odot}.
\end{align}
Notice the appearance of the new term proportional to $TX^2$, compared to the OT regime, is due to the regime $\eta\sim\epsilon$.

Taking a further $T$ derivative, we find the analogue of the acceleration equation \eqref{eq:acceleration} in this regime
\begin{multline}\label{near_ext_trans_eqn_self_force}
\frac{d^{2}X}{dT^{2}} = -X^{2} - T - (\eta/\epsilon)^{2/5}C_{1}XT + \\ \frac{1}{2(dX/dT)}\left(-2X - (\eta/\epsilon)^{2/5}C_{1}X^{2} + \frac{d\tilde{\Gamma}_{\odot}}{dT}\right)
\end{multline}
The time evolution of $\Gamma_\odot$ in \eqref{odot-2} has two contributions : one proportional to $dY/dT$, which can be computed using \eqref{app:A_Proper_Time_Evolution_dE} and a second one proportional to $Y(dX/dT)$. Altogether yields
\begin{equation}
\begin{aligned}
\frac{d\tilde{\Gamma}_{\odot}}{dT} = & 2X - (\eta/\epsilon)^{2/5}(\alpha\hat{\beta}\kappa)^{-3/5}\kappa\left(\frac{\partial\tilde{\Omega}}{\partial\tilde{r}}\frac{\partial^{2}G}{\partial\tilde{r}\partial\tilde{E}}\right)_{\isc} X^{2} \\
& + (\eta/\epsilon)^{2/5}\alpha^{1/5}(\hat{\beta}\kappa)^{-4/5}\left(\tilde{\Omega}\frac{\partial^{2}G}{\partial\tilde{r}\partial\tilde{E}}\right)_{\isc} Y\,\frac{dX}{dT}
\end{aligned}
\end{equation}
where we used \eqref{eq:circular-2} to simplify the first term. The latter cancels the $-2X$ term in \eqref{near_ext_trans_eqn_self_force}. Using the dominant contribution to the identity \eqref{eq:circular-3} evaluated at ISCO, the second term cancels the $C_1X^2$ term in \eqref{near_ext_trans_eqn_self_force}. Finally, the third term gives a non-trivial contribution to the acceleration equation
\begin{equation}\label{near-extremal-transition-eqna}
\frac{d^{2}X}{dT^{2}} = -X^{2} - T - (\eta/\epsilon)^{2/5}(C_{1}XT - C_{2}Y) 
\end{equation}
with constant defined by
\begin{align}
C_{2} = \frac{1}{2}
\alpha^{1/5}(\hat{\beta}\kappa)^{-4/5}\left(\tilde{\Omega}\frac{\partial^{2}G}{\partial\tilde{r}\partial\tilde{E}}\right)_{\isc}. 
\end{align}
and evolution equation for $Y$ such that
\begin{equation}\label{Evolution_Equation_Ya}
\frac{dY}{dT} = -\Lambda\,\frac{\partial \log\tilde{\Omega}}{\partial \tilde{r}}\bigg\rvert_{\isc}\,X, \ \text{with} \  \Lambda = \alpha^{-4/5}\kappa^{6/5}\hat{\beta}^{1/5}.
\end{equation}

In our treatment of the OT regime (non near-extremal spins), the terms in Eq.~\eqref{near-extremal-transition-eqna} were neglected since they scaled with $\eta^{2/5}$ and were subdominant. In the near-extremal case, one can clearly see that the $XT$ and $Y$ term are comparable to the (rescaled) radial acceleration provided $\eta \sim \epsilon$. As such, they \emph{must} be included in the analysis. Our final transition equation of motion differs from Eq.(43) in~\cite{kesden2011transition}, which correctly included the $Y$ term but missed the cross term $XT$, which is the same order as the terms being retained. Our analysis improves on~\cite{kesden2011transition} in two additional ways. Firstly, $Y$ was introduced in \cite{kesden2011transition} as a mathematical construct to ensure conservation of the four-velocity norm. The evolution equation for $Y$ was derived by forcing the equation of motion obtained from differentiation of the kinetic energy equation, Eq.~\eqref{CH:8 Radial Geodesic Equation}, to agree with that obtained by expansion of the left-hand-side of the acceleration equation, Eq.~\eqref{eq:acceleration}. This is equivalent to setting the radial self-force term to zero, which is equivalent to imposing the circular-to-circular condition. This physical interpretation of the procedure was not made clear in~\cite{kesden2011transition}, nor the interpretation of $Y$ as representing departures from the linear-in-proper-time evolution. Secondly, the scaling of the flux given in Eq.~(\ref{CH:8 Flux_Near_Extremal}) was not known at that time and this was left as an unspecified power of $\epsilon$. Now that we know this scaling we can do a more complete analysis of the near-extremal regime.

The quantities above can be computed in the near-extremal limit, $\epsilon \rightarrow 0$,
\begin{equation}
\begin{aligned}\label{near-extremal:constants_evaluated}
\Lambda & =  2^{52/15}(\tilde{C}_{H} + \tilde{C}_{\infty})^{6/5}/\sqrt{3} + \mathcal{O}(\epsilon^{2/3})\\ 
C_{1}& = 2^{8/5}(\tilde{C}_{H} + \tilde{C}_{\infty})^{2/5}+ \mathcal{O}(\epsilon^{2/3})\\
C_{2} & = 2^{-13/15}\cdot 3^{-1/2}(\tilde{C}_{H} + \tilde{C} _{\infty})^{-4/5}+ \mathcal{O}(\epsilon^{2/3}). 
\end{aligned}
\end{equation}
Equations \eqref{near-extremal-transition-eqna} and \eqref{Evolution_Equation_Ya} are a coupled set of ODEs which will link the adiabatic inspiral to a plunging geodesic.

As in the previous section we now consider the size of corrections to the transition equation. Corrections to the circular flux-balance law in the geodesic part of the transition equation scale like $\eta\,\dot{\tr}\,\epsilon^{\frac{2}{3}}$ according to \eqref{eq:correction-flux}. These are $\mathcal{O}(\epsilon)$ smaller than the terms kept in this regime~\footnote{Using the conservative assumption about the averaging timescale, these corrections are sub-leading by a factor of $\eta^{\frac{2}{3}}$.}. Similarly, corrections to the linear-in-proper-time angular momentum evolution enter through corrections to $\delta E$ and $\delta L$ and scale like $d\tr/d\tilde{\tau}$ times terms that are being retained. These are therefore subdominant since $d\tr/d\tilde{\tau} \sim \eta^{3/5} \epsilon^{2/5} \ll 1$. These corrections also contribute additional terms through corrections to the radial self-force part of the transition equation. These are of order $\eta \cdot \partial G/\partial \tilde{E}$ and $\eta \cdot \partial G/\partial \tilde{L}$, which scale like $\eta \epsilon^{2/3}$ and so are a factor of $(\eta/\epsilon)^{1/5} \epsilon^{1/3}$ smaller than the leading order terms in the transition equation and are therefore sub-dominant.

Eccentricity corrections enter like fractional $e^2$ corrections to the fluxes, and are only more important than the corrections described above if $e > R \sim \eta^{2/5}$ or $e^2 > \dot{\tr}$. In the near-extremal regime $\dot{\tr} \sim  \epsilon^{2/3}\,{\rm d}r/{\rm d}\tau \sim \eta^{5/3}$ and so eccentricity corrections become important when $e > \eta^{5/6}$. However, as shown in Section~\ref{sec:ecc}, for near-extremal inspirals eccentricity can only grow once $\tilde{r}-\tilde{r}_\isc \sim \mathcal{O}(\epsilon)$. In the transition zone $\tilde{r}-\tr_\isc \sim (\eta/\epsilon)^{2/5} \epsilon^{2/3} \gg \epsilon$ and so eccentricity has not started to grow when the transition zone is reached. Residual eccentricity from the adiabatic inspiral would be $\mathcal{O}(\eta)$ and eccentricity excited during the transition would be $\mathcal{O}(\eta^{4/5} \epsilon^{8/15})$ (or $\mathcal{O}(\eta^{7/10} \epsilon^{7/15})$ if it was coherently excited throughout the transition). These are smaller than the threshold $\eta^{5/6}$ at which the eccentricity corrections become more important than the non-geodesic past history corrections, which we have already shown to be sub-leading.

%We finally show that corrections to $\delta L$ are subdominant in Eq\eqref{near-extremal-scalings}. As we argued before, it is natural to think that the total angular momentum flux is a function of the radial coordinate (position) $\tilde{r}$. Notice that $\ddot{\tilde{L}} \sim \dot{\tilde{L}}\dot{r} \sim \eta^{8/5}\epsilon^{2/5}$ throughout the transition regime. Hence we can expand $\tilde{L}$ around the ISCO such that
%\begin{equation}
%    \delta L \sim \left(\frac{\eta}{\epsilon}\right)^{4/5}\epsilon^{2/3} + \left(\frac{\eta}{\epsilon}\right)^{6/5}\epsilon^{4/3} + \ldots
%\end{equation}
%so for $\eta\sim\epsilon$, the higher order corrections are subdominant. Thus we are fine to truncate to leading order in $T$.

\subsection{General Transition Equation - Very Near-Extremal}
\label{ssec:very-near-extremal}

The final regime concerns very rapidly rotating black holes, where $\epsilon \ll \eta$. Using the results in appendix~\ref{app:A:ISCO}, one can identify the a priori dominant contributions to the transition equation \eqref{app:B:General_Master_Equation} and \eqref{app:B_Gamma_odot_definition} to be (ignoring coefficients of $\mathcal{O}(1)$)
\begin{equation}
\begin{aligned}\label{V_NHEK_Master_Transition_Equation_Motion}
  \left(\frac{dR}{d\tilde{\tau}}\right)^{2} &\sim R^3 + R\,\delta L\,\epsilon^{2/3} + R^{2}\delta L + \delta L^2\,\epsilon^{4/3} + \Gamma_\odot \\
  \Gamma_\odot &\sim (\delta E- \delta L) \left(\epsilon^{2/3} + R +  \epsilon^{2/3}\delta L\right)\,.
\end{aligned}
\end{equation}

It is natural to expect that terms involving some explicit factors of $\epsilon$ should be sub-leading in this regime.
Assuming a scaling solution of the form $R\sim \eta^{\alpha}$ and $\tilde{\tau} \sim \eta^{\beta}$, we learn using \eqref{Linear_Time_Ang_Momenta} that $\delta L \sim \eta^{\beta + 1}$. Imposing the dominant terms $R^{3}$ and $R^{2}\delta L$ scale like $(dR/d\tilde{\tau})^2$ yields the scaling solutions $\alpha = 2/3$ and $\beta = -1/3$, so that
\begin{equation}
R = \eta^{2/3}\mathcal{R} \,, \quad \ttau = \eta^{-1/3}\mathcal{T}\,, \quad 
\delta L =  \eta^{2/3}\delta \mathcal{L}\,.
\label{eq:Very_extremal-scaling}
\end{equation}
As a consistency check, notice the terms $\epsilon^{2/3}R\delta L \sim \eta^{2}(\epsilon/\eta)^{2/3}$ and $\epsilon^{4/3}\delta L^{2} \sim \eta^{8/3}(\epsilon/\eta)^{4/3}$ are sub-dominant compared to the leading scaling $(d\tr/d\tilde{\tau})^{2} \sim \eta^{2}$.

The remaining question is whether $\Gamma_{\odot}$ is negligible in this regime or not. Using the scalings \eqref{eq:Very_extremal-scaling} together with \eqref{app:A_Proper_Time_Evolution_dE}, we infer that $\delta E - \delta L\sim\eta^{4/3}$. It follows $\Gamma_\odot\sim \eta^2$ from the term linear in $R$ in the second equation in  \eqref{V_NHEK_Master_Transition_Equation_Motion}. Introducing the finite variable $\mathcal{Y}$
\begin{equation}\label{app:B_Scaling_odot_near_ext}
    \Gamma_{\odot} = \eta^{2}\mathcal{Y}
\end{equation}
leads to the transition equation of motion
\begin{equation}
  \left(\frac{d\mathcal{R}}{d\mathcal{T}}\right)^{2} \sim \mathcal{R}^3 + \mathcal{R}^{2}\delta\mathcal{L} + \mathcal{Y}\,.
\end{equation}
Notice the radial velocity throughout the transition regime scales as $d\tr/d\tilde{\tau} \sim \eta$, as it were throughout the adiabatic inspiral regime and in the near-extremal case [see sec.(\ref{ssec:near-extremal})]. Thus the radial motion is fastest throughout the transition regime when the primary is of moderate spin: $\eta \ll \epsilon$.

 As a consistency check, we can substitute the scalings \eqref{eq:Very_extremal-scaling} and \eqref{app:B_Scaling_odot_near_ext} into the general transition equation \eqref{app:B:General_Master_Equation}
\begin{multline}\label{app:BTransition_Equation_OT}
\left(\frac{d\mathcal{R}}{d\mathcal{T}}\right)^{2} \sim \sum_{i=3}^{\infty}\eta^{2(i-3)/3}\mathcal{R}^{i} + (\epsilon/\eta)^{2/3}\delta \mathcal{L}\mathcal{R} + \\ \sum_{m=2}^{\infty}\eta^{2(m-2)/3}\mathcal{R}^{m}\delta\mathcal{L} + \epsilon^{2/3}(\epsilon/\eta)^{2/3}\delta\mathcal{L}^{2} + \\ \epsilon^{4/3}\delta\mathcal{L}^{2}\mathcal{R} + \sum_{n=2}^{\infty}\eta^{2(n-1)/3}\delta \mathcal{L}^{2}\mathcal{R}^{n} + \mathcal{Y}.
\end{multline}
Clearly the dominant terms occur when both $i = 3$ and $m = 2$ with the rest being subleading. 

The above scaling analysis proves the dominant terms in \eqref{app:B:General_Master_Equation} in the regime $\epsilon\ll \eta$ are captured by
\begin{equation}
\label{eq:regime-3}
\left(\frac{dR}{d\tilde{\tau}}\right)^{2} \simeq -\frac{2}{3}\alpha R^{3} + \gamma\, \delta L\,R^{2} + \Gamma_{\odot} + \dots
\end{equation}
where $\alpha$ and $\gamma$ are given in Eq.\eqref{near_extremal_beta} with
\begin{equation}\label{odot-3}
\Gamma_{\odot} \simeq \tilde{\Omega}_{\isc}(\delta E - \delta L)\,\frac{\partial^{2} G}{\partial\tilde{r}\partial \tilde{E}}\bigg\rvert_{\isc} R + \ldots \,.
\end{equation}

Keeping all coefficients of order one, the natural rescaled variables in this regime are
\begin{equation}
\begin{aligned}\label{scalings-3}
R &= \eta^{2/3}\alpha^{-3/5}\kappa^{2/5}X  \\
\ttau - \tilde{\tau}_{\isc} &= \eta^{-1/3}(\alpha\kappa)^{-1/5}T  \\
\delta E - \delta L &= \eta^{4/3} Y \\
\delta L &= -\eta^{2/3}\alpha^{-1/5} \kappa^{4/5} T . 
\end{aligned}
\end{equation}
In these variables, the radial velocity equation \eqref{eq:regime-3} can be expressed as
\begin{equation}
    \left(\frac{dX}{dT}\right)^{2} = -\frac{2}{3}X^{3} - K_{1}X^{2}T + \tilde{\Gamma}_{\odot}
\end{equation}
with 
\begin{equation}
\begin{aligned}
  K_{1} &= \gamma\alpha^{-3/5}\kappa^{2/5}\,, \\
  \tilde{\Gamma}_{\odot} &= \eta^{-2}\alpha^{4/5}\kappa^{-6/5}\Gamma_{\odot}\,, 
\end{aligned}
\end{equation}
and $\kappa$ as in \eqref{eq:kappa}.

Taking a further derivative with respect to $T$ yields the acceleration equation
\begin{equation}
\label{eq:acc-3}
\frac{d^{2}X}{dT^{2}} = -X^{2} - K_{1}XT + \frac{1}{2(dX/dT)}\left(\frac{d\tilde{\Gamma}_{\odot}}{dT}-K_{1}X^{2}\right).
\end{equation}
Using \eqref{app:A_Proper_Time_Evolution_dE} together with \eqref{scalings-3}, one finds that 
\begin{equation}
\begin{aligned}
    \frac{d\tilde{\Gamma}_{\odot}}{dT} &= \alpha^{1/5}\kappa^{-4/5}\tilde{\Omega}_{\isc}\frac{\partial^{2}G}{\partial\tilde{r}\partial\tilde{E}}\bigg\rvert_{\isc}\frac{dX}{dT}Y  \\ 
   & - \alpha^{-3/5}\kappa^{2/5}\left(\frac{\partial^{2}G}{\partial\tilde{r}\partial\tilde{E}}\frac{\partial\tilde{\Omega}}{\partial\tilde{r}}\right)_{\isc}X^{2}\,.
\end{aligned}
\end{equation}
Plugging this back in \eqref{eq:acc-3} and using the dominant contribution to the identity \eqref{eq:circular-3}, the $K_{1}X^{2}$ term cancels and one is left with 
\begin{equation}\label{Very_Near_Extremal_ODE}
\frac{d^{2}X}{dT^{2}} = - X^{2} - K_{1}XT + K_{2}Y
\end{equation}
together with the evolution equation for $Y(T)$ given by
\begin{equation}
\frac{dY}{dT} = -\alpha^{-4/5}\kappa^{6/5}\frac{\partial \log\tilde{\Omega}}{\partial \tilde{r}}\bigg\rvert_{\isc}X\,.
\end{equation}
where 
\begin{equation*}
  K_{2} = \frac{1}{2}\alpha^{1/5}\kappa^{-4/5}\tilde{\Omega}_{\isc}\frac{\partial^{2}G}{\partial \tilde{r}\partial \tilde{E}}\bigg\rvert_{\isc}\,.
\end{equation*}
In the limit $\epsilon \to 0$, the constants $K_1$ and $K_2$ approach the values
\begin{align*}
   K_{1} &\to 2^{6/5}3^{3/10}(\tilde{C}_{H} + \tilde{C}_{\infty})^{2/5} \\ 
   K_{2} & \to 2^{-7/5}3^{-1/10}(\tilde{C}_{H} + \tilde{C}_{\infty})^{-4/5}.
\end{align*}
As argued in previous sections, corrections to the circular flux-balance law contribute terms to the transition equation which scale like $d\tr/d\tilde{\tau}\sim\mathcal{O}(\eta)$ times terms that are being retained and like $\eta \epsilon^{2/3}$. Corrections to the linear-in-time angular momentum evolution enter with the same scaling as the former. The retained terms in the transition equation scale like $\eta^{4/3}$ in the very near extremal regime and so these corrections are both sub-leading. Eccentricity corrections enter like fractional $e^2$ corrections to the fluxes, but, as in the near-extremal case, eccentricity cannot grow until the transition zone has already been reached, and so these corrections are no larger than $\mathcal{O}(\eta^{5/3}\,\epsilon^{2/3})$ and are also sub-leading. 

%In the previous subsections, we showed that corrections to $\delta L$ are subdominant allowing us to linearly approximate $\tilde{L}$ throughout the transition regime. A similar argument here shows that
%\begin{equation}
%    \delta L \sim \eta^{2/3} + \eta^{4/3} + \ldots
%\end{equation}
%thus allowing us to truncate $\delta L$ linearly in $T$.

We conclude this subsection by noting that the transition equation of motion \eqref{Very_Near_Extremal_ODE} is perfectly well behaved in the limit $\epsilon \to 0$ and can therefore be used to compute an inspiral into a maximally spinning black hole with $a=1$. In this case, the horizon coincides with the ISCO in BL coordinates. However, the proper distance is $\Delta\ell = -\frac{M}{3}\ln\epsilon$ (see Fig.2 in \cite{1972ApJ...178..347B} together with explanations in \cite{1972ApJ...178..347B,Jacobson:2011ua} and more recently in appendix A of \cite{Gralla:2015rpa}). Hence, we terminate the integration of the ODE~\eqref{Very_Near_Extremal_ODE} at $\tilde{r} = \tilde{r}_{+}$, since our numerics are specific to BL coordinates. The presence of the horizon manifests itself in the transformation from proper time to coordinate time, which will be discussed, for non-extremal inspirals, in the next sub-section.

%We conclude this subsection by commenting that the transition equation of motion \eqref{Very_Near_Extremal_ODE} can be used to compute the full inspiral of a \emph{maximally} extremal black hole $a = 1$. In this situation, $\epsilon = 0 \ll \eta$ so, to begin, one would simply integrate the adiabatic inspiral solution to a suitable point (to be discussed in the next section). Then integrate \eqref{Very_Near_Extremal_ODE} forwards to $\tilde{r} = \tilde{r}_{+}$. This may be viewed as rather controversial since $|\tilde{r}_{\isc} - \tilde{r}_{+}| = 0$ so it is true that one cannot construct a neighborhood around the ISCO which does not contain the horizon. Although this is true, we feel that the transition equation of motion presented in this section will still present a physically relevant solution to the problem. Although the neighbourhood contains and surpasses the horizon in Boyer Lindquist coordinate, we simply terminate our solution once the horizon $r_{+}$ has been reached \ob{I like this and feel strongly about the validity of this paragraph... but feel free to remove it if it's obviously wrong}

\section{Results}\label{sec:results}

\subsection{Numerical Integration}\label{ssec:numerical_integration}

We now seek to compute a full worldline $\tilde{r}(\tilde{\tau})$ for $\infty > \tilde{r} \geq \tilde{r}_{+}$. Out of the three regimes just discussed, we restrict ourselves to the $\epsilon \sim \eta$ one. This is because the $\eta \ll \epsilon$ regime has already been considered in the literature \cite{2000PhRvD..62l4022O, kesden2011transition,Transition_Inspiral_Scott_Hughes,sundararajan2008transition} and the $\epsilon \ll \eta$ regime has been argued to be inaccessible throughout the transition regime in \cite{Geoffrey_Transition}. The latter conclusion follows from the observation that the waves emanating from the secondary produce a \emph{spin down} effect on the primary leading to a maximum attainable spin with $\epsilon\sim \eta$. For $\epsilon \sim \eta$, we try to find the solution to 
\begin{equation}
\begin{aligned}\label{coupled-near-extremal-transition-eqn-T}
\frac{d^{2}X}{dT^{2}} &= -X^{2} - T - (\eta/\epsilon)^{2/5}(C_{1}XT - C_{2}Y) \\ 
\frac{dY}{dT} &= -\frac{3}{4}\Lambda X
\end{aligned}
\end{equation}
which deviates off the past adiabatic inspiral and evolves into a geodesic plunge. The constants in Eq.\eqref{coupled-near-extremal-transition-eqn-T} are given by Eq.\eqref{near-extremal:constants_evaluated}. We can derive an equation for an adiabatic inspiral in proper time by using the quasi-circular approximation. Using our far-horizon expression for the energy flux defined by Eq.(\ref{CH:8 Flux_Thorne}) with both equations (\ref{CH:8 Circular Geodesic Equation Time}) and (\ref{CH:8 Circular Energy}), one derives
\begin{equation}\label{CH:8 True_Adiabatic_Inspiral}
\frac{d\tilde{r}}{d\tilde{\tau}} = -\eta\frac{64}{5}\tilde{\Omega}^{7/3}\frac{(2a - 3\tilde{r}^{1/2} + \tilde{r}^{3/2})\tilde{r}}{\tilde{r}^2 - 6\tilde{r} + 8a\tilde{r}^{1/2} - 3a^{2}}\dot{\mathcal{E}}(\tilde{r}).
\end{equation}
This equation diverges at the ISCO which is a break down of the quasi-circular approximation. We shall use Eq.(\ref{coupled-near-extremal-transition-eqn-T}) to smoothly transition from the adiabatic inspiral Eq.(\ref{CH:8 True_Adiabatic_Inspiral}) into a geodesic plunge to the horizon. We used a cubic spline to interpolate values for the relativistic correction $\dot{\mathcal{E}}(\tr)$ using exact flux data found in the \href{http://bhptoolkit.org}{BHPT}. We then numerically integrate Eq.(\ref{CH:8 True_Adiabatic_Inspiral}) by stepping forwards in proper time until $\tL(\tilde{\tau}) - \tL_{\isc} \sim \eta^{4/5}\epsilon^{-2/15}$. We feel this criteria is suitable for turning on the transition equation of motion since our model for the flux is well represented during the transition regime. When this criteria is met we can be sure that our model for flux evolution throughout the transition regime is correct to leading order. Once this is satisfied, we stop integrating our adiabatic inspiral solution and begin integrating our transition equation of motion \eqref{coupled-near-extremal-transition-eqn-T}.

Since we do not terminate our adiabatic inspiral solution at the ISCO, we do not know the precise proper time where the particle crosses the ISCO. As such, the variable $T$ is not a good choice of variable to integrate on the right hand side of \eqref{coupled-near-extremal-transition-eqn-T}. Instead, we substitute $T$ for $\delta L$ from Eq.\eqref{near-extremal-scalings} into our transition equation of motion, then
\begin{equation}
\begin{aligned}
\frac{d^{2}X}{dT^{2}} &= -X^{2} + B_{0}\delta L + (\eta/\epsilon)^{2/5}(C_{1}B_{0}\delta L + C_{2}Y) \\
\frac{dY}{dT} &= -\frac{3}{4}\Lambda X \label{near-extremal-transition-eqn} \\
\frac{d\delta L}{dT} &= B_{0}^{-1},\quad B_{0} = -\eta^{-4/5}\epsilon^{2/15}(\alpha\hat{\beta})^{1/5}\kappa^{-4/5}.
\end{aligned}
\end{equation}
We use initial conditions determined by the end of the adiabatic inspiral Eq.(\ref{CH:8 True_Adiabatic_Inspiral}) at some time $\ttau_{\text{init}}$.
\begin{equation}
\begin{aligned}\label{initial_conditions}
X(T_{\text{\text{init}}}) &= \eta^{-2/5}\epsilon^{-4/15}\alpha^{-3/5}(\hat{\beta}\kappa)^{-2/5}(\tilde{r}-\tr_{\isc})\\
\frac{dX}{dT}\bigg\rvert_{T_{\text{\text{init}}}} &= \eta^{-3/5}\epsilon^{-2/5}\alpha^{2/5}(\hat{\beta}\kappa)^{-3/5}\frac{d\tilde{r}}{d\tilde{\tau}}\bigg\rvert_{\tilde{\tau}_{\text{\text{init}}}}\\
Y(T_{\text{init}}) &= \eta^{-6/5}\epsilon^{-2/15}(\tilde{\Omega}_{\isc}^{-1
}\delta E_{\text{init}} - \delta L_{\text{init}})\\ 
\delta L(T_{\text{init}}) &= \tilde{L}_{circ}(\tilde{r}_{\text{init}}) - \tilde{L}_\isc. 
\end{aligned}
\end{equation}
Where $\tilde{L}_{circ}(\tr_{\text{init}})$ corresponds to the circular angular momenta evaluated at the end of the inspiral, $\tr_{\text{init}}$.  Using this prescription, we are able to integrate the coupled ODEs Eq.\eqref{near-extremal-transition-eqn} with initial conditions \eqref{initial_conditions} to obtain Fig.(\ref{fig:TransitionEOM_Plot}).
\begin{figure*}
\centering
\includegraphics[height = 8cm, width = 13cm]{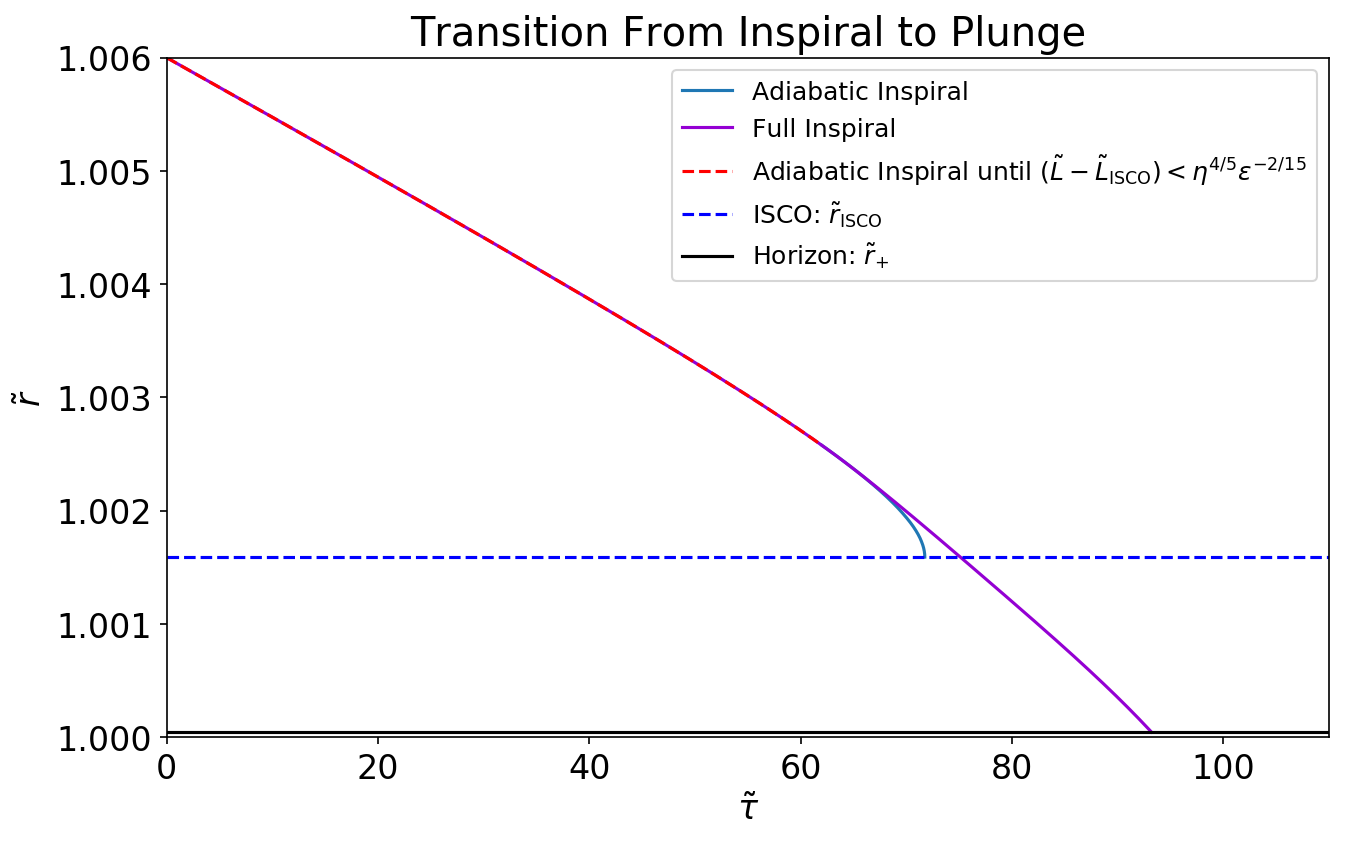}
\caption{In both plots we consider mass ratio $\eta = 10^{-5}$ and spin $a = 1-10^{-9}$. The transition regime begins at $\tr_{\text{init}} \approx 1.0026$ at $\ttau_{\text{\text{init}}}\approx 62.00$. The particle plunges into the horizon $\tr_{+}$ in proper time $\tilde{\tau}_{+} \approx 93.19$.}
\label{fig:TransitionEOM_Plot}
\end{figure*}
The transition solution smoothly deviates away from the adiabatic inspiral (blue curve), passes through the ISCO and reaches the horizon where the solution terminates. The plot on the right shows the full worldline in proper time $\tilde{r}(\tilde{\tau})$ where the inspiral starts at $\tilde{r} = 1.006$ and terminates at the horizon. This method ensures that $\tilde{r}(\tilde{\tau})$ is both continuous and once differentiable everywhere. 

Also, by our choice of integrating \eqref{near-extremal-transition-eqn} using the variable $\delta L$, we ensure continuity but \emph{not} differentiability in $\tilde{L}$  throughout the full inspiral. We note here that Apte and Hughes in \cite{Transition_Inspiral_Scott_Hughes} also found discontinuities in their evolution of both $\tL$ and $\tE$ and added corrections to ensure both (first order) differentiability and continuity at $\tilde{\tau}_{\text{init}}$. We consider a correction of the form  
\begin{equation}\label{L_transition_model}
\tL_{\text{trans}} =  \Delta \tL_{\text{cor}} + \tL_{\isc} + \frac{d\tilde{L}_{\isc}}{d\tilde{\tau}}(\tilde{\tau} - \tilde{\tau}_{\isc}).
\end{equation}
We have discussed previously that the leading order term in $\tilde{L}(\ttau) - \tL_{\isc} $ scales proportionally to $\eta^{4/5}\epsilon^{-2/15}$. So we choose to add a constant offset $\Delta \tL_{\text{cor}} \sim \eta^{6/5}\epsilon^{2/15}$ to the angular momenta evolution to ensure continuity in the $\tL_\text{trans}$ evolution. 

To calculate the evolution in $\tilde{E}$, one computes $\tilde{E}_{circ}$ given by Eq.\eqref{CH:8 Circular Energy} during the adiabatic inspiral regime. Then, during the transition regime, one integrates
\begin{equation}\label{E_transition_model}
\tilde{E} = \Delta \tE_{\text{cor}} + \tE_{\isc} + \int_{\tilde{\tau}_{\text{isco}}}^{\tilde{\tau}_{+}} \tilde{\Omega}(\tilde{r})\dot{\tilde{L}}_\text{trans} d\tilde{\tau}
\end{equation}
from the flux balance law $\dot{\tilde{E}} = \tilde{\Omega}(\tilde{r})\dot{\tilde{L}}$. The correction to $\Delta \tE_{\text{cor}}$ is chosen to ensure continuity with the end of the inspiral energy given by Eq.\eqref{CH:8 Circular Ang Mom} as previously discussed after equation \eqref{L_transition_model}.

\begin{figure}
\centering
\includegraphics[height = 6cm, width = 7.5cm]{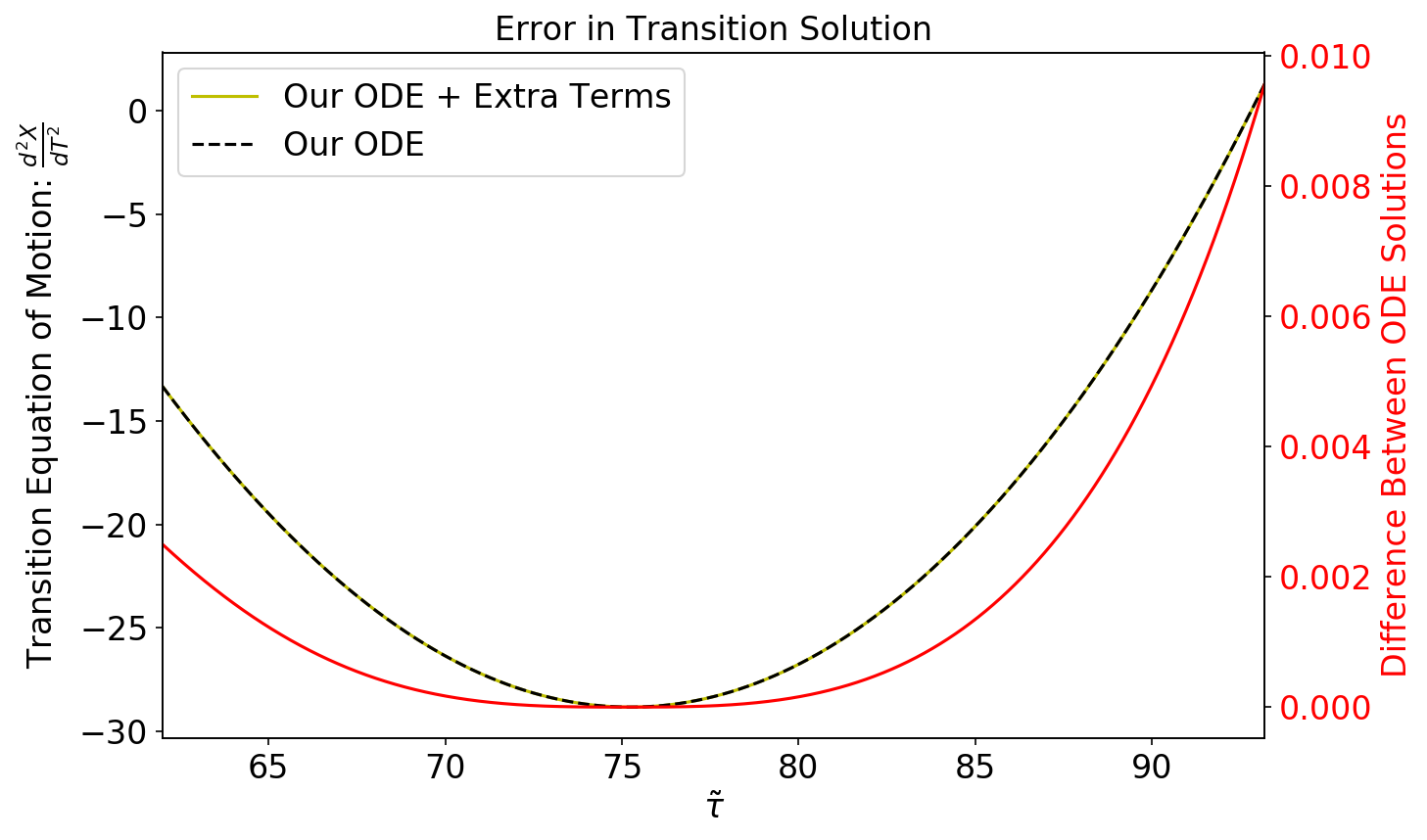}
\caption{Solution to~\eqref{near-extremal-transition-eqn} (black dashed line) and difference in solution when including the higher-order corrections given in Eq.~\eqref{extra_coefficients_near_extremal_ODE} (red solid line). The numerical difference is small throughout the transition regime reaching a maximum at plunge of $\sim 1\%$.}
\label{fig:ODE_Error}
\end{figure}

Notice here that this ensures that the energy flux obeys $\dot{\tilde{E}} = \tilde{\Omega}(\tilde{r})\dot{\tilde{L}}_{\isc}$ and is thus \emph{not} constant. This ensures that we are still granted a full cancellation of the dissipative part of the forcing term $\tilde{f}^{\tilde{r}}$ in Eq.\eqref{splitting_self_force}. This will yield a continuous evolution $\tilde{E}$ at the matching point with a discontinuous first derivative. At this point we will have a full trajectory $\tilde{r}(\tilde{\tau})$ with (continuous) integrals of motion in proper time $\tilde{E}(\tilde{\tau})$ and $\tilde{L}(\tilde{\tau})$. In each of ~\cite{2000PhRvD..62l4022O,Transition_Inspiral_Scott_Hughes,sundararajan2008transition}, the authors compute three separate worldlines in proper time; Adiabatic inspiral, transition, geodesic plunge. Apte \emph{et al} in \cite{Transition_Inspiral_Scott_Hughes}, provide an algorithm in which they freeze the constants of motion $\tE$ and $\tL$ when the extra terms in Eq.\eqref{Eq:OT_Eqn} exceed the leading order terms $X^{2}$ and $T$ by 5\%. As one would expect, as one ventures farther from the ISCO, the Taylor expansion used to derive these transition equations of motion will break down. As such, it is very natural for each of the aforementioned authors to compute a geodesic plunge to complete their worldlines in proper time $\tilde{r}(\tilde{\tau})$. Simply because, for moderate spins (non near-extremal), $|\tr_{+} - \tr_{\isc}| \sim \mathcal{O}(1) \nsim \eta^{2/5}$. For near-extreme black holes the ISCO is close to the horizon in Boyer Lindquist coordinates $|\tilde{r}_{\isc} - \tilde{r}_{+}| \sim \epsilon^{2/3}$. The scaling of the near-extremal transition zone is also $\epsilon^{2/3}$ and so the horizon is reached while the object is still in the transition zone. We therefore do not expect to need to add a geodesic plunge to compute full near-extremal inspirals. To verify this we numerically calculate  the extra terms in \eqref{near-extremal-transition-eqn}, which are 
\begin{equation}
\begin{aligned}\label{extra_coefficients_near_extremal_ODE}
C_{3} X^{3} &\Rightarrow C_{3} =\frac{1}{12} \left(\frac{\eta}{\epsilon}\right)^{2/5}\epsilon^{2/3} \frac{\partial^{4}G}{\partial \tilde{r}^{4}}\bigg\rvert_{\isc} \alpha^{-8/5}(\hat{\beta}\kappa)^{2/5} \\
C_{4}XY &\Rightarrow C_{4} = \frac{1}{2}\left(\frac{\eta}{\epsilon}\right)^{4/5}\epsilon^{2/3}\left(\tilde{\Omega}\frac{\partial^{3}G}{\partial\tilde{r}^{2}\partial\tilde{E}}\right)_{\isc}(\alpha\hat{\beta}\kappa)^{-2/5}.
\end{aligned}
\end{equation}
We compare the solution to \eqref{near-extremal-transition-eqn} when these terms are omitted or included in Figure~\ref{fig:ODE_Error}. The difference is at most $1\%$ even the horizon $\tr_{+}$. We conclude that we can use the solution from \eqref{near-extremal-transition-eqn} \emph{throughout} the plunging regime, for $\tilde{r}(\tilde{\tau}) \in [\tilde{r}_{+},\tilde{r}_{\isc}]$. It would be useful in the future to compare our results with the analytic geodesic plunges found in \cite{compere2018_NHEK}.

\subsection{Worldline in Boyer-Lindquist Coordinates}\label{ssec:BL_coords}

In the previous section, we computed the full worldline comprised of inspiral, transition and plunge parametrized as $\tilde{r}(\tilde{\tau})$. We now intend to do the same but in coordinate time so that our worldline is in Boyer-Lindquist coordinates $(\tilde{t},\tilde{r}(\tilde{t}),\theta = \pi/2,\phi(\tilde{t}))$. Loosely speaking, this is the time measured from Earth (at radial infinity) so is useful for observable purposes.

For the quasi-circular inspiral solution, we simply integrate the circular relation relating coordinate time to proper time via Eq.\eqref{CH:8 Circular Geodesic Equation Time}
\begin{equation}\label{t_2_tau_adiabatic_insp}
\tilde{t} = \int_{0}^{\tilde{\tau}_{\text{init}}}\frac{1 + a/\tilde{r}^{3/2}}{\sqrt{1 - 3/\tilde{r} + 2a/\tilde{r}^{3/2}}} d\tilde{\tau}
\end{equation}
where $\tilde{r}(\tilde{\tau})$ is the worldline constructed by integrating Eq.(\ref{CH:8 True_Adiabatic_Inspiral}) up to some suitable point to begin the transition solution, in our case, $\tilde{r}(\tilde{\tau}_{\text{init}}) = \tilde{r}_{\text{init}}$. To compute the trajectory in coordinate time $\tilde{r}(\tilde{t})$ throughout the transition regime, we must integrate 
\begin{equation}\label{CH:8 Final Coordinate Time Worldline GE}
\tilde{t} = \tilde{t}(\ttau_{\text{init}}) + \int_{\tilde{\tau}_{\text{init}}}^{\tilde{\tau}_{+}} T(\tilde{r},\tilde{E},\tilde{L},a)d\tilde{\tau}.
\end{equation}
where $T(\tilde{r},\tilde{E},\tilde{L},a)$ is given by \ref{CH:8 Time Geodesic Equation} and $\tilde{t}_{insp}$ is defined through $\tilde{t}(\tilde{\tau}_{\text{init}})$. Throughout the transition regime, we use the model for both $\tilde{E}(\tilde{\tau})$ and $\tilde{L}(\tilde{\tau})$ given by Eq.(\ref{E_transition_model}) and Eq.(\ref{L_transition_model}).  This will yield the  $\tilde{r}(\tilde{t})$ throughout the transition regime. Combining these results yield a full trajectory from radial infinity to the horizon in coordinate time $\tilde{r}(\tilde{t})$.

To then calculate the orbital velocity $d\phi/d\tilde{t} = \tilde{\Omega}$ in coordinate time we substitute $\tilde{r}(\tilde{t})$ found previously into Eq.\eqref{CH:8 Orbital Velocity}. This now gives $\tilde{\Omega}(\tilde{t})$ valid throughout the adiabatic inspiral regime. Using our solutions for $\tilde{E}(\tilde{\tau})$ and $\tilde{L}(\tilde{\tau})$ defined through Eq.\eqref{L_transition_model} and Eq.\eqref{E_transition_model} and $\tilde{r}(\tilde{t})$ throughout the transition regime, we calculate
\begin{equation}
\tilde{\Omega} = \frac{d\phi}{d\tilde{t}} = \frac{2a\tilde{E}\tilde{r} - a^2 \tilde{L} + \Delta(\tilde{r}) \tilde{L}}{\tilde{E}(\tilde{r}^2 + a^2)^2 - 2a\tilde{L}\tilde{r} - \Delta(\tilde{r}) a^2 \tilde{E}}.
\end{equation}
where $\Delta(\tilde{r}) = \tilde{r}^{2} - 2\tilde{r} + a^{2}$. This algorithm will provide a worldline in coordinate time $\tilde{r}(\tilde{t})$ which will be used for our waveforms. We stress here that $\tilde{r}(\tilde{t})$ is continuous and (once) differentiable.
\begin{figure}
\centering
\includegraphics[height = 6cm, width = 7.5cm]{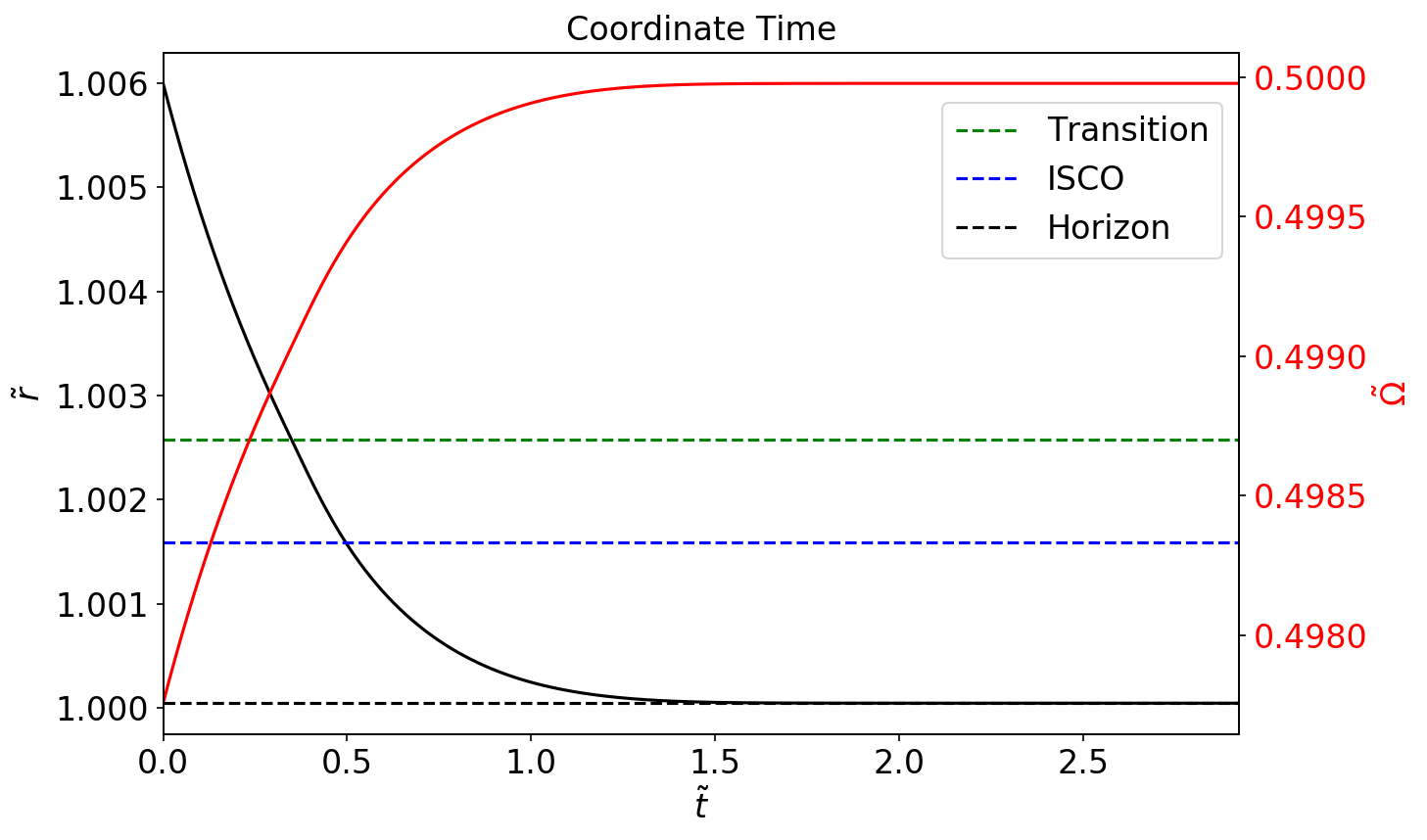}
\caption{The red curve shows the orbital velocity $\tilde{\Omega}$ and the black curve shows the trajectory in coordinate time $\tilde{r}(\tilde{t})$. Notice the smooth evolution of both $\tilde{r}(\tilde{t})$ and $\tilde{\Omega}$ during the start of the transition (green dashed curve). This smooth evolution continues through the ISCO (blue dashed curve) and evolves towards the horizon (black dashed curve). }
\label{fig:TransitionEOM_Plot}
\end{figure}

\subsection{Near-Extremal Waveform}\label{subsec:waveform}

Following \cite{2000PhRvD..62l4021F}, the root mean square (rms) amplitude of gravitational waves emitted towards infinity at harmonic $m$ is given by $h_{o,m} = \sqrt{\langle h_{+,m}^2 + h_{\times, m}^2\rangle}$. The plus and cross each represent individual transverse-traceless polarisations of the gravitational wave strain $h$. The amplitudes are averaged $\langle \cdot \rangle$ over the direction and over the period of the waves.  Furthermore, the rms amplitude 
generated by a particle on an equatorial circular orbit in the limit $\eta\to 0$ is related to the outgoing radiation flux in harmonic $m$ by 
\begin{equation}\label{RMstrain}
h_{o,m} = \frac{2M\sqrt{\eta\dot{\tilde{E}}_{\infty,m}}}{m\tilde{\Omega}D}
\end{equation}
for distance $D$ and outgoing fluxes defined by
\begin{equation}\label{energy_fluxes_harmonic}
\dot{\tilde{E}}_{\infty,m} = \eta\mathcal{A}_{m} \tilde{\Omega}^{2 + 2m/3}\dot{\mathcal{E}}_{\infty,m}
\end{equation}
where the amplitude $\mathcal{A}_{m}$ equals
\begin{equation*}
\mathcal{A}_{m} = \frac{8(m+1)(m+2)(2m)!\,m^{2m-1}}{(m-1)[2^{m}m!(2m+1)!!]^{2}}\,, \quad m\geq 2
\end{equation*}
and $\dot{\mathcal{E}}_{\infty,m}$ is the relativistic correction to $\dot{\tilde{E}}_{\infty,m}$ at each harmonic $m$. 

\begin{figure*}
\centering
\includegraphics[height = 10cm, width = 15cm]{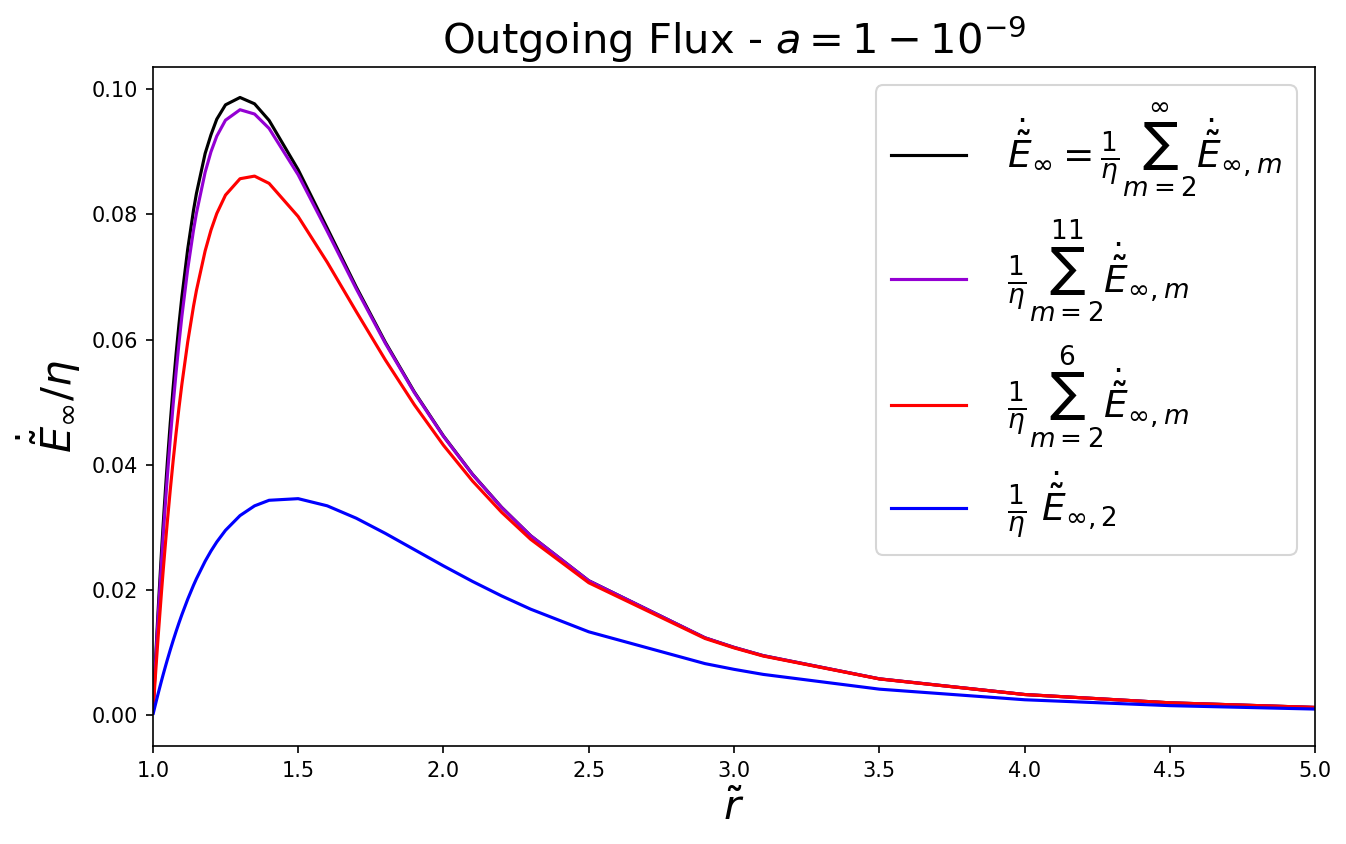}
\caption{Comparison of the total energy flux at infinity (black curve) including different harmonic $\dot{\tilde{E}}_{\infty,m}$ contributions. Note that at $\tr\approx 1.3$, the $m=2$ harmonic energy flux  $\dot{\tilde{E}}_{\infty,2}$ contributes $\sim$ 32\% of the total energy flux, whereas including the first 11 harmonics (violet curve) contributes $\sim 98\%$ at the least.}
\label{fig:HigherHarmonics}
\end{figure*}

An EMRI signal is a superposition of infinitely many harmonics of the fundamental frequency $\tilde{\Omega}$
\begin{equation}
h = \sum_{m=2}^{\infty}h_{o,m}\sin(2\pi \tilde{f}_{m} \tilde{t} + \phi),
\end{equation}
with $2\pi\,\tilde{f}_{m}= m\cdot \tilde{\Omega}$. Recall that the total emission of radiation through gravitational waves is related to the outgoing and ingoing flux by 
\begin{equation}\label{eq:fluxes_harmonics}
\begin{aligned}
\dot{\tE}_{GW} &= \dot{\tE}_{\infty} + \dot{\tE}_{H} \\
&= \sum_{m=2}^{\infty}\left(\dot{\tilde{E}}_{\infty,m} + \dot{\tilde{E}}_{H,m}\right)
\end{aligned}
\end{equation}
where $\dot{\tilde{E}}_{H,m}$ is the ingoing flux (towards the horizon) including the contribution from all $l$ for each harmonic $m$. Using the exact results from the \href{http://bhptoolkit.org}{BHPT} for a spin parameter of $a = 1-10^{-9}$, we constructed a cubic spline for each outgoing flux $\dot{\tE}_{\infty,m}$. Our results are plotted in figure \ref{fig:HigherHarmonics}. It is clear that including the higher order modes become increasingly important as the spin parameter increases towards unity. This has already been observed in \cite{compere2018_NHEK}. Hence, for near-extremal systems, only using the $m=2$ harmonic is not an accurate representation of the EMRI signal in general.

Figure \ref{fig:HigherHarmonics} suggests that truncating the sum at the eleventh harmonic in the outgoing flux \eqref{eq:fluxes_harmonics} is a good approximation to model a near-extremal waveform encapsulating quasi-circular inspiral, transition, and then plunge with suitable accuracy. The remaining difference from modes with $m > 11$ contributes a small difference in the amplitude of the waveform, but gravitational wave detectors are much less sensitive to amplitude corrections than corrections to the phase. The phase evolution is determined by the total flux, in which we are including all modes. We therefore believe that the approximate waveform with $11$ modes is a sufficiently accurate model for parameter estimation studies and will use this henceforth
\begin{equation}\label{Waveform}
h \approx \sum_{m=2}^{11}h_{o,m}\sin(2\pi \tilde{f}_{m} \tilde{t} + \phi).
\end{equation}

\begin{figure*}
\centering
\includegraphics[height = 10cm, width = 16cm]{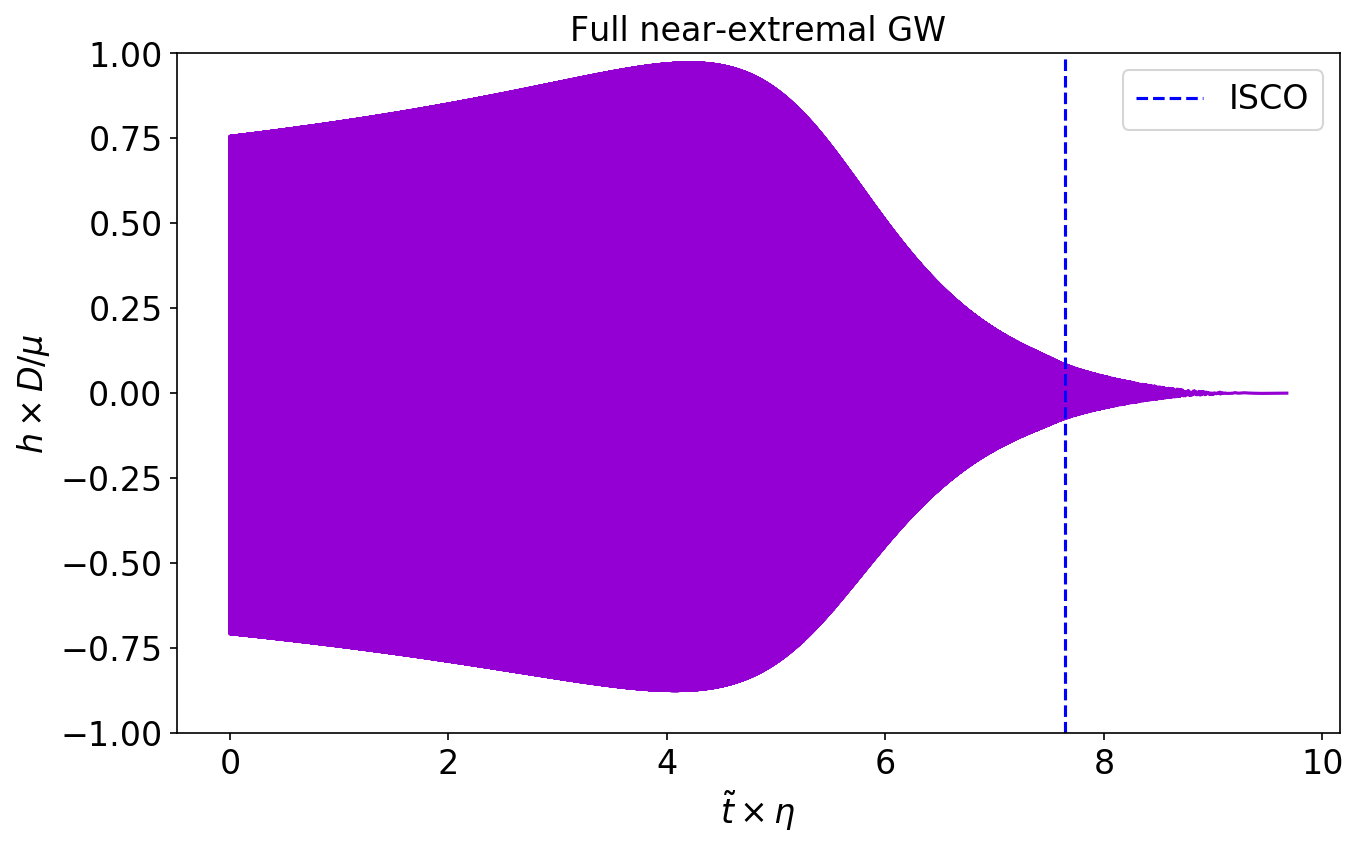}
\includegraphics[height = 10cm,width = 16cm]{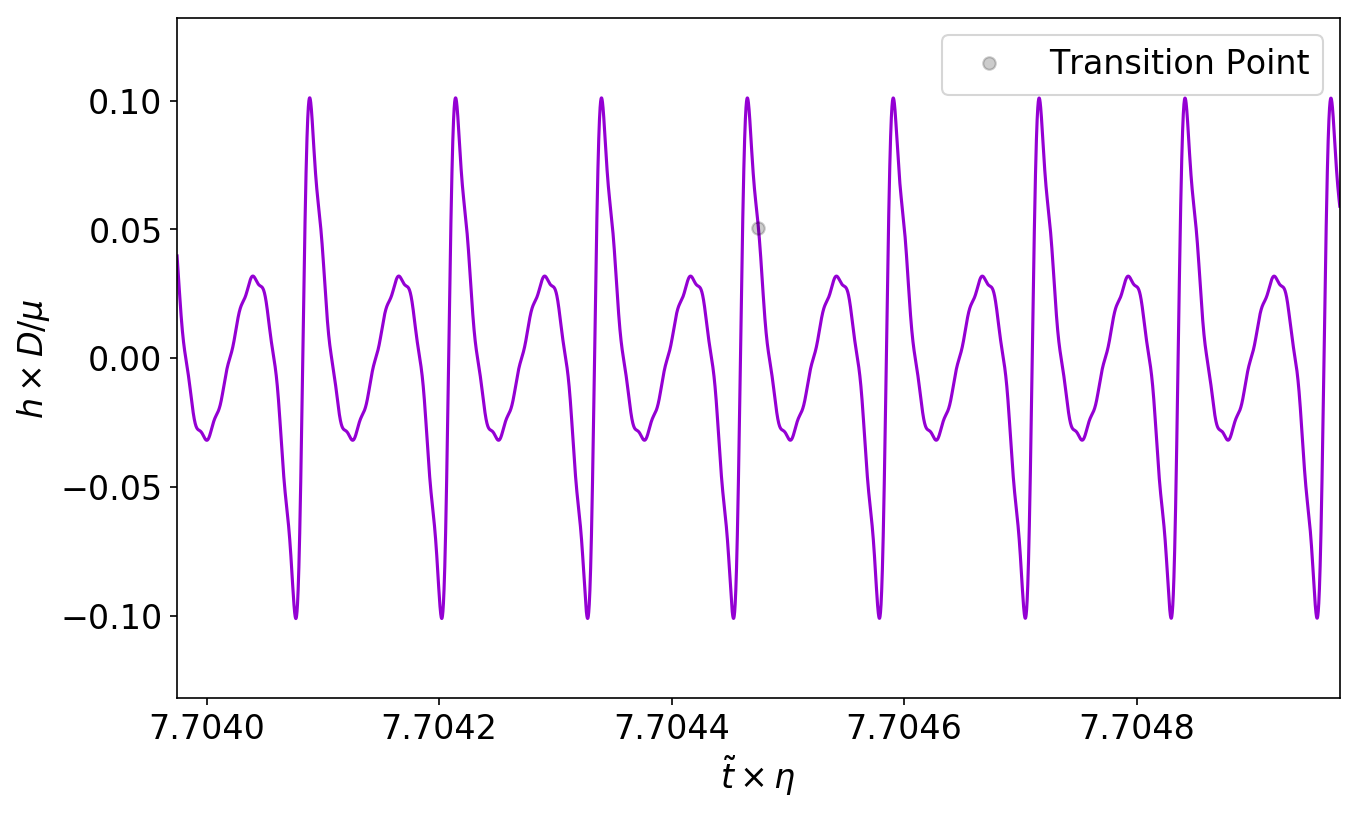}
\caption{(Top Plot) Here we plot the root mean square gravitational waveform for both inspiral, transition and plunge using the first eleven harmonics. Notice the smooth evolution of $h(\tilde{t})$. We terminate evolution of the waveform close to the plunge $\tilde{r} = \tilde{r}_{+} + \delta$ for suitably chosen $0 < \delta \ll 1$, otherwise the waveform will continue to decay for infinite coordinate time. This is obvious since the (point-like) particle as observed from infinity will never reach the horizon. In this example, we considered $a = 1-10^{-9}$ and $\eta = 10^{-5}$ so that we are in the $\eps \sim \eta$ regime. (Bottom Plot) We plot a zoomed in version of the top plot to show the reader the the smooth evolution of our adiabatic inspiral waves into the transition waves. The faded black dot indicates the moment the transition solution is turned on.}
\label{CH:8 Full Waveform in Coordinate Time Longer}
\end{figure*}

Once the ISCO is reached, we smoothly extrapolate each of the fluxes $\dot{\tilde{E}}_{\infty,m} \rightarrow 0$, as  $\tilde{r} \rightarrow \tilde{r}_{+}$.  
This is a similar approach to that found in Taracchini \emph{et al} in \cite{Small_Mass_Plunging}. Using \eqref{Waveform} and the results obtained in this paper, we plot a near-extremal waveform, including the transition from inspiral to plunge, in Fig.(\ref{CH:8 Full Waveform in Coordinate Time Longer}).
We notice that the waveform in Figure \ref{CH:8 Full Waveform in Coordinate Time Longer} exhibits the usual dampening before the ISCO is reached as seen by Gralla \emph{et al} in \cite{2016CQGra..33o5002G}. This, qualitatively, is a unique feature to near-extreme EMRIs as a gravitational wave source.

\section{Conclusion}
\label{sec:conc}

This paper has presented a solution to the problem of the transition from inspiral to plunge, \emph{for any} primary spin, for EMRIs on circular and equatorial orbits. This work has extended the treatment of Ori \& Thorne~\cite{2000PhRvD..62l4022O} which was the first analysis of this problem but did not apply to systems with near-extremal spins. This work also extended the analysis of~ \cite{kesden2011transition} which did consider near extremal spins, by providing a better physical interpretation of the procedure, identifying a missing term in the analysis and updating the treatment to use recent calculations of the near-extremal energy flux. We have also carefully identified the scaling of the various higher order terms arising from effects such as eccentricity and non-geodesic past-history to carefully demonstrate that these are all sub-dominant. Previous treatments have assumed that the quasi-circular assumption holds throughout the inspiral, but without rigorous justification. We have demonstrated that initial eccentricities excited during the adiabatic inspiral regime grow by the time the transition regime is reached, but are still sufficiently small to be sub-dominant. We have shown that corrections to the flux balance law \eqref{app:OT_Flux_Balance_Law} arising from eccentricity and from the non-geodesic past-history of the orbital evolution are also sub-dominant, if only marginally, but there are non-trivial deviations from the linear-in-proper-time evolution of energy and angular momentum in \eqref{Linear_Time_Ang_Momenta} that was assumed in OT. These deviations are encoded in the evolution of the parameter $\tilde{\Gamma}_{\odot}$ through the transition regime. 

Based on these arguments, we have derived a transition equation for each of the three scaling regimes: $\eta \ll \epsilon$, $\eta\sim\epsilon$ and $\eta\gg\epsilon$ and described a numerical scheme to generate a full inspiral trajectory in coordinate time, from radial infinity to the horizon. For near-extremal black holes, we found that there was no need to attach a geodesic plunge onto the transition solution as the inspiraling object reaches the horizon while still within the transition regime. Finally, we used these inspiral trajectories to construct a near-extremal waveform exhibiting the transition and plunging dynamics using results from the \href{http://bhptoolkit.org}{BHPT} \cite{BHPToolkit}. 

The OT procedure is straightforward, but with surprisingly rich phenomenology. Through semi-analytic means, one is able to derive an equation which describes the dynamics within the vicinity of the ISCO. However, \emph{in practice}, the OT theory has several shortcomings. The point at which the transition solution is taken to start has a significant influence on the time it takes the particle to reach the horizon and so the OT procedure does  \emph{not} define a unique worldline given a particular set of parameters for the source. This is clearly not physical behaviour. We argued in section~\ref{ssec:numerical_integration} that if the switch from the adiabatic inspiral to the transition equation is made when the constraint $\delta L \sim \eta^{4/5}\epsilon^{-2/15}$ is satisfied, the solution will be almost unique. This was verified numerically and we found it leads to plunge times consistent within $\pm 0.5M$. This very same problem was found in \cite{Transition_Inspiral_Scott_Hughes} but they saw no effect in their waveform analysis. For the $\eta \ll \epsilon$ case, there is a further degree of freedom as to when to attach the geodesic plunge. To do so, one must ``freeze" the integrals of motion $\tilde{E},\tilde{L}$ at the end of the transition regime and integrate the Kerr geodesic equations forward in coordinate time. Attaching the geodesic plunge is discussed, at length, in \cite{Transition_Inspiral_Scott_Hughes} but does not have a unique solution. Care must be taken as to when the transition solution and the geodesic plunge is attached or comparatively different radial trajectories will be produced. Fortunately for the $\epsilon \sim \eta$ case, there is no need to attach a geodesic plunge as shown earlier in section \ref{ssec:numerical_integration}.

Another issue with the OT method is that it can lead to discontinuities in the constants of motion $\tilde{E}(\tau)$ and $\tilde{L}(\tau)$ if the OT equations are integrated backwards from the ISCO rather than forwards from the point of the switch from the adiabatic inspiral to transition regime. Discontinuities in the constants of motion lead to discontinuities in the coordinate time trajectories and in the waveforms which must be avoided if these waveforms are to give physically reasonable results in parameter estimation studies. Our solution, which was to integrate forward not backwards, yields continuous, but not first order differentiable, trajectories. The procedure described in~\cite{Transition_Inspiral_Scott_Hughes} provides both. For parameter estimation studies we only require continuity of $\tilde{E}$ and $\tilde{L}$ and first order differentiability of $\tilde{r}(\tilde{t})$ and so our procedure should be sufficient, although this should be examined more carefully.
% We suspect that by the time LISA flies, more sophisticated techniques concerning the transition from inspiral to plunge will be developed and be in their late stages, thus replacing the OT procedure. 

There are natural extensions of this work. First, the waveforms constructed in this paper can be used to carry out a parameter estimation study to understand how well the parameters of near-extremal EMRIs can be measured with observations by LISA. Of particular interest is how well the spin can be determined, since the identification of an object that definitely has spin above the Thorne limit would be of profound significance. Second, our waveforms are missing the quasi-normal mode ringdown contribution. Hence, it would be very interesting to generate a \emph{full} waveform taking these into account, together with the plunging dynamics discussed here. Details on how to construct the waveform including this effect were discussed in \cite{Yang:2013uba}.  Finally, it would also be of interest to extend the current analysis to inspirals that are not circular and equatorial. The extension of OT was first performed in ~\cite{sundararajan2008transition} who attempted to solve the problem for generic orbits; both eccentric and inclined orbits. Sundararajan's treatment was corrected by~\cite{Transition_Inspiral_Scott_Hughes} in the case of arbitrary inclination. Hence, no one, as of yet, has considered the transition from inspiral to plunge in the case of eccentric orbits \emph{and} inclined orbits. These orbits are expected for EMRIs formed through standard astrophysical channels~\cite{amaro2007intermediate}. The extension to eccentric orbits will require more careful modelling of the self-force and the use of the (eccentricity-dependent) separatrix in place of the ISCO among other complications. A model of the transition for inspirals on \emph{generic} orbits into black holes \emph{arbitrary} spin will be invaluable  for the analysis of future LISA EMRI observations and is an important future topic of study. 

\begin{acknowledgments}
This work made use of data hosted as part of the Black Hole Perturbation Toolkit at \href{http://bhptoolkit.org/}{bhptoolkit.org}. We wish to thank Maarten van de Meent for his comments on the final stages of this manuscript. We also give our thanks to Maarten for pointing out useful literature concerning the conservative part of the self-force and giving direction on why we could neglect it. We also wish to thank Niels Warburton for his guidance using the toolkit and Peter Zimmerman for his useful discussions regarding the behaviour of near-extremal Kerr Black holes. Finally, we thank Geoffrey Comp\`{e}re for comments on an earlier version of this manuscript.
\end{acknowledgments}

\appendix

\section{The Innermost Stable Circular Orbit}
\label{app:A:ISCO}

In this appendix we review the main properties of the function $G(\tilde{r},\tE,\tL)$ determining the radial geodesic \eqref{CH:8 Radial Geodesic Equation}
\begin{equation}
\begin{aligned}
G(\tilde{r},\tilde{E},\tilde{L}) &= \tilde{E}^2 - 1 + \frac{a^2(\tilde{E}^2-1)-\tilde{L}^2}{\tilde{r}^2}\\
&+ \frac{2 (a \tilde{E}-\tilde{L})^2}{\tilde{r}^3} +\frac{2}{\tilde{r}}
\label{app:Potential}.
\end{aligned}
\end{equation}
together with its derivatives when evaluated at the ISCO orbit $\tr_\isc$. The spin dependence of these quantities will play a critical role in the identification of the different transition regimes discussed in section \ref{sec:master}.

Remember the ISCO radial coordinate $\tr_{\isc}$ is characterised by marginal stability
\begin{equation}\label{App:risco}
  G(\tilde{r},\tE_{\isc},\tL_{\isc})= \left. \frac{\partial G}{\partial \tilde{r}}\right|_{\isc} = \left. \frac{\partial^2 G}{\partial \tilde{r}^2}\right|_{\isc}   = 0\,.
\end{equation}
Labelling the energy and angular momentum of the ISCO orbit by $\tE_{\isc}$ and $\tL_{\isc}$, we can solve the second and third constraint equations by 
\begin{equation}
\begin{aligned}
  \tL_{\isc} &= \frac{\tilde{r}_\isc^2 - 3a^2 + 6\tilde{r}_\isc}{2\sqrt{3}\,\tilde{r}_\isc}\,, \\ 
  \tE_{\isc} &= \frac{6\tilde{r}_\isc - 3a^2 - \tilde{r}_\isc^2}{2\sqrt{3}\,a\,\tilde{r}_\isc}\,.
\label{eq:ELisco}
\end{aligned}
\end{equation}
Plugging these into $G(\tilde{r}_\isc,\tE_{\isc},\tL_{\isc})= 0$, one derives the relation 
\begin{equation}
  \frac{2}{3\tilde{r}_\isc} = 1-\tE_{\isc}^2\,,
\label{eq:erisco}
\end{equation}
which combined with \eqref{eq:ELisco} yields
\begin{equation}\label{App:r_ISCO identity}
  \tilde{r}_\isc^2 - 6\tilde{r}_\isc + 8a\sqrt{\tilde{r}_\isc} - 3a^2 =0\,,
\end{equation}
whose solution $r_{0}(a)$ reproduces \eqref{CH:8 ISCO location} \cite{1972ApJ...178..347B}. This equality allows to simplify the energy \eqref{CH:8 Circular Energy} and angular momentum \eqref{CH:8 Circular Ang Mom}
of equatorial circular orbits when evaluated at ISCO to
\begin{align}
\tE_\isc &= \frac{1-2/\tr_{\isc} + a/\tr_{\isc}^{3/2}}{\sqrt{1-3/\tr_{\isc} + 2a/\tr_{\isc}^{3/2}}} = \frac{4\sqrt{\tr_{\isc}} - 3a}{\sqrt{3}\tr_{\isc}} \label{app:Simple_Energy}\\
\tL_\isc &= \tilde{r}^{1/2}\frac{1 - 2a/\tilde{r}^{3/2} + a^{2}/\tilde{r}^{2}}{\sqrt{1 - 3/\tilde{r} + 2a/r^{3/2}}} = 2\sqrt{3} - \frac{4a}{\sqrt{3\tr_{\isc}}}\,.\label{app:Simple_Ang_Mom}
\end{align}

Armed with these identities, we move towards the evaluation of the derivatives controlling the expansions \eqref{eq:gexpansion} relevant to the transition regime. First, we introduce some notation
\begin{equation}
\begin{aligned}
  A_n &= \frac{\partial^{n}G}{\partial \tilde{r}^{n}}\bigg\rvert_{\isc}\,,\\
  B_n &= \left(\frac{\partial^{n+1} G}{\partial \tilde{r}^{n} \partial \tE}\tilde{\Omega} + \frac{\partial^{n+1}G}{\partial \tilde{r}^{n} \partial \tilde{L}}\right)_{\isc}\,, \\
   C_n &= \left(\frac{\partial^{n+2} G}{\partial \tilde{r}^{n} \partial \tE^{2}}\tilde{\Omega}^{2} + 2\frac{\partial^{n+2} G}{\partial \tilde{r}^{n} \partial \tilde{L}\partial \tilde{E}}\tilde{\Omega} + \frac{\partial^{n+2} G}{\partial \tilde{r}^{n} \partial \tL^{2}}\right)_{\isc}\,
\end{aligned}
\label{eq:Taylor-coe}
\end{equation}
with $\tilde{\Omega}$ as in\eqref{CH:8 Orbital Velocity}. Either by explicit calculation or by induction, one can prove for any integer $n\geq 0$
\begin{widetext}
\begin{eqnarray*}
  \frac{\partial^{n}G}{\partial \tr^{n}} &=&  (-1)^{n}\bigg(\frac{(n+2)!(a\tilde{E} - \tilde{L})^{2}}{\tilde{r}^{n+3}} + \frac{(n+1)!(a^2(\tilde{E}^2 - 1) - \tilde{L}^{2})}{\tilde{r}^{n+2}} + \frac{2n!}{\tilde{r}^{n+1}}\bigg) - \delta_{n0}(1 - \tilde{E}^{2})\,, \\
   \frac{\partial^{n+1}G}{\partial \tilde{r}^{n}\partial \tilde{E}} &=& (-1)^{n}\left(\frac{2(n+2)!(a^2\tilde{E} - a\tilde{L})}{\tilde{r}^{n+3}} + \frac{2(n+1)!a^{2}\tilde{E}}{\tilde{r}^{n+2}}\right) + 2\delta_{n0}\tilde{E}\,, \\
\frac{\partial^{n+1}G}{\partial \tilde{r}^{n}\partial \tilde{L}} &=& -(-1)^{n}\left(\frac{2(n+2)!(a\tilde{E} - \tilde{L})}{\tilde{r}^{n+3}} + \frac{2(n+1)!\tilde{L}}{\tilde{r}^{n+2}}\right)\,, \\
  \frac{\partial^{n+2}G}{\partial \tilde{r}^{n}\partial \tilde{L}^{2}} &=& (-1)^{n}\left(\frac{2(n+2)!}{\tilde{r}^{n+3}} - \frac{2(n+1)!}{\tilde{r}^{n+2}}\right)\,, \\
\frac{\partial^{n+2}G}{\partial \tilde{r}^{n}\partial \tilde{L}\partial \tilde{E}} &=& -(-1)^{n}\left(\frac{2a(n+2)!}{\tilde{r}^{n+3}}
\right)\,, \\
\frac{\partial^{n+2}G}{\partial \tilde{r}^{n}\partial \tilde{E}^{2}} &=& (-1)^{n}\left(\frac{2a^{2}(n+2)!}{\tilde{r}^{n+3}} - \frac{2a^{2}(n+1)!}{\tilde{r}^{n+2}}\right) + 2\delta_{n0}\,  
\end{eqnarray*}
where $\delta_{n0} = 1$ for $n=0$ and zero otherwise.  Finally, evaluating these derivatives at $(\tr_\isc,\tE_\isc,\tL_\isc)$ and using the properties \eqref{App:risco}-\eqref{app:Simple_Ang_Mom}, we can derive the exact results
\begin{eqnarray}
  A_n &=&  (1 - \delta_{n0})\frac{(-1)^{n}(n-1)(n-2)n!}{3\tr_{\isc}^{1+n}}\,, \label{App:General_Radial_Derivative} \\
  B_n &=& 2(1 - \delta_{n0}) (-1)^n (n+1)!\,\frac{n(a-\sqrt{\tr_{\isc}}) + a - 2\sqrt{\tr_{\isc}} +\tr_{\isc}^{3/2}}{\tr_{\isc}^{n}\sqrt{3\tr_{\isc}} \left(a-\sqrt{\tr_{\isc}}\right) \left(a+\tr_{\isc}^{3/2}\right)}\,,\label{App:General_Mixed_Radial_First_E_L} \\
  C_n &=& 2\cdot\frac{\delta_{0n} -(-1)^{n}(2a + \sqrt{\tilde{r}_\isc}[\tilde{r}_\isc - 2 - n])(n+1)! }{\tilde{r}_\isc^{(2n+1)/2}(a + \tilde{r}_\isc^{3/2})^{2}}\,,
\label{App:General_Mixed_Radial_First_E2_EL_LL} 
\end{eqnarray}
\end{widetext}
Notice equations \eqref{App:General_Radial_Derivative}-\eqref{App:General_Mixed_Radial_First_E_L} recover the familiar identities for circular orbits 
\begin{align*}
    &\left(\frac{\partial G}{\partial \tilde{E}}\tilde{\Omega} + \frac{\partial G}{\partial \tilde{L}}\right)_{\isc} = 0, \\
    & \qquad G_{\isc} = \frac{\partial G}{\partial \tilde{r}}\bigg\rvert_{\isc} = \frac{\partial^{2}G}{\partial \tilde{r}^{2}}\bigg\rvert_{\isc} = 0.
\end{align*}

Let us study the behaviour of these derivatives for near extremal black holes, i.e. in the limit $\epsilon\to 0$ as introduced in section \ref{CH:8 Sec: Preliminaries}. Remember $\tr_\isc$ is given by 
\begin{equation}
\tr_{\isc} \to 1 + 2^{1/3}\epsilon^{2/3} + \frac{7}{4\cdot 2^{1/3}} \epsilon^{4/3} + \mathcal{O}(\epsilon^{2}) 
\label{app: Expansion_r_ISCO}\,.
\end{equation}
Using this expansion together with $a = \sqrt{1-\epsilon^{2}}$, we can evaluate the leading terms of all previous derivatives to be 
\begin{widetext}
\begin{align}
  A_n & \to (1 - \delta_{n0})(n-2) (n-1)\left(\frac{1}{3} (-1)^n \Gamma(n+1) + \mathcal{O}(\epsilon^{2/3})\right)\,,\label{App:General_Radial_Extremal} \\
  B_n & \to (1 - \delta_{n0})\frac{(-1)^{n}\Gamma(n+2)}{\sqrt{3}}\left(n-1 - \frac{4n^2 + n + 1}{2^{5/3}}\epsilon^{2/3}\right) + \mathcal{O}(\epsilon^{4/3})\,,\\ 
\label{App:General_Mixed_E_L_Extremal} 
C_{n} & \to  -\frac{1}{4} (-1)^n (n-1) \left(-2 + 2^{1/3}\epsilon ^{2/3}[2 n +3]\right) (n+1)! + \frac{(-1)^n (4n^2 -3n -3) (n+2)!}{2^{10/3}} \epsilon ^{4/3} \nonumber \\
 &+ p_{n0} \\
  \frac{\partial^{n+1}G}{\partial\tilde{r}^{n}\partial\tilde{E}}\bigg\rvert_{\isc}  &\to  \frac{2}{\sqrt{3}}(-1)^{n+1}(n+1)![(n+1) - 2^{1/3}(n^2 + 3n + 3)\epsilon^{2/3}] + \frac{2}{\sqrt{3}}(1 + 2^{1/3}\epsilon^{2/3})\delta_{n0} + \mathcal{O}(\epsilon^{4/3})\label{app:ISCO_E_Expansion} \\
  \left(\frac{\partial^{n+2}G}{\partial \tilde{r}^{n}\partial\tilde{E}} + \tilde{\Omega}\frac{\partial^{n+2}G}{\partial\tilde{r}^{n}\partial\tilde{E}^{2}}\right)_{\isc} &\to (\delta_{0n} - (-1)^{n}(n+1)^{2}n!) + \frac{(-1)^{n}(7 + 13n + 4n^{2})(n+1)! - 3\delta_{0n}}{2^{5/3}}\epsilon^{2/3} + \mathcal{O}(\epsilon^{4/3})\label{app:gamma_odot_middle_term}\\
  \frac{\partial^{n+2}G}{\partial\tilde{r}^{n}\partial\tilde{E}^{2}}\bigg\rvert_{\isc} &\to  2((-1)^{n}(3+n)(n+1)! + \delta_{0,n})   -2^{4/3}(-1)^{n}(4+n)(n+2)!\epsilon^{2/3} + \mathcal{O}(\epsilon^{4/3})  \label{app:ISCO_EE_Expansion} 
\end{align}
\end{widetext}
where we defined 
\begin{equation}
p_{n0} = \frac{2-3\cdot 2^{1/3} \epsilon ^{2/3}}{4} \delta _{0n} + \mathcal{O}(\epsilon^{2}).
\end{equation}

What we learn is that $A_{n}\sim\mathcal{O}(1)$ for all $n\geq3$, $B_{1} \sim \epsilon^{2/3}$, $B_{n} \sim \mathcal{O}(1)$ for $n\geq 2$, $C_{0}\sim C_{1} \sim \epsilon^{4/3}$ and $C_{n} \sim \mathcal{O}(1)$ for $n\geq 2$. Furthermore,
\eqref{app:ISCO_E_Expansion} and \eqref{app:gamma_odot_middle_term} are $\mathcal{O}(\epsilon^{2/3})$ for $n=0$ and $\mathcal{O}(1)$ for $n\geq 1$, whereas \eqref{app:ISCO_EE_Expansion} is always $\mathcal{O}(1)$.

%%%%%%%%%%  retrograde orbits %%%%%%%%%%%%%

\section{Retrograde Orbits}
\label{App:Retrograde_Orbits}

\begin{figure}
\centering
\includegraphics[height = 6cm, width = 8cm]{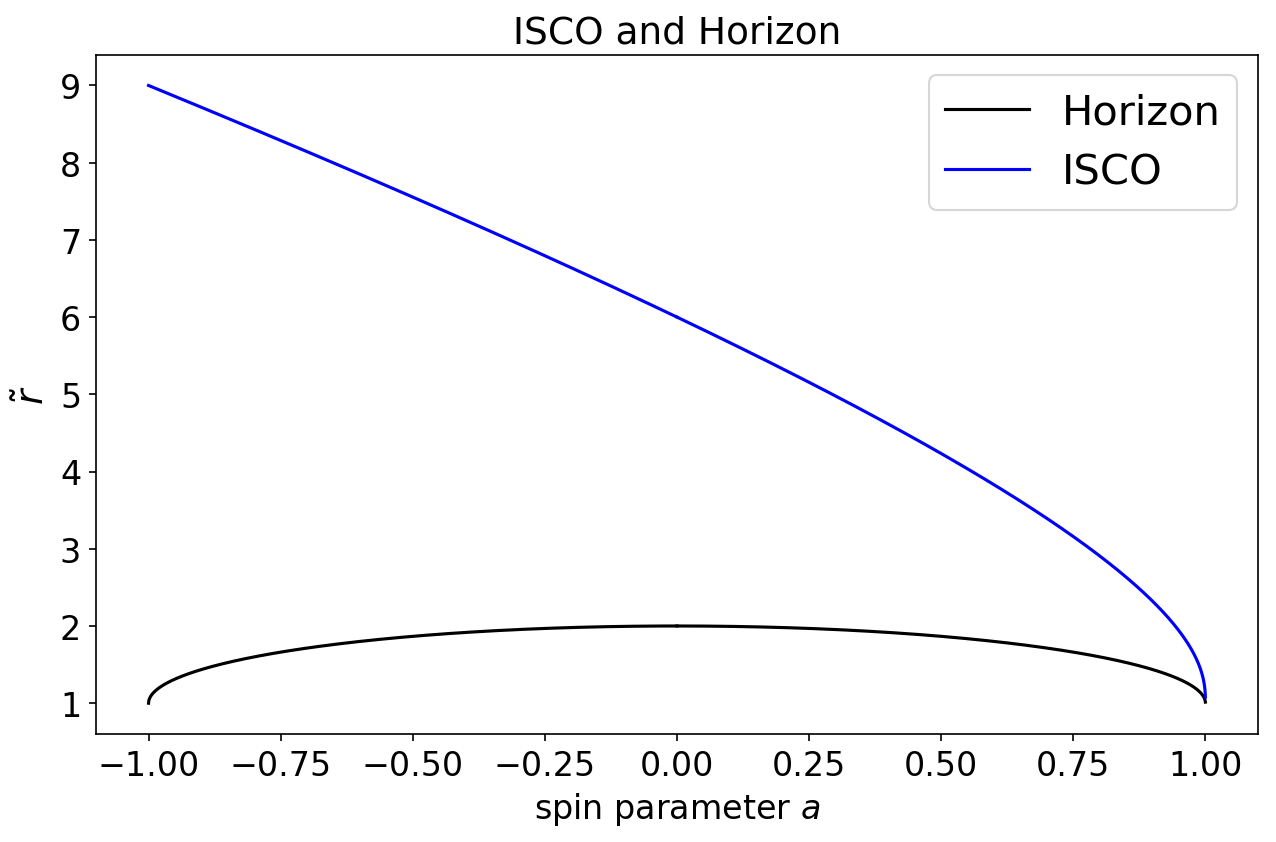}
\caption{This plot shows the relationship between $\tilde{r}_{\isc}$ and $\tilde{r}_{+}$ with the spin parameter $a\in[-1,1]$. Notice that for $a>0$ (prograde orbits), the ISCO and horizon locations coincide in B-L coordinates, whereas for $a<0$ (retrograde orbits), these remain at a finite B-L coordinate distance.}
\label{ISCO_vs_Horizon}
\end{figure}

In this section, we will restrict our attention to retrograde orbits. That is, orbits opposing the direction with the primaries angular momenta. These orbits are of interest because the ISCO is much further away from the horizon, which implies that the radial distance travelled during plunge time is much longer. We plot the location of the ISCO as a function of spin $a$ in figure (\ref{ISCO_vs_Horizon}).
Due to frame-dragging, we expect the ISCO to be farther from the hole since the space is dragged in the opposite direction to the compact objects orbital direction. In our conventions, retrograde orbits correspond to $a<0$ and $\tL>0$. Hence, near-extremal ones are characterised by $a \to -1$, or equivalently, by

\begin{equation}\label{CH:8_Retrograde_Parameter}
a \rightarrow -\sqrt{1-\epsilon^{2}}, \ \text{where} \ \epsilon \ll 1.
\end{equation}

notice that the horizon takes the same form as in the case of prograde orbits
\begin{align*}
\tilde{r}_{+} = 1 + \sqrt{1-a^{2}} =  1 + \epsilon
\end{align*}
as to be expected. Using a spin parameter of negative parity, the expressions for $\tilde{E},\tilde{L}, \tilde{\Omega}$ and $\tilde{r}_{\isc}$ remain the same. However, each quantity will be different at the ISCO of a retrograde orbit. By substituting Eq. (\ref{CH:8_Retrograde_Parameter}) into Eqs.(\ref{CH:8 Circular Ang Mom}), (\ref{CH:8 Circular Energy}), (\ref{CH:8 ISCO location}) and (\ref{CH:8 Orbital Velocity}) for small $\epsilon \ll 1$, one finds
\begin{align}
\tilde{r}_{\isc} &= 9 - \frac{45}{32}\epsilon^{2} +\mathcal{O}(\epsilon^{4}) \label{eq:isco-retro} \\
\tilde{E}_{\isc} &= \frac{5}{3 \sqrt{3}}-\frac{1}{96 \sqrt{3}}\epsilon ^2 +\mathcal{O}(\epsilon^{4})\\
\tilde{L}_{\isc} &= \frac{22}{3 \sqrt{3}}-\frac{3 \sqrt{3}}{16}\epsilon ^2 +\mathcal{O}(\epsilon^{4})\\
\tilde{\Omega}_{\isc} &= \frac{1}{26} + \frac{373}{43264}\epsilon ^2 +\mathcal{O}(\epsilon^{4}).
\end{align}
Notice here that the expansion in $\epsilon$ is no longer increasing in powers of $\epsilon^{2/3}$ and now in $\epsilon^{2}$. Also notice that $|\tr_{\isc} - \tilde{r}_{+}|\sim\mathcal{O}(1)$ rather than of order $\epsilon^{2/3}$ like in the case of near-extremal prograde orbits. Like we have done previously, we consider the Kerr radial velocity expanded around the ISCO
\begin{equation}\label{765_Expansion_ISCO_New_Variables_Retrograde}
\left(\frac{dR}{d\tilde{\tau}}\right)^{2} \simeq -\frac{2}{3}\alpha R^{3} + 2\beta\delta LR + \gamma \delta LR^{2} + \Gamma_{\odot} + \ldots.
\end{equation}
with small variables 
\begin{align}
\tilde{E} - \tE_{\isc} &= \tilde{\Omega}_{\isc}\delta E \\
\tilde{L} - \tL_{\isc} &= \delta L \\
\tilde{r} - \tr_{\isc} &= R.
\end{align}
The coeffcients in \eqref{765_Expansion_ISCO_New_Variables_Retrograde} can be approximated for $\epsilon \rightarrow 0$ under the retrograde condition Eq.\eqref{CH:8_Retrograde_Parameter}
\begin{align*}
\alpha &= -\frac{1}{4}\frac{\partial^{3}G}{\partial \tilde{r}^{3}}\bigg\rvert_{\isc} \rightarrow  \frac{1}{6561}\\
\beta &=  \frac{1}{2}\left(\frac{\partial^{2}G}{\partial\tilde{r}\partial{\tilde{L}}} + \tilde{\Omega}\frac{\partial^{2}G}{\partial\tilde{r}\partial\tilde{E}}\right)_{\isc} \rightarrow \frac{4}{351\sqrt{3}} \\
\gamma &= \frac{1}{2}\left(\frac{\partial^{3}G}{\partial\tilde{r}^{2}\partial{\tilde{L}}}+ \tilde{\Omega}\frac{\partial^{3}G}{\partial\tilde{r}^{2}\partial\tilde{E}}\right)_{\isc}\rightarrow -\frac{1}{351\sqrt{3}}. 
\end{align*}
and $\Gamma_{\odot}$ in \eqref{765_Expansion_ISCO_New_Variables_Retrograde} defined through equation~\eqref{app:B_Gamma_odot_definition}. Notice that none of the coefficients in our transition equation of motion depend on the extremality parameter $\epsilon$. This gives us no reason to introduce any scalings on $\tilde{r},\tilde{\tau}$ and $\delta L$ like we did for prograde orbits around rapdily rotating black holes. As such, let us introduce similar scalings to OT
\begin{align}\label{scalings_OT_retrograde}
R &= \eta^{2/5}\alpha^{-3/5}(\beta\kappa)^{2/5}X \\
\ttau - \tilde{\tau}_{\isc} &= \eta^{-1/5}(\alpha\beta\kappa)^{-1/5}T \\
\delta E - \delta L &= \eta^{6/5} Y. \label{deviation_circularity_parameter} \\
\delta L &= -\eta^{4/5}(\alpha\beta)^{-1/5} \kappa^{4/5} T  
\end{align}
Substituting these results into Eq.(\ref{765_Expansion_ISCO_New_Variables_Retrograde}) we find that 
\begin{equation}\label{retrograde_trans_equation}
\left(\frac{dX}{dT}\right)^{2} = -\frac{2}{3}X^{3} - 2XT + \alpha^{4/5}(\eta \beta\kappa)^{-6/5}\Gamma_{\odot}.
\end{equation}
Since $R\sim \eta^{2/5}$, we only need the first term of $\Gamma_{\odot}$ 
\begin{equation}
\Gamma_{\odot} = \eta^{6/5}\left(\tilde{\Omega}\frac{\partial G}{\partial \tilde{E}}\right)_{\isc}Y.
\end{equation}
and taking derivatives of Eq.(\ref{retrograde_trans_equation}) and following an identical procedure to subsection (\ref{ssec:OT}),
\begin{equation}
\frac{d^{2}X}{dT^{2}} = -X^{2} - T
\end{equation}
with evolution equation for $Y$
\begin{equation}
\frac{dY}{dT} = -\frac{\partial \log \tilde{\Omega}}{\partial \tilde{r}}\bigg\rvert_{\isc}(C_{1}K_{0})^{-1} X
\end{equation}
for $K_{0} = \alpha^{4/5}(\beta\kappa)^{-6/5}$. Which is precisely the equation of motion for the transition regime derived by OT in \cite{2000PhRvD..62l4022O}. Although the quantities $\alpha,\beta$ and $\kappa$ present in the change of coordinates are different, the physics and ultimate end goal are the same. As a result, we stop our analysis of retrograde orbits here since we feel that this problem has already been solved by the community for smaller spin values $a \geq -0.999$. We conclude that, for near-extremal retrograde orbits, there is nothing new to learn about the transition regime. It can be solved in the matter of OT in \cite{2000PhRvD..62l4022O}. We do remark that the quantity $\kappa$ can no longer be computed using the near-extremal formula defined by $\dot{\tilde{E}}_{GW} = (\tilde{C}_{H} + \tilde{C}_{\infty})(\tilde{r} - \tilde{r}_{+})/\tilde{r}_{+}$. This is because the transition region is far from the horizon of the primary hole [see figure (\ref{CH:8 fig:Compare_Flux_Formulas})]. Instead we have to use the numerical quantity
\begin{align*}
\kappa  &= \left(\tilde{\Omega}^{-1}\frac{d\tilde{t}}{d\tilde{\tau}}\frac{d\tilde{E}}{d\tilde{t}}\right)_{\isc} \\
 &= \left(-\frac{32}{5}\tilde{\Omega}^{7/3}\frac{1 + a/\tilde{r}^{3/2}}{\sqrt{1 - 3/\tilde{r} + 2a/\tilde{r}^{3/2}}}\dot{\mathcal{E}}(\tilde{r})\right)_{\isc}.
\end{align*}
Various results are tabulated (including retrograde orbits) in \cite{2000PhRvD..62l4021F}. The downside of this equation is that it can only be evaluated numerically. 

\section{Osculating Elements Equations}
\label{App:D_Osculating_Elements_Equations}

The proper time derivative of the radial geodesic equation \eqref{CH:8 Radial Geodesic Equation} yields \eqref{eq:acceleration}
\begin{equation}
\frac{d^{2}\tilde{\tr}}{d\tilde{\tau}^{2}} - \frac{1}{2}\frac{\partial G}{\partial \tilde{r}} = \frac{1}{2}\left(\frac{d\tE}{d\tilde{\tau}}\frac{\partial G}{\partial \tE} + \frac{d\tL}{d\tilde{\tau}}\frac{\partial G}{\partial \tilde{L}}\right)\left(\frac{d\tilde{r}}{d\tilde{\tau}}\right)^{-1}\, \label{our_equation}.
\end{equation}
The purpose of this appendix is to review how this equation is equivalent to the radial component of a forced geodesic equation
\begin{equation}\label{eq:forced}
u^{\nu}\nabla_{\nu}u^{\tilde{r}}= \frac{d^{2}\tilde{x}^{\tilde{r}}}{d\tilde{\tau}^{2}} + \Gamma^{\tilde{r}}_{\rho\sigma}\frac{d\tilde{x}^{\rho}}{d\tilde{\tau}}\frac{d\tilde{x}^{\sigma}}{d\tilde{\tau}}= \tilde{f}^{\tilde{r}}\,,
\end{equation}
where $\tilde{x}^{\mu} = (\tilde{r},\tilde{t},\theta,\phi)$, $\nabla_{\nu} = \nabla_{\tilde{x}^{\nu}}$, $u^{\mu} = d\tilde{x}^{\mu}/d\tilde{\tau}$ is the four-velocity of the particle and $\tilde{f}^{\mu}$ a forcing term driving deviations from geodesic motion.

To show the equivalence between \eqref{our_equation} and \eqref{eq:forced}, we use the osculating elements formulation \cite{pound2008osculating}. Since this method does not take into account conservative effects arising from the self-force 
\cite{barack2018self}, the component $\tilde{f}^{\tilde{r}}$ in this appendix will only account for the dissipative piece in \eqref{splitting_self_force}. Its conservative piece is treated in more detail in the main text (See section \ref{sec:self-force}).

Since the four velocity $u^\alpha$ is normalised, it follows $\tilde{f}^{\alpha}$ is normal to it by proper time differentiation 
\begin{equation}
 u^\alpha u_\alpha = -1 \,\,\Longrightarrow\,\, \tilde{f}^\alpha u_\alpha = 0\,.
\label{eq:f-id}
\end{equation}
Evaluating \eqref{eq:forced} along the radial direction, solving \eqref{eq:f-id} for $\tilde{f}^{\tr}$ and plugging it into \eqref{eq:forced} yields
\begin{equation}\label{762_Oscullating_Elements_Eqn}
\frac{d^{2}\tilde{r}}{d\tilde{\tau}^{2}} + \Gamma^{r}_{\rho\sigma}\frac{d\tilde{x}^{\rho}}{d\tilde{\tau}}\frac{d\tilde{x}^{\sigma}}{d\tilde{\tau}} = -\frac{\tilde{f}^{\phi}\tilde{u}_{\phi} + \tilde{f}^{\tilde{t}}\tilde{u}_{\tilde{t}}}{\tilde{u}_{\tilde{r}}}\,.
\end{equation}
The left hand side
\begin{multline}
\Gamma^{\tilde{r}}_{\rho\sigma}\frac{d\tilde{x}^{\rho}}{d\tilde{\tau}}\frac{d\tilde{x}^{\sigma}}{d\tilde{\tau}} =  \Gamma^{\tilde{r}}_{\ \tilde{r}\tr}\left(\frac{d\tilde{r}}{d\tilde{\tau}}\right)^{2} + \Gamma^{\tilde{r}}_{\ \phi\phi}\left(\frac{d\phi}{d\tilde{\tau}}\right)^{2} + \\ \Gamma^{\tr}_{\ \tilde{t}\tilde{t}}\left(\frac{d\tilde{t}}{d\tilde{\tau}}\right)^{2} + 2\Gamma^{\tr}_{\ \tilde{t}\phi}\frac{d\tilde{t}}{d\tilde{\tau}}\frac{d\phi} {d\tilde{\tau}} 
\end{multline}
is computed using the Kerr Christoffel symbols and the geodesic equations \eqref{CH:8 Radial Geodesic Equation}-\eqref{CH:8 Time Geodesic Equation}
\begin{align*}
\Gamma^{\tilde{r}}_{\rho\sigma}\frac{dx^{\rho}}{d\tilde{\tau}}\frac{d\tilde{x}^{\sigma}}{d\tilde{\tau}}  & = \frac{3(a\tilde{E} - \tilde{L})^{2}}{\tilde{r}^{4}} - \frac{a^{2}(\tilde{E}^{2} - 1) - \tilde{L}^{2}}{\tilde{r}^{3}} + \frac{1}{r^{2}} \\ 
& = -\frac{1}{2}\frac{\partial G}{\partial \tilde{r}}
\end{align*}
To evaluate the right hand side, $\tilde{f}^{\tilde{r}}$, we first notice the existence of two Killing vectors : $\xi^{\mu} = \partial/\partial \tilde{t}$ and $\psi^{\mu} = \partial/\partial\phi$, associated with time and angular translational invariance, respectively. There exists a conserved charge associated with each :
\begin{equation}\label{conserved_quantities}
\tilde{E} = -\xi^{\mu}u_{\mu}, \quad \tilde{L} = \psi^{\mu}u_{\mu}.
\end{equation}
It follows from Eq.\eqref{conserved_quantities} that $u_{\phi} = \tilde{L}$ and $u_{\tilde{t}}= -\tilde{E}$. Finally, we relate the proper time derivatives of these charges with the forcing terms in \eqref{eq:forced}. For example, consider  the proper time derivative of $\tilde{E}$ 
\begin{equation}
\begin{aligned}
    -\frac{d\tilde{E}}{d\tilde{\tau}} &= u^\beta\nabla_\beta(\xi^{\alpha}u_{\alpha}) \\
    &= \xi^{\alpha}(u^{\beta}\nabla_{\beta}u_{\alpha}) + u^{\alpha}u^{\beta}(\nabla_{\beta}\xi_{\alpha})\\
    & = \tilde{f}_{\tilde{t}}\,,
\end{aligned} 
\label{eq:dEdtau}
\end{equation}
where Killing's equation was used in the last step. A similar calculation leads to $d\tilde{L}/d\tilde{\tau} = \tilde{f}_{\phi}$. Solving the two equations $d\tilde{L}/d\tilde{\tau}$ and $d\tilde{E}/d\tilde{\tau}$ for $\tilde{f}^{\phi}$ and $\tilde{f}^{\tilde{t}}$ gives
\begin{align*}
    \tilde{f}^{\phi} &= -\frac{1}{\Delta}\left(g_{\tilde{t}\tilde{t}}\frac{d\tilde{L}}{d\tilde{\tau}} + g_{\phi \tilde{t}}\frac{d\tilde{E}}{d\tilde{\tau}}\right)\,, \\
     \tilde{f}^{\tilde{t}} &= \frac{1}{\Delta}\left(g_{\tilde{t}\phi}\frac{d\tilde{L}}{d\tilde{\tau}} + g_{\phi \phi}\frac{d\tilde{E}}{d\tilde{\tau}}\right). 
\end{align*}
where we used the identity $(g_{\phi \tilde{t}})^{2} - g_{\phi\phi}g_{\tilde{t}\tilde{t}} = \Delta$ for $\Delta = \tilde{r}^{2} - 2\tilde{r} + a^2$. Since $u_{\tilde{r}} = g_{\tilde{r}\tilde{r}}(d\tilde{r}/d\tilde{\tau})$, it follows the right hand side of \eqref{762_Oscullating_Elements_Eqn} is
\begin{equation*}
\tilde{f}^{\tilde{r}} = \frac{1}{\Delta g_{\tilde{r}\tilde{r}}}\bigg(\frac{d\tilde{E}}{d\tilde{\tau}}[g_{\phi \tilde{t}}\tilde{L} + g_{\phi\phi}\tilde{E}] +  \frac{d\tilde{L}}{d\tilde{\tau}}[g_{\tilde{t}\tilde{t}}\tilde{L} + g_{\tilde{t}\phi}\tilde{E}]\bigg)\left(\frac{d\tilde{r}}{d\tilde{\tau}}\right)^{-1}
\end{equation*}
Noticing that
\begin{align}
    \frac{1}{\Delta g_{\tilde{r}\tilde{r}}}(g_{\phi \tilde{t}}\tilde{L} + g_{\phi\phi}\tilde{E}) &= \frac{1
}{2}\frac{\partial G}{\partial \tilde{E}}\\
    \frac{1}{\Delta g_{\tilde{r}\tilde{r}}}( g_{\tilde{t}\tilde{t}}\tilde{L} + g_{\tilde{t}\phi}\tilde{E}) &= \frac{1}{2}\frac{\partial G}{\partial \tilde{L}}
\end{align}
we reach the desired conclusion
\begin{equation}
\tilde{f}^{\tr} = \frac{1}{2}\left(\frac{d\tE}{d\tilde{\tau}}\frac{\partial G}{\partial \tE} + \frac{d\tL}{d\tilde{\tau}}\frac{\partial G}{\partial \tilde{L}}\right)\left(\frac{d\tilde{r}}{d\tilde{\tau}}\right)^{-1}.
\end{equation}

\bibliographystyle{IEEEtran}
\bibliography{Transition_Inspiral_Paper}% Produces the bibliography via BibTeX.

\end{document}